\documentclass[12pt]{article}
\pdfoutput=1
\usepackage{epsf,amsfonts,amssymb,epsfig,amsmath,graphics,cite}
\usepackage[dvipsnames]{xcolor}
\usepackage{graphicx,subfig}
\usepackage{booktabs}
\usepackage{cancel}
\usepackage{float}
\usepackage{verbatim}
\usepackage{mathrsfs}
\usepackage{empheq}
\usepackage{enumerate}
\usepackage{bbm}
\usepackage{soul}
\usepackage{slashed}

\usepackage[colorlinks=true
,urlcolor=blue
,anchorcolor=blue
,citecolor=blue
,filecolor=blue
,linkcolor=blue
,menucolor=blue
,linktocpage=true
]{hyperref}

\usepackage{tikz}
\usetikzlibrary{cd}
 
\makeatletter

\newcommand{\Rmnum}[1]{\expandafter\@slowromancap\romannumeral #1@}
\makeatother

\newcommand{\nn}{\notag }
\def\be{\begin{equation}}
\def\ee{\end{equation}}

\newcommand{\ii}{\mathrm{i}}
\newcommand{\ex}{\mathrm{e}}
\newcommand{\diff}{\mathrm{d}}
\newcommand{\dd}{\mathrm{d}}

\newcommand{\vol}{\mathrm{vol}}

\newcommand{\cN}{\mathcal{N}}
\newcommand{\cE}{\mathcal{E}}
\newcommand{\cI}{\mathcal{I}}

\newcommand{\hm}{r}
\newcommand{\hs}{\bar s}

\newcommand{\hook}{\mathbin{\rule[.2ex]{.4em}{.03em}\rule[.2ex]{.03em}{.9ex}}}

\newcommand{\g}{\gamma}
\newcommand{\s}{\sigma}

\newcommand\cO{\mathcal{O}}

\newcommand{\tN}{{t_N}}
\newcommand{\tS}{{t_S}}

\newcommand{\nN}{{n_N}}
\newcommand{\nS}{{n_S}}

\newcommand{\vpvar}{\rho}

\newcommand{\newc}{x}

\numberwithin{equation}{section}       

\begin{document}

\begin{titlepage}

\vfill

\begin{flushright}
CCTP-2025-11\\
ITCP-2025-11
\end{flushright}


\begin{center}
   \baselineskip=16pt
   {\Large\bf 
Spindle solutions, hyperscalars and smooth uplifts}
  \vskip 1cm
Igal Arav$^1$,  Jerome P. Gauntlett$^2$, \\
Matthew M. Roberts$^{3}$ and Christopher Rosen$^4$\\
     \vskip 1cm     
                                                    \begin{small}
                                \textit{$^1$Instituut voor Theoretische Fysica, KU Leuven,\\
                                Celestijnenlaan 200D, B-3001 Leuven, Belgium}
        \end{small}\\
        \begin{small}\vskip .3cm
      \textit{$^2$Abdus Salam Centre for Theoretical Physics, 
  Imperial College\\ Prince Consort Rd., London, SW7 2AZ, U.K.}
        \end{small}\\
                \begin{small}\vskip .3cm
      \textit{$^3$    Science Institute, University of Iceland\\
    Dunhaga 3, 107 Reykjav\'ik, Iceland }
        \end{small}\\
              \begin{small}\vskip .3cm
           \textit{$^4$Crete Center for Theoretical Physics, Department of Physics, University of Crete,\\
71003 Heraklion, Greece}
        \end{small}\\
                       \end{center}
\vfill

\begin{center}
\textbf{Abstract}
\end{center}

\begin{quote}
We construct $AdS_3\times Y_7$ solutions of type IIB supergravity, where $Y_7$ is a smooth $S^5$ bundle over 
a spindle $\Sigma(n_N,n_S)$,
which are dual to $\mathcal{N}=(0,2)$ SCFTs in $d=2$. The solutions are constructed using the
$D=5$ STU $U(1)^3$ gauged supergravity theory coupled to a hyperscalar charged under $U(1)_B$.
We investigate spindle solutions with two new features: first, we allow $(n_N,n_S)$ to be non-coprime integers, including orbifolds of the round $S^2$, which can lead to non-unique, inequivalent uplifts, distinguished by the hyperscalar spectra, for given magnetic flux through the spindle.
Second, we also allow the hyperscalar to vanish at the poles leading to solutions
carrying non-vanishing $U(1)_B$ flux.
The new hyperscalar $AdS_3$ solutions can naturally arise as the endpoint of RG flows,
triggered by relevant hyperscalar deformations of the $AdS_3$ solutions of the STU model.
\end{quote}

\vfill

\end{titlepage}

\tableofcontents

\section{Introduction}
Spindles can be used to construct novel examples of the AdS/CFT correspondence preserving supersymmetry \cite{Ferrero:2020laf}.
Solutions of the form $AdS\times \Sigma$, where $\Sigma$ is a spindle, can be constructed in a lower-dimensional gauged
supergravity and then uplifted to $D=10,11$. Two particularly interesting features are
first, that the uplifted solutions can be completely regular, despite the orbifold singularities at the poles of 
the spindle. Second, that supersymmetry can be realised in two distinct ways: a twist and an
anti-twist \cite{Ferrero:2021etw}. For the twist, the Killing spinors
at the two poles have the same chirality, while for the anti-twist  they have opposite chiralities. The twist class is 
a generalisation of the standard topological twist on a two-sphere \cite{Maldacena:2000mw}, 
while the anti-twist class is new.

Many solutions have now been found, mostly in the context of gauged supergravity models
in $D=4,5,7$. 
Previous constructions have focussed\footnote{Some non-coprime examples
were considered in table 1 of
\cite{BenettiGenolini:2024kyy}, but we will see later that they do not uplift to regular solutions.}
on ``coprime spindles" where
the spindle  $\Sigma=\Sigma(n_N,n_S)$, which locally looks like $\mathbb{R}^2/\mathbb{Z}_{n_{N,S}}$ at the two
poles, has $n_N$ and $n_S$ coprime. Here we analyse ``non-coprime
spindles" with $n_N$ and $n_S$ having a common factor, which includes as a special case, orbifolds
of $S^2$ when $n_N=n_S$. We
analyse when the uplifted solutions can be regular orbibundles, with well defined spinors, generalising the analysis of \cite{Ferrero:2021etw}. In the coprime case, the orbibundles are uniquely determined by the suitably quantised 
magnetic fluxes
through the spindle. However, in the non-coprime case there is additional discrete flux data that needs to be specified
to fix the bundle. This feature leads to rich
new classes of uplifted non-coprime solutions that have been overlooked in previous constructions.

There are several constructions of spindle solutions in $D=4,5$ gauged supergravity models
that are coupled to vector multiplets e.g. \cite{Hosseini:2021fge,Boido:2021szx,Ferrero:2021etw,Ferrero:2021ovq,Couzens:2021rlk},
and there have also been some constructions with
charged hypermultiplets \cite{Arav:2022lzo,Suh:2022pkg,Suh:2023xse,Amariti:2023mpg,Hristov:2023rel}.
In previous work it has been assumed that any hyperscalar is a non-vanishing constant at both poles of the spindle.
If the hyperscalar is charged with respect to a ``broken" $U(1)_B$, then one can show that this 
restriction implies that the $U(1)_B$ magnetic flux through the spindle is necessarily zero, $p_B=0$ \cite{Arav:2022lzo}. This assumption turns out to be overly restrictive. Here we show that there are rich classes of solutions
in which 
the hyperscalars smoothly approach, in the orbifold sense, zero at the poles.  
Furthermore, such solutions have non-vanishing 
$U(1)_B$ magnetic flux through the spindle, $p_B\ne 0$, with its value determined by the boundary conditions of the hyperscalar at the poles.

In this paper we will revisit the $D=5$ gauged supergravity model considered in \cite{Arav:2022lzo,BenettiGenolini:2024kyy}, which is a consistent truncation of type IIB supergravity on $S^5$. It consists of the STU model, with $U(1)^3$ symmetry,
coupled to an additional complex hyperscalar, which comprises half of a hypermultiplet, and is charged under $U(1)_B\subset U(1)^3$.
The theory contains two $AdS_5$ vacua, one dual to $\mathcal{N}=4$ SYM theory, with the $U(1)^3$ symmetry a subgroup of the 
$SO(6)$ R-symmetry.
The second is dual to the LS $\mathcal{N}=1$ SCFT, for which $U(1)_B$ is no longer a symmetry and the remaining
$U(1)^2$ symmetry is a subgroup of the $SU(2)_F\times U(1)_R$ global symmetry.
There are known $AdS_3\times \Sigma$ solutions of the STU model (with vanishing hyperscalar) which exist in
both the anti-twist class \cite{Ferrero:2020laf,Hosseini:2021fge,Boido:2021szx,Ferrero:2021etw} and the twist class \cite{Ferrero:2021etw}. Here we will see that there is a rich landscape of new non-coprime spindles of the STU model, both
in the twist and the anti-twist classes. 
Our constructions also include orbifolds of the $S^2$ factor in the 
$AdS_3\times S^2$ solutions of \cite{Benini:2013cda}.
For the STU solutions we analyse BPS
fluctuations of the hyperscalar, which correspond to chiral operators in the dual $d=2$ SCFT. We will see that
there is a significant difference between the twist and the anti-twist solutions: for the twist solutions there
are, at most, a finite number of such modes, while for the anti-twist solutions we find an infinite number
of such modes.

In addition, we also construct, numerically, new $AdS_3\times \Sigma$ solutions
with non-vanishing hyperscalar, both in the coprime and the non-coprime class.
We find that they only exist in the anti-twist class and, moreover, we find that
the hyperscalar can vanish at just one of the poles of the spindle.

\begin{figure}[htbp!]
\begin{center}
\begin{tikzcd}[row sep=2cm, column sep=2cm]
\begin{tabular}{c}
$AdS_5$ \\
$\cN=4$~SYM
\end{tabular}
 \arrow[d] \arrow [rd,"\mathrm{Anti-Twist}"] 
 \arrow[r] 
 & \begin{tabular}{c}
$AdS_5$ \\
$\cN=1$~LS
\end{tabular} 
\arrow[d, "\substack{\mathrm{Anti-Twist, } \\  p_B=0}"]  \\
\begin{tabular}{c}
$AdS_3\times\Sigma$ \\
STU \\
Twist \& Anti-Twist
\end{tabular}
  \arrow[r, "\mathrm{Anti-Twist}" ]& \begin{tabular}{c}
$AdS_3\times\Sigma$ \\
Hyperscalar~$\neq 0$\\Anti-Twist
\end{tabular}\end{tikzcd}

\caption{
Possible RG flows between various solutions. We argue that all of the
$AdS_3\times \Sigma$ solutions with non-vanishing hyperscalar 
can be obtained from an RG flow from an $AdS_3\times \Sigma$ solution
of the STU model in the anti-twist class, with the same orbifold data.
We have also indicated how the solutions
could be related by RG flows across dimensions after compactifying $\mathcal{N}=4$ SYM or LS on a spindle with the same spindle data, with some subtleties noted in the text. 
}
\label{figone}
\end{center}
\end{figure}
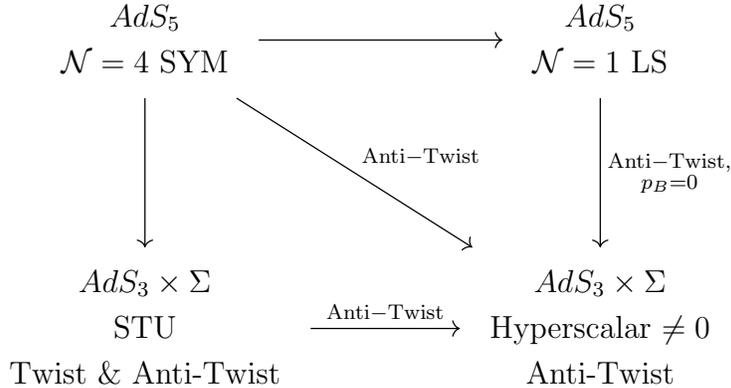

In figure \ref{figone} we indicate how the various $AdS_3$ solutions can be understood to arise from
RG flows. The top
horizontal line corresponds to the homogeneous RG flow between 
$\mathcal{N}=4$ SYM theory and the LS $\mathcal{N}=1$ SCFT \cite{Freedman:1999gp}.
The left vertical line corresponds to the RG flow across dimension when compactifying
$\mathcal{N}=4$ SYM theory on a spindle; the non-coprime STU solutions in the lower left corner are
discussed in this paper and are new. 
The lower right hand corner are the new 
$AdS_3\times \Sigma$ solutions with non-vanishing hyperscalar; 
 if the hyperscalar is non-vanishing at
both poles these have $p_B=0$, and in the coprime case
are the same as in \cite{Arav:2022lzo}, while the non-coprime cases are new.
If the hyperscalar vanishes at one of the poles, then $p_B\ne0$ and these solutions
are all new. We argue
that these hyperscalar solutions
can all be obtained from an RG flow, along the bottom horizontal line, starting from an STU model solution in the anti-twist class with the same
spindle data and driven by a relevant operator dual to the hyperscalar. 
In principle, the new solutions could also arise from a flow across dimensions along
a diagonal line coming from the $\mathcal{N}=4$ SYM fixed point, again with the same spindle data and a source 
for the operator dual to the hyperscalar. 
We should note, however, that there are subtleties\footnote{Specifically, constructing suitable
boundary Killing spinors is problematic unless additional boundary deformations are added; we will not have more to
say on this topic here.} regarding RG flows across dimensions for the anti-twist class \cite{Arav:2022lzo,Ferrero:2020twa}. 
The $AdS_3$ solutions with non-vanishing hyperscalar and $p_B=0$, might also arise from a flow across dimension starting from the LS fixed point, as indicated by the right-hand vertical line \cite{Arav:2022lzo}.  It is less clear that this will be possible for
the new $AdS_3$ solutions with $p_B\ne0$, since there is no $U(1)_B$ symmetry for the LS fixed point.

In a separate development, equivariant localisation has emerged as a powerful principle to extract
physical information from supersymmetric solutions of supergravity without having an explicit solution \cite{BenettiGenolini:2023kxp}.
In \cite{BenettiGenolini:2024kyy} this technique was applied to coprime $AdS_3\times \Sigma$ solutions of $D=5$ supergravity, including hyperscalar solutions with non-vanishing hyperscalar at the poles (i.e. $p_B=0$),
and it was shown how the central charge of the dual $\mathcal{N}=(0,2)$ SCFT can be obtained from an extremization
principle that extremizes a trial central charge over the weights of the R-symmetry Killing vector and other data at the poles of
the spindle. 

We reexamine all of the $AdS_3\times \Sigma$ solutions using the equivariant localization point of view. This
enables us to obtain an off-shell expression for the central charge, which after extremization gives the central charge
as well as the values of the vector multiplet scalars at the poles,
without solving the BPS equations. There are some positivity conditions which give rise to necessary conditions
for the existence of the solutions. Here, when $p_B\ne 0$, unlike in previous examples, these conditions are not sufficient to imply
the existence of the hyperscalar solutions, which we show by solving the BPS equations numerically.
We also provide strong numerical evidence
that sufficient conditions for the existence of hyperscalar solutions are obtained by demanding that, associated with the hyperscalar, there is a relevant 
operator
in the $AdS_3\times \Sigma$ solutions of the STU model that could drive an RG flow to 
the new $AdS_3\times \Sigma$ solutions with non-vanishing hyperscalar as indicated in the lower part of
figure \ref{figone}.

The plan of the rest of the paper is as follows. In section \ref{sec:d5sugramodelhypers} we introduce the supergravity model. In section \ref{secbpseqs} we analyse the BPS
equations directly, as well as discuss the regularity conditions for the uplifted spindle solutions in the non-coprime case,
for vanishing hyperscalar. Section \ref{sec:stusols} discusses STU solutions with vanishing hyperscalar, emphasising
new classes of
regular non-coprime solutions. In section \ref{seclinfluct} we analyse hyperscalar fluctuations about the STU solutions, for
both the twist and the anti-twist class, which gives rise to chiral operators in the dual SCFT, and this section also discusses
the additional conditions that need to be imposed to ensure the hyperscalar is a section of a regular line bundle.
Section \ref{sechyperscsol} analyses the $AdS_3\times \Sigma$ solutions with non-vanishing hyperscalar.
In section \ref{exasmplessec} we summarise some specific examples of $AdS_3\times \Sigma$ solutions with hyperscalars
as well as the spectrum of hyperscalar modes of $AdS_3\times \Sigma$ STU solutions. 
In section \ref{eqlocalsec} we analyse the solutions using equivariant localization and we conclude in section \ref{fincomms} with some discussion. We have included
six appendices with additional material: 
appendix \ref{appbpseqns} analyses the BPS equations and also discusses $AdS_3\times S^2$ solutions;
appendix \ref{app:WCP} analyses the conditions for regular circle orbibundles by viewing them as Seifert fibrations;
appendix \ref{appCmingaugedsugra} considers STU solutions of minimal gauged supergravity;
appendices \ref{appa} and \ref{s3app} comment on the conditions required for the hyperscalar to be a section of a smooth
line bundle over the spindle. Finally, appendix \ref{plotssols} includes some plots associated with some
representative examples of the new $AdS_3\times \Sigma$ hyperscalar solutions.

\section{The setup}\label{sec:d5sugramodelhypers}

\subsection{The $D=5$ supergravity model}
Our analysis will be in the context of a $D=5$ gauged supergravity theory, whose solutions can be uplifted on
$S^5$ to obtain exact solutions of type IIB supergravity.
It consists of the STU model, which can be viewed as an $\mathcal{N}=2$ gauged supergravity coupled to two vector multiplets, 
coupled to a complex, hyperscalar  which is half of a hypermultiplet. This theory was also considered in \cite{Bobev:2014jva,Arav:2022lzo,BenettiGenolini:2024kyy}, where more details can be found.

The bosonic Lagrangian in a mostly plus signature 
is given\footnote{\label{footconvs}
We have obtained this from \cite{Arav:2022lzo} by taking
$\alpha\to \frac{1}{2\sqrt{6}}\varphi^1$, $\beta\to -\frac{1}{2\sqrt {2}}\varphi^2$,
$\varphi\to\frac{1}{2}\vpvar$, $g_{\mu\nu}\to -g_{\mu\nu}$, $A\to\frac{c_1}{2 }A$,
$g\to 2c_2$, $\gamma^\mu\to \ii c_3\gamma^\mu$
with $c_i=\pm 1$ and $c_1 c_2=-1$, $ c_2 c_3=+1$. We will choose $c_3=+1$. We have also redefined $W\to-\frac{1}{2}W$, $\mathcal{P}\to \mathcal{V}/4$, $Q\to \frac{1}{2}Q$.}
 by
\begin{align}\label{d3lagoverall}
\mathcal{L}\, = \, \, 
\frac{1}{16\pi G}\sqrt{-g}&\Big[R-\mathcal{V}-\frac{1}{2}\sum_{i=1}^2(\partial \varphi^i)^2
-\frac{1}{4} \sum_{I=1}^{3}\left(X^{I}\right)^{-2}(F^{I})^{2}
\nn\\
&-\frac{1}{2}(\partial\vpvar)^2-\frac{1}{2}\sinh^2\vpvar (D\theta)^2\Big]\,.
\end{align}
Here $A^{I}$ are three $U(1)$ gauge fields, $I=1,2,3$, with field strengths $F^{I}=\diff A^{I}$. 
It will be convenient to define another basis for $U(1)^3\subset SO(6)$ with the associated gauge fields given by
\begin{align}\label{altbasis}
A_R=A^1+A^2+A^3\,,\qquad
A_B=A^1+A^2-A^3\,,\qquad
A_F=A^1-A^2\,,
\end{align}
which we refer to as the ``R-symmetry", the ``broken symmetry" and the ``flavour symmetry", respectively.
We will later consider solutions that also preserve $SU(2)_F\times U(1)^2\subset SO(6)$ symmetry which necessarily have 
$A_F=0$.

The $X^I$ are parametrised by two real scalars, $\varphi^1, \varphi^2$ in the vector multiplets via
\begin{align}\label{X5d}
X^{1}\, = \, \ex^{-\frac{\varphi^1}{\sqrt{6}}-\frac{\varphi^2}{\sqrt{2}}}\,, \qquad
X^{2}\, = \, \ex^{-\frac{\varphi^1}{\sqrt{6}}+\frac{\varphi^2}{\sqrt{2}}}\,, \qquad
X^{3}\, = \, \ex^{\frac{2\varphi^1}{\sqrt{6}}}\,,
\end{align}
and they satisfy the constraint 
\begin{align}\label{prepotdef}
\mathcal{F}(X^I)\equiv X^1 X^2 X^3=1\,,
\end{align}
where $\mathcal{F}$ is the prepotential.
The potential is 
\begin{align}\label{P5dhpyer}
\mathcal{V}=2\bigg[\sum_{i=1}^2(\partial_{\varphi^i} W)^2+(\partial_\vpvar W)^2\bigg]-\frac{4}{3}W^2\,,
\end{align}
where the real superpotential $W$ is 
\begin{align}
\label{superpottextLS}
W=\sum_{I=1}^3 X^I+\sinh^2\frac{\vpvar}{2}\, (\zeta_IX^I)\,, 
\end{align}
and the FI parameters are given by $\zeta_I=(1,1,-1)$. The hyperscalar, $\rho e^{i\theta}$,
is charged with respect to the broken symmetry $U(1)_B$ and
\begin{align}
D\theta=  d\theta -\zeta_I A^I=d\theta -A_B\,,
\end{align}
which is a gauge invariant quantity.

For a bosonic solution to preserve supersymmetry, we require 
\begin{align}\label{5d_susy1}
\Big[\nabla_\mu-\frac{\ii}{2}Q_\mu+\frac{1}{6}W\Gamma_\mu
+\frac{\ii}{24}\sum_I(X^I)^{-1}F_{\nu\rho}^{I}(\Gamma_\mu{}^{\nu\rho}-4\delta^\nu_\mu\Gamma^\rho)\Big]\epsilon&=0\,,\nn\\
\Big[\Gamma^\mu{\partial}_\mu\varphi^i-2\partial_{\varphi^i} W+\frac{\ii}{2}\sum_{I=1}^3\partial_{\varphi^i} \left(X^{I}\right)^{-1}\,{F}^{(I)}_{\mu\nu}\Gamma^{\mu\nu}\Big]\epsilon&=0 \,,\nn\\
\Big[\Gamma^\mu{\partial}_\mu\vpvar-2\partial_\vpvar W+2 \ii \partial_\vpvar Q_\mu\Gamma^\mu\Big]\epsilon&=0 \,,
\end{align}
where $\epsilon$ is a Dirac spinor, with the one-form $Q$ given by
\begin{align}\label{Qdefexp}
Q=A_R-\sinh^2\frac{\rho}{2}D\theta\,.
\end{align}
In particular, we see that the Killing spinor has charge 1/2 with respect to the R-symmetry $U(1)_R$ gauge field
$A_R$.

The model admits an $AdS_5$ vacuum solution with unit radius and vanishing scalar fields, which uplifts to the 
maximally supersymmetric $AdS_5\times S^5$ solution dual to $\mathcal{N}=4$, $SU(N)$ SYM theory. The $D=5$ Newton constant is given by $\frac{1}{G_5}=\frac{2N^2}{\pi}$ and the central charge is $a_{\mathcal{N}=4}=\frac{N^2}{4}$.
Associated with this solution, there is a consistent truncation to minimal gauged supergravity obtained by setting all
of the scalars to zero and setting $A^1=A^2=A^3$.
There is another $AdS_5$ vacuum with scalars given by
\begin{align}\label{LSvacvals}
\ex^{\frac{\sqrt{3}}{\sqrt{2}}\varphi^1}=2\, ,\qquad \varphi^2=0\, ,\qquad \ex^\vpvar=3\,,
\end{align}
and radius $L_{LS}=3/2^{5/3}$. After uplifting on $S^5$ to type IIB supergravity \cite{Pilch:2000ej}
this solution is dual to the $d=4$, $\mathcal{N}=1$ Leigh--Strassler (LS) SCFT \cite{Leigh:1995ep}; the latter 
arises as the IR limit of an RG flow from $\mathcal{N}=4$ SYM theory deformed by a mass deformation and the corresponding holographic solution was found in 
\cite{Freedman:1999gp}. The central charge of the LS SCFT, in the large $N$ limit, is given by $a_{LS}=\frac{27}{32}a_{\mathcal{N}=4}=\frac{27}{128}N^2$. Associated with the LS solution, there is another consistent truncation to minimal gauged supergravity obtained by setting all
of the scalars to their constant values \eqref{LSvacvals} and setting $A^1=A^2=\frac{1}{2}A^3$ (i.e. $A_B=A_F=0$).

\subsection{The \texorpdfstring{$AdS_3$ ansatz and BPS equations}{AdS3} }

 We consider the ansatz given by
 \begin{align}\label{ansatz5d2}
     \dd s_5^2 &= \ex^{2V}\dd s^2(AdS_3)+f^2dy^2+ h^2 dz^2\,,\nn\\
     A^I&=a^Idz\,,
\end{align}
with $V,f,h,a^I$ all functions of $y$ and we take $\Delta z=2\pi$. 
The two vector multiplet scalars $\varphi^1,\varphi^2$ are functions of $y$ while
the hyperscalar is of the form $\rho(y)e^{i\bar\theta z}$ with constant $\bar \theta$.
We will also utilise the following Poincar\'e coordinates for 
$AdS_3$ as well as the $D=5$ orthonormal frame
\begin{align}
\label{eq:5d_frame_ansatz}
e^0=e^V \frac{dt}{u},\quad
e^1=e^V \frac{d\phi}{u},\quad
e^2=e^V \frac{du}{u},\quad
e^3=f dy\,,\quad
e^4=h dz\,,
\end{align}
with $f,h\ge 0$. We are interested in solutions in which $y,z$ parametrise a two-dimensional spindle denoted by $\Sigma$.

We assume the Killing spinor $\epsilon$ has the form
\begin{align}\label{epsilon3plus2}
\epsilon = \psi\otimes \chi\, ,
\end{align}
where $\psi$ is a Killing spinor on $AdS_3$ and
$\chi$ is a spinor on $\Sigma$.
The $D=5$ gamma matrices can be written $\Gamma_i=\beta_i\otimes \gamma_3$, $\Gamma_{a+2}=1\otimes \gamma_a$
where $\beta_i$ and $\gamma_a$ are the $D=3$ and $D=2$ gamma matrices and $\gamma_3=-i\gamma_1 \gamma_2$.
The $D=3$ spinor satisfies 
\begin{align}\label{eq:3d_cks_eqt}
D_i \psi = \frac{\kappa}{2} \beta_i\psi,
\end{align}
with $\kappa=\pm 1$ determining the chirality of the preserved Poincar\'e supersymmetry of the 
SCFT dual to the $AdS_3\times \Sigma$ solution i.e. $\mathcal{N}=(2,0)$ or $ (0,2)$. 
After substituting this into the Killing spinor equations \eqref{5d_susy1} one obtains a set
of $D=2$ Killing spinor equations for $\chi$.

  Analysing in the same way as in \cite{Arav:2022lzo,Arav:2024exg}, we can determine the BPS equations for this ansatz. Some details are included in appendix \ref{appbpseqns}.
The spinor $\chi$ takes the form
\begin{align}\label{spinorsbps}
\chi = e^{\frac{i\hs z}{2}}e^\frac{V}{2} \begin{pmatrix}\sin\frac\xi 2 \\ \cos \frac \xi 2 \end{pmatrix}
\equiv e^\frac{V}{2}\zeta\,,
\end{align}
where $\hs$ is a constant. 
After an integration of the BPS equations, one finds the following expression for $h$:
\begin{align}\label{eq:h_int_solnt}
h=ke^V \sin\xi\,,
\end{align}
with $k$ a constant. For later use we note the
form of the following bilinears
\begin{equation}\label{5dbilinears}
    S \equiv \zeta^\dagger\zeta=1\,, \qquad P \equiv \zeta^\dagger\gamma_3\zeta=-\cos\xi\,, 
    \qquad \xi^\mu \equiv  -\ii\zeta^\dagger\gamma^\mu\gamma_3\zeta=\frac{1}{k}(\partial_z)^\mu\,.
\end{equation}

The remaining BPS equations can be written as
\begin{align}\label{bulkbps}
f^{-1}{\xi'}  =& \frac{Q_z-\hs}{ke^{V}}\,, \nn \\
f^{-1}{V'} = & -\frac{W}{3}\sin\xi\,,  \nn\\
f^{-1}{\varphi_i'} =& 2 \partial_{\varphi_i} W \sin\xi\,,  \nn\\
f^{-1}{\rho'}= & \frac{2 \partial_\rho W}{\sin\xi}\,,
\end{align}
along with the constraint equations 
\begin{align}\label{bpsconsts}
Q_z-\hs &= k \left(2 \kappa-We^V \cos\xi \right)\,,\nn\\
\partial_\rho Q_z&=-k e^V\partial_\rho W \cos\xi\,.
\end{align}

Furthermore, one also finds that the field strengths for the gauge fields can be written in the form (no sum on $I$)
\begin{align}\label{expsforFs}
(X^I)^{-2}F^I_{34}=\frac{2\kappa}{e^VX^I}-2\cos\xi-2\cos\xi\sinh^2\frac{\rho}{2}\zeta_I\,.
\end{align}
Using the BPS equations one can then deduce the important relations
\begin{align}\label{eq:gauge_bpst}
(a^I)' =(\cI^I)'\,,
\end{align}
where we have defined
\begin{align}\label{eq:gauge_bpst2}
 \cI^I\equiv -k x^I\,,
\end{align}
and the ``dressed scalars" $x^I$ are given by
\begin{align}\label{dressed}
x^I\equiv \cos\xi e^V X^I\,.
\end{align}

By analysing the equations of motion for the gauge fields, given by
\begin{align}
(e^{3V}(X^I)^{-2}F^I_{34})'=-fh^{-1}e^{3V}\sinh^2\rho (D\theta)_z\zeta_I\,,
\end{align}
(no sum on $I$), we can obtain further information.
For vanishing hyperscalar, all three can be immediately integrated and using \eqref{expsforFs}
we deduce 
\begin{align}\label{threeconschges}
\rho = 0:\qquad  \frac{1}{2\kappa}\cE^I  =\cos\xi {e^{3V}}\left(\frac{1}{x^I} - \kappa\right)\,,
\end{align}
 where $\cE^I$ are constants. For non-vanishing hyperscalar, only two linear combinations of
 the gauge field equations of motion can be integrated and we find
 \begin{align}\label{twoconschges}
\frac{1}{2\kappa}\cE_R &= \cos\xi e^{3V}\left[ \left(\frac{1}{x^1}+\frac{1}{x^2}+\frac{2}{x^3} \right)-4\kappa \right]\,,\\ \nn
\frac{1}{2\kappa}\cE_F &= \cos\xi e^{3V} \left(\frac{1}{x^1}-\frac{1}{x^2} \right)\,,
\end{align}
where $\cE_R=\cE^1+\cE^2+2\cE^3$ and $\cE_F =\cE^1-\cE^2$ are constants.

\section{Analysis of BPS equations}\label{secbpseqs}
In this section we discuss some aspects of the boundary conditions for the BPS equations and obtain
an expression for the central charge.
We also discuss necessary conditions for the uplifted solutions associated with 
non-coprime spindles to give rise to regular solutions with well-defined spinors. There
are additional regularity conditions when the hyperscalar is non-trivial, which are discussed in later sections.

\subsection{Some boundary conditions and the central charge}\label{somebcs}
We now restrict to conformal gauge
\begin{align}\label{confgauge}
f=e^V\,,
\end{align}
and then also using \eqref{eq:h_int_solnt}, find the $D=5$ metric can be written as
\begin{align}\label{confgaugemetric}
 \dd s_5^2 &= \ex^{2V}\left(\dd s^2(AdS_3)+dy^2+ k^2\sin^2\xi dz^2\right)\,.
\end{align}

We will assume $y\in[y_N,y_S]$ with $y_N<y_S$, so that $y,z$ parametrise a compact spindle.
The boundary conditions for the $\mathbb{R}^2/\mathbb{Z}_{n_{N,S}}$ orbifold singularities at the poles of the 
spindle implies
\begin{align}\label{firstbcsec}
k\sin\xi &= \frac{1}{ n_{N}}(y-y_{N})+\dots\,,\qquad k\sin\xi = \frac{1}{ n_{S}}(y_S-y)+\dots\,,\nn\\
\cos\xi|_{N,S} &= (-1)^{t_{N,S}} ,\qquad \qquad \quad \qquad\text{$t_{N,S}=0,1$}\,.
\end{align} 
Here we have taken $k\sin\xi\ge 0$, but no sign choice is made for $k$.
This can be accomplished by taking
\begin{align}
\xi = t_N \pi + \frac{(-1)^{t_N}}{k n_N} (y-y_N) +\dots\,,\qquad
\xi = t_S \pi + \frac{(-1)^{t_S}}{k n_S} (y_S-y) +\dots\,,
\end{align}
at the two poles.
From the first equation of \eqref{bulkbps} we deduce
\begin{align}\label{eq:s_Qz_bc}
(Q_z-\hs)|_{N} = \frac{(-1)^{t_N}}{n_N}\,,\qquad
(Q_z-\hs)|_{S} = \frac{(-1)^{t_S+1}}{n_S}\,,
\end{align}
while \eqref{bpsconsts} implies
\begin{align}\label{eq:genl_Wbdy}
\frac{(-1)^{t_N}}{k n_N} = 2 \kappa - (-1)^{t_N} e^V W|_{N}\,,\qquad
\frac{(-1)^{t_S+1}}{k n_S} = 2 \kappa - (-1)^{t_S} e^V W|_{S}\,,
\end{align}
which constrains the value of the dressed scalar fields at the poles.

From the values of dressed scalars \eqref{dressed} at the poles, $x^I_{N,S}$, we can determine the values of 
$e^{V}$ at the poles (since $\cos^3\xi e^{3V}=\mathcal{F}(x^I)\equiv x^1 x^2 x^3$) as well as the vector multiplet
scalars in \eqref{X5d} at the poles.
We can also obtain expressions for the central charge and fluxes.
For the former, first recall the central charge of the $d=2$ SCFT can be obtained from the $D=3$ Newton constant via
$c = \frac{3}{2}\frac{1}{G_3}$. In turn $G_3$ can be obtained from the $D=5$ Newton constant
 \begin{align}
\frac{1}{G_3} =\frac{1}{G_5} \Delta z \int_N^S e^V f h dy, \qquad 
\end{align}
where $\Delta z=2\pi$, and we have used $f,h>0$. We have $\frac{1}{G_5}=\frac{2N^2}{\pi}$ so that the $D=5$ $AdS_5$ vacuum is dual to $SU(N)$ 
$\cN=4$ Yang-Mills theory. Also, the BPS equations imply
\begin{align}
e^V f h= -\frac{k\kappa}{2} (e^{3V} \cos\xi)'\,,
\end{align}
and so we obtain 
\begin{align}\label{eq:central_charge_Vxi}
c={3k\kappa }[  \mathcal{F}(x^I_N)- \mathcal{F}(x^I_S) ] N^2\,.
\end{align}

The quantised magnetic field fluxes of the three gauge fields are defined by
\begin{align}\label{infbrflcon}
\frac{1}{2\pi}\int _\Sigma F^I\equiv \frac{p^I}{n_Nn_S}\,,
\end{align}
with $p^I\in \mathbb{Z}^3$, as we discuss further in section \ref{smoothupliftssec}. 
From \eqref{eq:gauge_bpst}, \eqref{eq:gauge_bpst2} we have
$\frac{p^I}{n_Nn_S}=\left. \cI^I\right|^S_N$
and so we can write
\begin{align}\label{eq:pi_xit}
 \frac{p^I}{n_N n_S} = k \left( x^I_N -  x^I_S \right).
\end{align}
For fixed $p^I$ this constrains the value of the dressed scalar fields at the poles and $k$.

The above results have direct analogues in the analysis of section \ref{eqlocalsec} that uses equivariant localization, generalising
\cite{BenettiGenolini:2024kyy}, and leads to a result for the off-shell central charge that can be extremized to get the on-shell result. 
It is also possible to utilise the gauge equations of motion to get an on-shell expression for the central charge directly, without
an extremization principle, as we explain later. To do so one exploits the fact
that there are additional conserved charges.
For vanishing hyperscalar, $\rho=0$, we have three conserved charges \eqref{threeconschges},
which have the same value at the two poles, and
so we get three more constraints on $x^I_{N,S}$:
\begin{align}\label{eq:constraint_eqt}
\rho=0:\qquad\frac{\cE^I}{2 \kappa} = \mathcal{F}(x^I_N) \left( \frac{1}{x^I_N }- \kappa \right)
= \mathcal{F}(x^I_S) \left( \frac{1}{x^I_S }- \kappa \right)\,.
\end{align}
For the associated STU model solutions one can then obtain an expression for the
central charge in terms of $n_{N,S}$, $t_{N,S}$ and the fluxes $p^I$ (see \eqref{stufullsoln}).
For $\rho\ne 0$, we have two conserved charges $\cE_R $, $\cE_F $ \eqref{twoconschges},  giving two additional constraints. For the corresponding
solutions one can obtain an analogous expression for the central charge (given in \eqref{localcetc}) that, in addition,
also depends on the behaviour
of the hyperscalar at the poles of the spindle, as we shall see.

\subsection{Smooth uplifts}\label{smoothupliftssec}

We are interested in obtaining smooth solutions of type IIB supergravity after uplifting
on $S^5$. The type IIB solutions will be of the form $AdS_3\times Y_7$, with
$Y_7$ an $S^5$ bundle over the spindle $\Sigma[n_N, n_S]$. This bundle can be constructed by first
considering the construction of smooth circle bundles over the spindle. Doing this
for each of the three $U(1)$'s in the $D=5$ gauged supergravity, we can then consider the three
associated complex line bundles. The $S^5$ can be embedded in the $\mathbb{C}^3$ fibre of the
latter and we may then
form the associated $S^5$ bundle over $\Sigma[n_N, n_S]$ by using the same $U(1)^3$ 
transition functions of the three line bundles.

Previous work has focussed on ``coprime spindles" with $(n_N,n_S)$ coprime integers. Here we also consider ``non-coprime spindles" with $\text{hcf}(n_N,n_S)\ne 1$. The presence of the hyperscalar also introduces additional features, which we discuss in later sections.

In appendix \ref{app:WCP} we carry out an analysis of circle bundles over both 
coprime spindles and non-coprime spindles, extending the analysis of \cite{Ferrero:2021etw}.  
Here we consider two patches for a $U(1)^3$ orbibundle, covering the north and south poles, with
$\psi^I_{N,S}$, coordinates on the $S^1$ fibres in each patch
with $\Delta \psi_{N,S}^I=2\pi$. We choose a gauge where the connection one-forms, $A^I$, are 
not regular at the poles of each patch, but instead  have flat connection pieces that capture the
orbifold data. Evaluating the one-form at the poles in the two patches we have
\begin{align}\label{Agaugemn}
A^I|_N\to \frac{m^I_N}{n_N}dz\,,\qquad
A^I|_S\to \frac{m^I_S}{n_S}dz\,,
\end{align}
with $m^I_N\in\mathbb{Z}_{n_N}$, $m^I_S\in\mathbb{Z}_{n_S}$. 
Furthermore, the gauge fields in the two patches are related by a $U(1)^3$ gauge transformation on the overlap of the patches via
\begin{align}\label{gtgfs}
A^I|_{\text{$N$ patch}}=A^I|_{\text{$S$ patch}}+\gamma^I dz\,,
\end{align}
with $\gamma^I\in \mathbb{Z}^3$. 
On the total space of the orbibundle, $(d\psi^I+A^I)$ are smooth global one-forms; the gauge transformation
\eqref{gtgfs} is implemented by identifying the angular coordinates $(\psi^I_N,z)$ with $(\psi^I_S-\gamma^I z,z)$ on the overlap (and reversing the orientation).
Using Stokes' theorem, \eqref{Agaugemn} and \eqref{gtgfs} imply that
the flux of the gauge field through the spindle defined in \eqref{infbrflcon}
is given by\footnote{Note that $p^{here}=\lambda^{there}$ and $\gamma^{here}=p^{there}$ in
\cite{Ferrero:2021etw}.}
\begin{align}\label{condfluxsmooth2}
p^I=n_Nm^I_S -n_Sm^I_N  +\gamma^I n_N n_S \in\mathbb{Z}\,.
\end{align}
Importantly, for the total space of the circle orbibundle to be smooth it is necessary and sufficient 
that the coprime conditions:
\begin{align}\label{coprimecondsms}
\mathrm{hcf}(m^I_N,n_N)=1 \qquad \text{and}\qquad \mathrm{hcf}(m^I_S,n_S)=1\,,
\end{align}
are satisfied \cite{Ferrero:2021etw} for each $I$. As a consequence we must have
\begin{align}\label{coprimelamndannns}
\mathrm{hcf}(p^I,n_N)=\mathrm{hcf}(p^I,n_S)=\mathrm{hcf}(n_N,n_S)\,,
\end{align}
for each $I$.

We highlight that the key condition \eqref{condfluxsmooth2},
with $m^I_N\in\mathbb{Z}_{n_N}$, $m^I_S\in\mathbb{Z}_{n_S}$ and
satisfying \eqref{coprimecondsms}, can also be written in
the form
\begin{align}\label{condfluxsmoothagain}
p^I=n_N(m^I_S+\gamma^I n_S) -n_Sm^I_N  =n_Nm^I_S -n_S(m^I_N -\gamma^I n_N)
\,.
\end{align}
Thus, by considering $m^I_N\in\mathbb{Z}$, $m^I_S\in\mathbb{Z}$, and satisfying \eqref{coprimecondsms},
we can effectively eliminate $\gamma^I$ and demand
\begin{align}\label{condfluxsmoothagain2}
p^I=n_Nm^I_S -n_Sm^I_N\,.
\end{align}
However, in this section we continue with 
$m^I_N\in\mathbb{Z}_{n_N}$, $m^I_S\in\mathbb{Z}_{n_S}$.

We first recall the most studied case of coprime spindles with $\mathrm{hcf}(n_N,n_S)=1$. In this case smoothness is equivalent to the condition that $\mathrm{hcf}(p^I,n_N)=\mathrm{hcf}(p^I,n_S)=1$.
Moreover, if one specifies $(n_N,n_S,p^I)$, then the discrete fluxes $m^I_N\in\mathbb{Z}_{n_N}$ and $m^I_S\in\mathbb{Z}_{n_S}$ satisfying \eqref{condfluxsmooth2} are uniquely fixed and, moreover,
satisfy \eqref{coprimecondsms}. Thus, for given 
$\mathrm{hcf}(n_N,n_S)$, the bundle is uniquely determined by specifying the magnetic fluxes $p^I$.
If one focusses on just a single $U(1)$, the total space of the orbibundle is a specific Lens space, which was identified in \cite{Ferrero:2020twa} and also discussed in appendix \ref{app:WCP}.
 
 We next consider the case of non-coprime spindles with $\mathrm{hcf}(n_N,n_S)=h\ne 1$.
 If the fluxes for the spindles is $p^I$ then we can show that they necessarily arise as the 
 $h$-fold ``flux quotient" of a coprime spindle in the following sense.
 We must have 
\begin{align}\label{noncoprimecaseh}
(n_N,n_S,p^I)=h(\hat n_N, \hat n_S,\hat p^I)\,,
\end{align}
with 
$(\hat n_N, \hat n_S,\hat p^I)$ specifying a smooth orbibundle for a coprime spindle
with the conditions $\mathrm{hcf}(\hat n_N,\hat n_S)=\mathrm{hcf}(\hat p^I, \hat n_S)=\mathrm{hcf}(\hat p^I, \hat n_N)=1$, for each $I$. 
Notice that the fluxes for this non-coprime spindle can be expressed as
\begin{align}\label{fluxexpgensd}
\frac{1}{2\pi}\int _\Sigma F^I=\frac{p^I}{n_Nn_S}=\frac{\hat p^I}{h \hat n_N \hat n_S}\,,
\end{align}
which is the origin of the term ``flux quotient". Importantly, we find that for given 
$(n_N,n_S,p^I)$, the discrete fluxes $m^I_N\in\mathbb{Z}_{n_N}$ and 
$m^I_S\in\mathbb{Z}_{n_S}$ satisfying \eqref{condfluxsmooth2} and \eqref{coprimecondsms}
are, in general, no longer uniquely fixed. Thus, in addition to the fluxes $p^I$, we must also 
give the specific values of $m^I_N\in\mathbb{Z}_{n_N}$, $m^I_S\in\mathbb{Z}_{n_S}$ to specify
the bundle. The total space of the orbibundles for non-coprime spindles for a single $U(1)$ are again
Lens spaces, which are identified in appendix \ref{app:WCP} by viewing them as Seifert fibrations.
In general, for given fluxes $p^I$, different 
$m^I_{N,S}$ satisfying \eqref{condfluxsmooth2}, \eqref{coprimecondsms} give rise to inequivalent Lens spaces.

Interestingly, not every  $h$-fold flux quotient of a smooth orbibundle for a coprime spindle 
 will be smooth i.e. there are no solutions for the $m^I$ satisfying \eqref{condfluxsmooth2}, \eqref{coprimecondsms}.
 In particular, if $(\hat n_N, \hat n_S, \hat p^I)$ is a smooth coprime spindle 
 then there is no smooth $h$-fold flux quotient with $(n_N,n_S,p^I)=h(\hat n_N, \hat n_S,\hat p^I)$ when
$\hat n_N$, $\hat n_S$ and any one of the $\hat p^I$ are all odd and $h$ is even,
as we show in appendix \ref{app:WCP}. Although we have not proven
it, we believe that this is the only case that is obstructed. 

 In addition we need to take into account that we have spinors that are charged under the R-symmetry gauge field
 $A_R=A^1+A^2+A^3$, have definite charge with respect to the Killing vector on the spindle, $\partial_z$, and are non-vanishing.
 As explained in \cite{Ferrero:2021etw}
 this implies that supersymmetry can preserved in one of two ways called the ``twist" and the ``anti-twist". The spinors are
 necessarily chiral at the poles and in the twist case they have the same chirality (i.e. $(t_N,t_S)=(0,0)$ or $(1,1)$ in \eqref{firstbcsec})
 while the anti-twist case they have opposite chirality (i.e. $(t_N,t_S)=(0,1)$ or $(1,0)$ in \eqref{firstbcsec}).
 This places a constraint on the R-symmetry flux $p_R=p^1+p^2+p^3$, 
 \begin{align}\label{susyprcond}
\frac{p^R}{n_N n_S} = \frac{(-1)^{t_N+1}}{n_N}+ \frac{(-1)^{t_S+1}}{n_S}\,,
\end{align}
which also arises from the BPS equations, as we discuss below.  
As discussed in \cite{Ferrero:2021etw} we should again introduce two patches to 
discuss regularity of the spinors. First recall the BPS
equation \eqref{eq:s_Qz_bc}. We also recall the definition of 
$Q$ in \eqref{Qdefexp}: when $\rho=0$ at the poles clearly $Q=A_R$ at the poles.
As in \cite{Arav:2022lzo}, and as we discuss later, when $\rho\ne 0$ at a
pole we must have
$D\theta=0$ at the pole and so again $Q=A_R$ at the pole. Thus, using the gauge
choice \eqref{Agaugemn} that we used for describing the $U(1)^3$ bundle, in the 
$N$ and $S$ patches we can take
\begin{align}\label{rsymms}
\frac{m^R_N}{n_N}-\bar s_N=\frac{(-1)^\tN}{n_N}\,,
\qquad
\frac{m^R_S}{n_S}-\bar s_S=\frac{(-1)^{\tS+1}}{n_S}\,,
\end{align}
where $m^R\equiv \sum_Im^I$ and $\bar s_N$, $\bar s_S$ are the phases of
the Killing spinor in the two patches and $\bar s_N=\bar s_S+\gamma^R$ 
on the 
overlap of the two patches, with $\gamma^R=\sum_I\gamma^I$ and $\gamma^I$
as in \eqref{gtgfs}. 
Note that to be consistent with the orbifold identifications 
we should take $\bar s_{N,S}\in \mathbb{Z}$ (this was not emphasised in
\cite{Ferrero:2021etw}  but follows from the discussion around (2.27) and (2.28) of
\cite{Ferrero:2021etw}.)

For coprime spindles with $\text{hcf}(n_N,n_S)=1$, \eqref{rsymms} is not an extra condition that
needs to be satisfied, as we now show. From \eqref{condfluxsmooth2} we have 
$p^R=n_N(m^R_S+\gamma^R n_S) -n_Sm^R_N$. Then using the lemma below \eqref{condfluxsmooth3}, we know that with $\text{hcf}(n_N,n_S)=1$ there are always solutions
for $m_{N,S}$ to this equation with\footnote{The lemma also says $\text{hcf}(m^R_N,n_N)=\text{hcf}(p^R,n_N)$ and $\text{hcf}(m^R_S,n_S)=\text{hcf}(p^R,n_S)$. Now for coprime $n_{N}, n_S$
we have $\text{hcf}(p^I,n_N)=\text{hcf}(p^I,n_S)=1$ for each $I$, but it doesn't then follow that
$\text{hcf}(p^R,n_N)=\text{hcf}(p^R,n_S)=1$. However, the argument we make here shows that
\eqref{susyprcond} implies $m^R_N={(-1)^{t_N}} $ mod $n_N$ and 
$m^R_S={(-1)^{t_S+1}} $ mod $n_S$ and so in fact we do have
$\text{hcf}(m^R_N,n_N)=\text{hcf}(p^R,n_N)=1$ and 
$\text{hcf}(m^R_S,n_S)=\text{hcf}(p^R,n_S)=1$.} $m^R_N$ unique mod $n_N$ and 
$m^R_S$ unique mod $n_S$.
Then the imposition of the condition 
\eqref{susyprcond}, $p^R={(-1)^{t_N+1}}{n_S}+ {(-1)^{t_S+1}}{n_N}$, implies that we must have 
$m^R_N={(-1)^{t_N}} $ mod $n_N$ and 
$m^R_S={(-1)^{t_S+1}} $ mod $n_S$, which
is equivalent to \eqref{rsymms}.
On the other hand for non-coprime spindles, in general, 
\eqref{rsymms} is an additional condition that needs to be imposed to have well-defined spinors.

For solutions with non-vanishing hyperscalar, we also need to ensure that the hyperscalar, 
which is charged with respect to the gauge field
$A_B=A^1+A^2-A^3$, is a section on the associated line orbibundle over the spindle.
This gives an additional constraint on the flux $p_B$ and, in the non-coprime case, also constrains
$m^I_{N,S}$ via a constraint on
$m^B\equiv m^1+m^2-m^3$ at the poles, as we discuss in later sections.

\section{STU solutions}\label{sec:stusols}
Solutions of the STU model are obtained when the hyperscalar is set to zero. 
Analytic solutions for the STU model are given in \cite{Ferrero:2021etw} (see also \cite{Ferrero:2020laf,Hosseini:2021fge,Boido:2021szx}).
We now discuss how one can obtain the central charge without the explicit solution and without extremizing
(as in section \ref{eqlocalsec} and in \cite{BenettiGenolini:2024kyy}). We also comment on the new
non-coprime spindles.

When $\rho=0$ we have $\cos\xi e^V W_{STU} =\sum_I x^I$,
so that \eqref{eq:genl_Wbdy} implies that the dressed scalars are constrained via
\begin{align}\label{eq:stu_k_x123t}
\sum_I x^I_{N}=2 \kappa -\frac{(-1)^{t_N}}{kn_N}\,,
\qquad
\sum_I x^I_{S}=2 \kappa +\frac{(-1)^{t_S}}{kn_S}\,.
\end{align}
Using this and \eqref{eq:pi_xit} we deduce that the R-symmetry flux is indeed given by
\begin{align}\label{conststuflux}
\frac{p^R}{n_N n_S} = \frac{(-1)^{t_N+1}}{n_N}+ \frac{(-1)^{t_S+1}}{n_S}\,,
\end{align}
as noted in the previous section.

We first consider coprime spindles. In this case, for given $\kappa$, 
the spindle data can be taken to be
$n_{N,S}$, $t_{N,S}$ and the fluxes $p^I$, constrained by \eqref{conststuflux} and all coprime to both $n_N$ and $n_S$, so we have
7 undetermined quantities given by $x^I_{N,S}$ and $k$. We can eliminate $x^I_{S}$ from \eqref{eq:pi_xit}
and then there is one constraint on $x^I_{N}$ and $k$ from \eqref{eq:stu_k_x123t}, leaving three undetermined parameters (as in
section \ref{eqlocalsec}). The expressions 
for the conserved charges \eqref{eq:constraint_eqt} gives us three more constraints which allows us to solve\footnote{One finds 
three (nonlinear) equations relating $x^I_N$ to $p^I,~n_{N,S},~k$. These can be solved similarly to \cite{Arav:2024exg}.} 
for $x^I_{N,S}$ and $k$. 
Defining 
\begin{align}\label{defs}
s=n_N^2+n_S^2-\sum_I(p^I)^2\,,
\end{align}
we find the following results for the dressed scalars at the poles, the central charge and $k$:
\begin{align}\label{stufullsoln}
x^I_N&=    -\frac{2\kappa}{s}((-1)^{t_S}n_N+ p^I) p^I                 \,,\nn\\
x^I_S&=    -\frac{2\kappa}{s}((-1)^{t_N}n_S+ p^I) p^I                 \,,\nn\\
c&=6\kappa N^2 \frac{ p_1p_2p_3}{\nN \nS s}\,,\nn\\
k&=\frac{\kappa s}{2 n_N n_S(n_S(-1)^{t_N}-n_N(-1)^{t_S})}\,,
\end{align}
in agreement with the explicit solutions found in \cite{Ferrero:2021etw}. 

We briefly pause to note that $AdS_3\times S^2$ solutions of the STU model, with a homogeneous metric on $S^2$, 
can also be obtained, recovering the results of \cite{Benini:2013cda}. 
The sphere solutions only exist in the twist class with
$t_N=t_S$. Note if we set
$n_N=n_S=1$ in \eqref{stufullsoln},
we see that $1/k\to 0$. However, if one takes this limit carefully (see section 2.4 of \cite{BenettiGenolini:2024kyy}),
or explicitly solves the BPS equations as in appendix \ref{spherecasebps},
one finds that the sphere case only exists in the twist class $t_N=t_S$, with $x_N^I=x_S^I$ constant on $S^2$ 
and
\begin{align}\label{stufullsolnS2}
x^I&=    -\frac{2\kappa}{s}((-1)^{t_N}+ p^I) p^I                 \,,\nn\\
c&=3{\kappa}N^2(p^1x^2 x^3 +p^2 x^1x^3 +p^3 x^1 x^2 )\nn\\
&=
6\kappa N^2 \frac{ p_1p_2p_3}{2-(p_1^2+p_2^2+p_3^2)}\,,
\end{align}
along with the topological twist constraint $p_R\equiv p^1+p^2+p^3=2(-1)^{t_N+1}$.
If $\kappa=+1$, for example, we have solutions with $(-1)^t x^I>0, c>0$, provided that two
$p^I>0$ and $p_R=2(-1)^{t_N+1}$.

Returning to spindles, necessary conditions for an STU solution to exist are given by the conditions
\begin{align}
\label{eq:stu_positivity}
(-1)^{t}x^{1,2,3}|_{N,S} > 0 , \qquad c > 0.
\end{align}
Focussing on $\kappa=+1$, for example,
we find the following necessary conditions for anti-twist solutions:
\begin{align}
\text{Anti-twist}:\qquad &(t_N,t_S)=(1,0):\qquad p^I>0\,,\quad p_R=n_S-n_N\,,\nn\\
&(t_N,t_S)=(0,1):\qquad p^I>0\,,\quad p_R=n_N-n_S\,.
\end{align}
We also find that necessary conditions for twist solutions 
are given by
\begin{align}\label{twistexamples}
\text{Twist}:\qquad \qquad &(t_N,t_S)=(0,0):\qquad \text{two $p^I>0$}\,,\quad p_R=-n_S-n_N\,,\nn\\
&(t_N,t_S)=(1,1):\qquad \text{no solutions}\,.
\end{align}
This is in agreement with the explicit solutions constructed\footnote{For the twist solutions
\eqref{twistexamples}, we have $e^{3V_{N,S}}=\mathcal{F}(x^I_{N,S})$. 
When $\kappa=+1$ \eqref{stufullsoln} implies that $s<0$ and the sign of $k$
is the same as the sign of $n_N-n_S$. Then from \eqref{eq:central_charge_Vxi} we see that
the sign of  $e^{3V_N}-e^{3V_S}$ is the same as the sign of $n_N-n_S$.
This is the origin of the extra condition $n_1<n_2$ in (3.31) of \cite{Ferrero:2021etw}, which
assumed, in the setup there, that $e^{3V_1}<e^{3V_2}$.}
in \cite{Ferrero:2021etw}. In other words, the necessary conditions \eqref{eq:stu_positivity} are in fact sufficient for the
existence of solutions in the case of the STU model.

 We now consider non-coprime spindles. In this case the spindle data can be taken to be
$n_{N,S}$, $t_{N,S}$ and fluxes $p^I$, constrained by \eqref{conststuflux}, as well as 
$m^I_N\in\mathbb{Z}_{n_N}$, coprime to $n_N$,
and $m^I_S\in\mathbb{Z}_{n_S}$ coprime to $n_S$, and also consistent with
\eqref{rsymms} to have regular spinors. Otherwise, the analysis is exactly as above and in particular, the solutions
with the same $n_{N,S}$, $t_{N,S}$ and $p^I$, but different $m^I_N$ and $m^I_S$ will have exactly
the same central charge (in the large $N$ limit that we are studying). The different non-coprime spindle
solutions are obtained from the construction of the explicit solutions in \cite{Ferrero:2021etw} by just inserting suitable
discrete fluxes into the gauge fields. 
The solutions with the same central charge but different fluxes uplift to different type IIB solutions and are
dual to different SCFTs; one way in which they can be distinguished is by considering their spectrum.
In the next section we will determine the scaling dimension of operators dual to the charged hyperscalar for these
different solutions. 
Recall from \eqref{noncoprimecaseh} that for a non-coprime spindle we have
$(n_N,n_S,p^I)=h(\hat n_N, \hat n_S,\hat p^I)$ with $(\hat n_N, \hat n_S,\hat p^I)$ specifying a coprime spindle. Hence
from the expression for the central charge in \eqref{stufullsoln} we deduce that the central charges are related by
\begin{align}
c(n_N,n_S,p^I)=\frac{1}{h}c(\hat n_N, \hat n_S,\hat p^I)\,.
\end{align}

We now illustrate with some examples. Before doing so, we note that generically
the STU solutions will preserve $U(1)^3$ symmetry. However, a subset will preserve $SU(2)_F\times U(1)^2$
symmetry, after uplifting on $S^5$. Necessarily these solutions have fluxes with $p^1=p^2$ so that $p_F=0$. For coprime spindles.
we necessarily have the uniquely specified $m_N^I$, $m_S^I$ satisfying 
$m_N^1=m_N^2$ and $m_S^1=m_S^2$. However, for non-coprime cases it is possible that
the $m_N^I$, $m_S^I$ do not satisfy this condition and as a result the $SU(2)_F$ symmetry is broken
in a subtle way. It is also possible for solutions to preserve $SU(3)_F\times U(1)$ symmetry,
when $p^1=p^2=p^3$ and when both $m_N^I$ and $m_S^I$ are all equal, as we discuss below.

We now consider anti-twist STU solutions with, for definiteness and no loss of generality,
\begin{align}\label{atassump}
\kappa=1\,,\qquad t_N=1\,, \qquad t_S=0\,.
\end{align}
Notice, in particular, that the regularity condition for the spinors in \eqref{rsymms} requires that
$m^R_N=-1$ mod $n_N$ and $m^R_S=-1$ mod $n_S$.
We focus on solutions with $p_F=0$, both for simplicity and to illustrate the point about the possibility of breaking
$SU(2)_F$ that we just discussed.
We start with the coprime case $(n_{N}, n_S)=(1,5)$ and $p^I=(1,1,2)$ which has $p_F=0$ and also $p_B=0$ (also see table \ref{table1}). In this case we have $m_N^I=(0,0,0)$ and
$m_S^I=(1,1,2)$. Notice that $m^R_N=0=-1$ mod $n_N$ and $m^R_S=4=-1$ mod $n_S$, so we are satisfying the condition \eqref{rsymms}: in fact as we are in a coprime case this
was guaranteed (recall the argument in the paragraph below \eqref{rsymms}).

We now consider the possibility of $h$-fold flux quotients.
The simplest case would be a 2-fold flux quotient with
$(n_{N}, n_S)=2(1,5)$ and $p^I=2(1,1,2)$. However, this case 
is obstructed in the sense that there 
are no solution for $m_N^I$ and $m_S^I$, satisfying
the coprime conditions \eqref{condfluxsmooth2}, \eqref{coprimecondsms}. 
Indeed in appendix \ref{app:WCP} we prove that there is no $h$-fold flux quotient of
any coprime spindle case with $h$ even when
$n_N, n_S$ are both odd and one of the $p^I$ is odd. Furthermore, from extensive checks, it seems that this
is the only class that is obstructed, as we conjectured one paragraph below \eqref{fluxexpgensd}.

Next consider a 3-fold flux quotient with
$(n_{N}, n_S)=3(1,5)$ and $p^I=3(1,1,2)$ (also see table \ref{table2p1}). 
In this case we find a unique solution for $m_{N,S}^I$
 satisfying
the coprime conditions \eqref{condfluxsmooth2}, \eqref{coprimecondsms}: $m_N^I=(2,2,1)$ and
$m_S^I=(11,11,7)$, which in this case does
satisfy the condition for regularity of the spinors \eqref{rsymms}. Therefore, this is an example of a non-coprime spindle that uplifts to a regular, supersymmetric solution and moreover it preserves $SU(2)_F$ symmetry. We will also see later
that this STU solution has a relevant operator consistent with it flowing under RG to a new hyperscalar spindle solution,
with the same spindle data, including the $m_{N,S}^I$.

Next consider a 5-fold flux quotient with
$(n_{N}, n_S)=5(1,5)$ and $p^I=5(1,1,2)$ (also see table \ref{table2p1}).
In this case we find 64 solutions for $m_N^I$ and $m_S^I$, satisfying
the coprime conditions \eqref{condfluxsmooth2}, \eqref{coprimecondsms}. Of these, 13 satisfy the regularity condition on the spinors \eqref{rsymms} i.e. the remaining 51 solutions uplift to regular geometries but the spinors are not well defined. Of the 13 regular solutions 10 of them break the $SU(2)_F$ symmetry:
for example one of the 13 has $m_N^I=(2,1,1)$ and $m_S^I=(11,6,7)$, while another
has $m_N^I=(3,4,2)$ and $m_S^I=(16, 21,12)$ . The 3 regular solutions that preserve the
$SU(2)_F$ symmetry have
(i) $m_N^I=(4,4,1)$ and $m_S^I=(21,21,7)$;
(ii) $m_N^I=(1,1,2)$ and $m_S^I=(6,6,12)$;
(iii) $m_N^I=(3,3,3)$ and $m_S^I=(16,16,17)$.
We will see later 
that of these, only the STU solution (ii)
has a relevant operator consistent with it flowing under RG to a new hyperscalar spindle solution.

We can continue in a similar way and construct infinite classes of new anti-twist solutions in the non-coprime case.
It is also interesting to consider STU solutions with $p^1=p^2=p^3\equiv p$, which only exist in the anti-twist class. When in addition $m_{N,S}^1=m_{N,S}^2=m_{N,S}^3\equiv m_{N,S}$, they preserve $SU(3)$ flavour symmetry and they are also solutions of minimal $D=5$ gauged supergravity 
(with $A^1=A^2=A^3$), which were first discussed in 
 \cite{Ferrero:2020laf} in the case of coprime spindles; here we can further comment on the 
 case of non-coprime spindles. We continue assuming \eqref{atassump} for definiteness.
For smooth uplifts on $S^5$ we have $p=(n_S-n_N)/3$, so $n_S-n_N$ must be divisible by 3.
For smooth uplifts that preserve $SU(3)$ flavour symmetry we should take
$m_{N,S}^1=m_{N,S}^2=m_{N,S}^3\equiv m_{N,S}$. Smoothness requires
$p=n_Nm_S -n_S m_N$, with $\text{hcf}(m_N,n_N)=1$, $\text{hcf}(m_S,n_S)=1$,
and, recalling \eqref{rsymms}, $3 m_N=-1$ mod $n_N$, $3 m_S=-1$ mod $n_S$ for well-defined spinors. As we show in appendix \ref{appCmingaugedsugra}, a unique
uplift on $S^5$ preserving $SU(3)$ exists if and only if $n_S$ and $n_N$ are not divisible by 3 and 
$n_S-n_N$ must be divisible by 3.
Moreover, the total space is then a Lens space $L(p,1)$ fibred over the spindle. These results are valid both for the coprime case,
in precise alignment with
\cite{Ferrero:2020laf,Gauntlett:2006af} (see also \cite{Ferrero:2020twa}), and also for the non-coprime case. It is also interesting to highlight that there are STU solutions with 
$p^1=p^2=p^3\equiv p$ and $m_{N}^I$, $m_S^I$ not all equal, which break the $SU(3)$ flavour symmetry. Indeed there are solutions with $n_S$ and $n_N$ not divisible by 3,
and both $SU(3)$ invariant and $SU(3)$ breaking smooth BPS solutions. There are also
solutions with $n_S$ and $n_N$ which are divisible by 3, and hence there are no $SU(3)$ invariant
solutions, but there are $SU(3)$ breaking smooth BPS solutions. Examples are given in tables \ref{tableat1}, \ref{tableat2}.

For the twist case we can illustrate with 
 \begin{align}
\kappa=1\qquad \Rightarrow\qquad  t_N=0\,, \qquad t_S=0\,.
\end{align}
Some representative coprime STU solutions are presented in table \ref{table7}.
 Interestingly, we can also consider $h$-fold flux quotients of the $AdS_3\times S^2$ solutions of
 \cite{Benini:2013cda}, which only exist in the twist class. Specifically, we consider
 $(n_N,n_S,p^I)=h(1,1,\hat p^I)$ with the $\hat p^I$ associated with a topologically twisted 
 $AdS_3\times S^2$ solution as in \eqref{stufullsolnS2}. When $m_N^I$, $m_S^I$, satisfying
 \eqref{condfluxsmooth2}, \eqref{coprimecondsms} and \eqref{rsymms} exist, we obtain regular $AdS_3\times Y_7$ solutions with $Y_7$ a smooth $S^5$ bundle
 over the (good) orbifold of $S^2$.
For example, we can consider $h$-fold flux quotients for the case of $S^2$ with $\hat p^I= (1,1,-4)$.
For $h=3$ we find a unique smooth solution with $m_N^I = (1,1,2)$ and $m_S^I = (2,2,1)$, 
which has well defined spinors and also preserves $SU(2)_F$ symmetry.
For $h=5$ there are 27 smooth uplifts, 7 of which admit regular killing spinors and thus preserve supersymmetry. Of these 7, only one preserves the $SU(2)_F$ symmetry, with $m_N^I = (2,2,2)$, $m_S^I = (3,3,3)$. The remaining 6 break the 
$SU(2)_F$ symmetry and are given by $m_N^I = (1,2,3)$, $m_S^I = (2,3,4)$ and their permutations. More details of these solutions can be found in table \ref{3folds2}.

 \section{Hyperscalar fluctuations for the STU solutions}\label{seclinfluct}
We now analyse linearized, supersymmetric perturbations of the hyperscalar $\rho e^{i\theta}$ about the STU solutions. There are two reasons for doing this. The first is that it allows us to distinguish
some of the STU model solutions with non-coprime spindles with different values of $m^I_{N,S}$
from each other. As we discussed in the last subsection, these STU solutions, with the same values of $(n_N,n_S,p^I)$ and different 
values of $m^I_{N,S}$, are dual to $d=2$ SCFTs with the same central charge. After uplifting on $S^5$
they can be distinguished by the fact that the topologies of the $S^5$ bundles over the spindles
$\Sigma(n_N,n_S)$
are different, in general. They can also be distinguished in $D=5$, in some cases, by computing the conformal
scaling dimension, $\Delta$, of the operator dual to the hyperscalar.
 
 The second reason concerns possible supersymmetric RG flows from STU 
 $AdS_3\times \Sigma$  solutions to new supersymmetric $AdS_3\times \Sigma$ solutions with non-vanishing
 hyperscalar, which we can construct numerically.
 As we will see, the latter hyperscalar solutions have magnetic fluxes $p_R, p_B, p_F$, with $p_R, p_B$ constrained by \eqref{constfluxes}.
 A natural question is whether these solutions can arise as the IR limit of an RG flow
 starting from an $AdS_3\times \Sigma$ solution of the STU model with the same magnetic fluxes
 and, in the non-coprime case the same $m^I_{N,S}$,
 and perturbed by a relevant operator dual to
 the charged hyperscalar. Clearly two necessary conditions for such an RG flow to exist is
 that there is a suitable relevant deformation that can drive the flow and also
 that the central charge should decrease under the RG flow. Remarkably
 we find that these two conditions seem to be precisely correlated i.e.
 for any specific hyperscalar solution that exists, such an RG flow from an STU solution is always possible.

We now analyse linearized perturbations of the hyperscalar $\rho e^{i\theta}$ about an STU 
$AdS_3\times \Sigma$ solution
of the form 
\begin{align}\label{linpertt}
\rho=w(y)u^\delta\,,\qquad
\theta = \bar\theta z\,.
\end{align}
Here $u$ is the Poincar\'e AdS radial coordinate, as in \eqref{eq:5d_frame_ansatz}, and so we have assumed the perturbation has a fixed scaling dimension $\Delta=2-\delta$. If $0< \Delta< 2$ then this is associated with having a source dual to a relevant scalar operator. The case of $\Delta=0$ is the unitary bound and if $\Delta>2$ the mode is dual to an irrelevant operator.\footnote{The case of $\Delta=2$ is potentially interesting as it corresponds to a classically marginal operator. However, we have found
no STU solutions with smooth uplifts and regular spinors that have $\Delta=2$ modes.
}
The perturbation breaks the azimuthal rotational symmetry of the spindle, as well as the $U(1)_B$ symmetry of the STU solution, but preserves a diagonal subgroup of the two.
We also recall that the gauge-invariant one form $D\theta$ is given by $D\theta=(\bar\theta- \zeta_I a^I) dz$.
The hyperscalar is a section of a line bundle over the spindle that is associated with the $U(1)_B$ orbibundle.
To ensure that we have a globally defined section, again we need to use a north and south pole patch as we did previously.

We demand that the perturbation preserves all of the $\mathcal{N}=(0,2)$ (or $\mathcal{N}=(2,0)$)
Poincar\'e Killing spinors of the STU solutions. Analysing the BPS equations, as in appendix \ref{linpertapp}, we find
\begin{align}\label{eq:delta_from_Dtheta_zetaIt}
\delta =-\frac{\kappa}{k} \left({D\theta_z+\zeta_I \cI^I }\right)\,,
\end{align}
with the function $w(y)$ satisfying the following ODE (in conformal gauge \eqref{confgauge}):
\begin{align}\label{wponweqn}
\frac{w'}{w} = \frac{1}{k} \left( \frac{D\theta_z}{\tan\xi} - \zeta_I \cI^I \tan\xi \right)\,,
\end{align}
We now assume that in the two patches, the hyperscalar behaves like 
\begin{align}\label{eq:Dtheta_poles}
w &\sim (y-y_{N})^{\hm_{N}}\,,\qquad
w\sim (y_S-y)^{\hm_{S}}\,,
\end{align}
with $r_{N,S}\ge 0$ as we approach the poles, and, recalling \eqref{firstbcsec}, smoothness
implies that we should take $\hm_N, \hm_S\in\mathbb{Z}_{\ge 0}$.
Then from \eqref{wponweqn} we deduce that
\begin{align}\label{eq:Dtheta_poles2}
D\theta_z|_N&=(-1)^\tN \frac{\hm_N}{n_N}\,, 
\qquad
D\theta_z|_S=-(-1)^\tS \frac{\hm_S}{n_S}\,.
\end{align}

Next, evaluating \eqref{eq:delta_from_Dtheta_zetaIt} at the two poles, with $\delta$ the same constant at both, 
and using
\eqref{eq:pi_xit}, we deduce that for the perturbation to be regular we
also require that the STU solutions satisfies
\begin{align}\label{eq:pb_constraint}
p_B = (-1)^\tN \hm_N \nS + (-1)^\tS \hm_S \nN\,.
\end{align}
If $ \hm_N =\hm_S=0$, the hyperscalar perturbation would be a non-vanishing constant at each pole, as studied
in \cite{Arav:2022lzo}, and $p_B=0$. We see that by allowing the hyperscalar to vanish at one or both of
the poles then we can have $p_B\ne 0$. Clearly, $p_B\in\mathbb{Z}$, as required for regularity
of the circle orbibundle, if we have $\hm_N, \hm_S\in\mathbb{Z}_{\ge 0}$.

Using the explicit result for $\zeta_I \cI^I$ in the STU solutions discussed in the previous section, 
we can express the scaling dimension, $\Delta$, of the operator dual to this perturbation as
$\Delta=2-\delta$ with 
\begin{align}\label{eq:delta_final_general}
\delta = \frac{1}{4s}\left[p_R^2+p_B^2-4p_F^2-2(\hm_S n_N^2-3(-1)^{t_N+t_S}n_N n_S(\hm_N+   \hm_S)+\hm_N n_S^2)
\right]\,,
\end{align}
where $s$ is given in \eqref{defs}, $p_R$ as in \eqref{conststuflux} and $p_B$ as in \eqref{eq:pb_constraint}.

We now return to \eqref{eq:Dtheta_poles2}. We want the hyperscalar $\rho e^{i\theta}$ to be a section of
a line bundle on the spindle, with unit charge with respect to the connection one-form $A_B=\zeta_IA^I$. As discussed in appendix \ref{appa} 
we should take a $N$ and a $S$ pole patch, with
\begin{align}\label{hsregconds}
A_B|_N\to \frac{\zeta_I m^I_N}{n_N}dz\,,\qquad 
A_B|_S\to \frac{\zeta_I m^I_S}{n_S}dz\,,
\end{align}
and then consider the phase of the complex scalar to be $\theta=\bar\theta_N z$ and
$\theta=\bar\theta_S z$ in the two patches.
The patches are then glued together with a $U(1)_B$ gauge transformation
with 
\begin{align}
A_B|_{\text{$N$ patch}}=
A_B|_{\text{$S$ patch}}+\zeta_I\gamma^I dz\,,
\end{align}
with $\gamma^I\in \mathbb{Z}$, 
so that $\bar\theta_N=\bar\theta_S+\zeta_I\gamma^I$ (as the hyperscalar has unit charge with respect to $A_B$).
In this
gauge we have
\begin{align}
D\theta_z|_N=\bar\theta_N-\frac{\zeta_I m^I_N}{n_N}\,,\qquad
D\theta_z|_S=\bar\theta_S-\frac{\zeta_I m^I_S}{n_S}\,,
\end{align}
and the orbifold identifications for the line bundle imply that $\bar\theta_{N,S}\in\mathbb{Z}$ (a point that was not emphasised
in \cite{Arav:2022lzo}).
Comparison with \eqref{eq:Dtheta_poles2} then reveals that we must have
\begin{align}\label{rmrels}
(-1)^\tN {\hm_N}&=n_N\bar\theta_N-{\zeta_I m^I_N}\,,\nn\\
-(-1)^\tS {\hm_S}&=n_S\bar\theta_S-{\zeta_I m^I_S}\,,
\end{align}
which, in particular, is consistent with the smoothness condition $r_{N,S}\in \mathbb{Z}_{\ge 0}$, 
noted above. Also note that one of these conditions along with \eqref{eq:pb_constraint}
implies the second condition, as we explain below.

This analysis also shows that in the non-coprime case, where the values of
$m_{N,S}^I$ are extra data that we
need to specify the orbibundle, the value of $\delta$ in \eqref{eq:delta_final_general}
will explicitly depend on the specific values
of  $m_N^I$ and $m_S^I$. 
Thus, two non-coprime STU solutions with the same $n_{N,S}$, $t_{N,S}$ and $p^I$, but different $m^I_N$ and $m^I_S$ will 
have the same central charge but, in general, different spectra. We illustrate this below.

It is convenient to define the ``Kaluza-Klein" integer $n_{KK}$ for the hyperscalar modes via 
\begin{align}
n_{KK}\equiv (-1)^\tN \bar\theta_N\,,
\end{align}
and then we can write
\begin{align}\label{rmrels2}
\frac{r_N}{n_N}&= n_{KK}-(-1)^\tN\frac{\zeta_I m^I_N}{n_N}\,,\nn\\
\frac{\hm_S}{n_S}&=(-1)^{\tN+\tS+1} \frac{r_N}{n_N}+(-1)^\tS \frac{p_B}{n_N n_S}\,.
\end{align}
The first line comes from the first line of \eqref{rmrels}, while the second comes from 
\eqref{eq:pb_constraint}. We also highlight that using \eqref{condfluxsmoothagain2} we can then also
write $r_S$ in the form of \eqref{rmrels} with $\bar \theta_S=(-1)^{t_N}n_{KK}$.
Now we require the integers $r_{N,S}\ge 0$, for regularity, and
so we can now discuss a key difference between the spectrum of BPS hyperscalar fluctuations about
STU solutions for the twist and the anti-twist class.  

For the twist class we have $t_N=t_S$ and hence we require 
\begin{align}
0\le {r_N}\le (-1)^\tS\frac{p_B}{ n_S}\,,
\end{align}
with $r_N$ as in \eqref{rmrels2}.
By choosing different values of $n_{KK}$
we see that at most there will be a \emph{finite} number
of fluctuations of the hyperscalar with $r_{N,S}\ge 0$.
For example, consider the special case of a tear drop spindle with $n_N=1$, which is a coprime spindle. We can choose $m^I_N=(0,0,0)$ and obtain a smooth supersymmetric uplift. Then, provided 
$(-1)^\tS{p_B}\ge 0$ we obtain $\lfloor \frac{|p_B|}{n_N n_S} \rfloor+1$ solutions
with $n_{KK}=0,1,\dots,\lfloor \frac{|p_B|}{n_N n_S} \rfloor$. 
On the other hand if $(-1)^\tS{p_B}< 0$ we don't obtain any solutions.

For the anti-twist class we have $t_N+t_S=1$, and so we need to satisfy
\begin{align}
r_N\ge 0,\qquad \text{and} \qquad r_N\ge (-1)^{t_S+1}\frac{p_B}{n_S}\,,
\end{align}
with $r_N$ as in \eqref{rmrels2}, and this leads to an \emph{infinite} number of solutions
for $n_{KK}$. Thus, the spectrum of the BPS hyperscalar fluctuations in the anti-twist
class is substantially different to the twist class.

We can make some additional observations. 
We first use \eqref{eq:Dtheta_poles2} and \eqref{eq:stu_k_x123t} to rewrite $\delta$ in \eqref{eq:delta_from_Dtheta_zetaIt} in the form
\begin{align}
(-1)^t\kappa\delta=-2\kappa(-1)^t r+(-1)^tx^1(1+r)+(-1)^tx^2(1+r)+(-1)^tx^3(r-1)\,,
\end{align}
which is valid at both poles. Since $(-1)^tx^I>0$ we notice that if $r\ge 1$ then the last three
terms on the right hand side are all positive. We can now consider this expression for 
the two anti-twist cases, setting $\kappa=+1$ for definiteness. Demanding that $\delta>0$, in order to get a relevant mode, we find that for the anti-twist case with $t_N=1$ and $t_S=0$ we necessarily have $r_N=0$, while for the anti-twist case with $t_N=0$ and $t_S=1$ we necessarily have $r_S=0$. 

Next, we can make some further observations regarding the STU solutions in the
anti-twist class which admit relevant modes, focussing for definiteness on the class
\begin{align}
\kappa=+1\,,\qquad t_N=1\,,\quad t_S=0\,,\qquad r_N=0\,.
\end{align}
To do so it is convenient to introduce the rescaled variables
\begin{align}\label{newvs}
u=\frac{n_S}{n_N}\,,\qquad v=\frac{p_F}{n_N}\,.
\end{align}
We demand the conditions $(-1)^t x^I>0$ at both poles, $c>0$, using \eqref{stufullsoln}, with
$p_R$ given by \eqref{conststuflux} for the STU solution, and $p_B$ given by \eqref{eq:pb_constraint}. 
With $r_S\ge 0$ we find these conditions require
\begin{align}\label{stuconstratintsone}
u>1+r_S\,,\qquad r_S+u>1+2|v|\,.
\end{align}
If we first consider $r_S=0$, these algebraic conditions for the STU solution
with $p_B=0$ are equivalent to
\begin{align}\label{rnrszerocondsstu}
r_N=r_S=0:\qquad n_S-n_N>2|p_F|\ge 0\,.
\end{align}
Furthermore, these conditions automatically imply that $\delta>0$ and that the hyperscalar mode is relevant.
Thus, when $r_N=r_S=0$ there is a relevant mode that can induce an RG flow from the STU solution in the UV 
to the hyperscalar solutions with $p_B=0$ that were already constructed in \cite{Arav:2022lzo}.
We next consider $r_N=0$ and $r_S\ge 1$. The conditions \eqref{stuconstratintsone} on the STU solutions
now do not imply $\delta>0$. Demanding that in addition $\delta>0$, with
$\delta$ given in \eqref{eq:delta_final_general},
so we have a relevant hyperscalar mode, implies
\begin{align}\label{cons1hsmodes}
&1\le \hm_S<1+3u-2^{3/2}\sqrt{u+u^2}\,,\nn\\
&|v|<\frac{1}{2}\sqrt{1-2\hm_S+\hm_S^2-2u-6\hm_S u +u^2}\,.
\end{align}
In particular, the first condition implies that $u\ge 8$. In these variables, and for these solutions, the fluxes $p^I$ can be written
\begin{align}\label{cons2hsmodes}
p^1&=\frac{n_N}{4}(\hm_S+u-1+2v)\,,\nn\\
p^2&=\frac{n_N}{4}(\hm_S+u-1-2v),\nn\\
p^3&=-\frac{n_N}{2}(\hm_S-u+1),
\end{align}
and demanding that these are integers provides additional constraints. For such solutions
we expect that the RG flow induced by the relevant hyperscalar deformation will
end up at a new hyperscalar solution in the IR, as we discuss in the next section.

Note that for the anti-twist STU solutions with $p^1=p^2=p^3$, some of which are
solutions of minimal gauged supergravity (the ones with $m^I_N$ and $m^I_S$ all equal and hence $SU(3)$ invariant after uplift on $S^5$), there
are no relevant hyperscalar modes. 
Indeed, from \eqref{cons2hsmodes}, we should take $v=0$ and $u=1+3r_S$,
but the latter is not consistent with the first condition in
\eqref{cons1hsmodes}.

Finally, we consider the twist solutions. Recall that when $\kappa=+1$ there are only STU twist solutions with 
$t_N=t_S=0$. Demanding that $(-1)^t x^I>0$ at both poles, $c>0$, 
using \eqref{stufullsoln}, with $p_R$ given by \eqref{conststuflux} for the STU solution, and $p_B$ given by \eqref{eq:pb_constraint},  then we obtain  solutions. However, if we also demand that $\delta\ge 0$, with $\delta$ as in
\eqref{eq:delta_final_general}, we find that there are no solutions. 
Thus, there are no relevant hyperscalar deformations for 
any of the STU twist solutions.
 
For the special case of $AdS_3\times S^2$ twist solutions, and their $h$-fold flux quotients, we should set $n_N=n_S$. In
 this case the expression for $\Delta$ can be written as
 \begin{align}\label{deltas2case}
 \Delta=\frac{8 p_B(2n_N+p_B)}{-4n_N^2+4n_N p_B+3 p_B^2+4p_F^2}\,,
 \end{align}
 and, in particular, only depends on $r_{N,S}$ via their sum, since from \eqref{eq:pb_constraint}
 we have $p_B=n_N(r_N+r_S)$. This leads to a degeneracy of the scaling dimensions for the modes.
When $n_N=n_S=1$, associated with the round $S^2$, this is to be expected since the $S^2$ has an enhanced $SO(3)$ isometry group. Interestingly, the degeneracy can also persist for $h$-fold flux quotients of the $S^2$ (sometimes with reduced degeneracy).

 \section{Hyperscalar solutions}\label{sechyperscsol}
 We now consider $AdS_3\times \Sigma$ solutions with non-vanishing hyperscalar. 
Solutions with the hyperscalar non-vanishing at both poles were constructed in \cite{Arav:2022lzo}.
Here we allow the hyperscalar to vanish at the poles and we also allow for the possibility of non-coprime
spindles. 
We only find anti-twist solutions and these necessarily have the hyperscalar vanishing at only one pole, as we explain below.
We continue using the conformal gauge \eqref{confgauge}.
 
We can write the superpotential \eqref{superpottextLS} as
\begin{align}
\cos\xi e^V W =\sum_I x^I+\sinh^2\frac{\rho}{2}\zeta_I x^I\,,
\end{align}
and hence
\begin{align}
\label{eq:dW_for_I}
\cos\xi e^V\partial_\rho W = \frac{1}{2}{\zeta_I x^I} \sinh\rho\,.
\end{align}
We assume that near the poles the hyperscalar behaves as
\begin{align}\label{falloffhs}
\rho = C_N(y-y_{N,S})^{\hm_{N}}+\cO(y-y_{N})^{\hm_{N}+1}\,,
\quad
\rho = C_S(y_{S}-y)^{\hm_{S}}+\cO(y_S-y)^{\hm_{S}+1}\,,
\end{align}
with the constants $C_{N,S}>0$. 
As we discussed in the analysis of the linearised perturbation of the BPS equations in
section \ref{seclinfluct}, and also in appendix \ref{appa}, smoothness of the hyperscalar
(in the orbifold sense), requires that 
\begin{align}\label{rnrsintpos}
 {\hm_{N,S}}\in\mathbb{Z}_{\ge 0}\,.
\end{align}
We will shortly see that
this is consistent with the BPS equations.

Analysing the BPS equation for $\rho$ \eqref{bulkbps}, 
near the pole we find that the dressed scalars satisfy the constraint
\begin{align}\label{eq:m_pole_relation}
\zeta_I x^I_N=\frac{(-1)^{t_N}\hm_N}{kn_N}\,,\qquad 
\zeta_I x^I_S=\frac{(-1)^{t_S+1}\hm_S}{kn_S}\,.
\end{align}
Similarly, the second constraint in \eqref{bpsconsts} can be written, for $\rho\ne0$,
\begin{align}
\zeta_I x^I=\frac{1}{k}D\theta\,.
\end{align}
Hence, at the poles we deduce
\begin{align}\label{eq:zeta_scI_m}
D\theta_z|_{N}=\frac{\hm_N}{n_N} (-1)^{t_N}\,,\qquad
D\theta_z|_{S}=-\frac{\hm_S}{n_S} (-1)^{t_S}\,.
\end{align}
As discussed in appendix \ref{appa}, and as in the previous section, the orbifold 
identifications for the line bundle imply that in the gauge \eqref{Agaugemn}
we have
\begin{align}\label{deethetassecsix}
D\theta_z|_N=\bar\theta_N-\frac{\zeta_I m^I_N}{n_N}\,,\qquad
D\theta_z|_S=\bar\theta_S-\frac{\zeta_I m^I_S}{n_S}\,,
\end{align}
with $\bar\theta_{N,S}\in\mathbb{Z}$ which, as noted above,
is consistent with \eqref{rnrsintpos}.

Next consider evaluating $W$ at the poles. If $\hm_{N,S} >0$ then $\rho$ vanishes, while if $\hm_{N,S}=0$ then \eqref{eq:m_pole_relation} implies $\zeta_Ix^I=0$. So, in both cases 
\begin{align}
e^V W|_{N,S} =\sum_I (-1)^{t_{N,S}}x^I_{N,S}\,.
\end{align}
Then, considering the first constraint in \eqref{bpsconsts} and using \eqref{eq:s_Qz_bc} we deduce the following
constraints on the dressed scalars at the poles (as for the STU solutions):
\begin{align}\label{eq:scI_sum}
 \sum_I x^I_N= 2 \kappa -\frac{(-1)^{t_N} }{kn_N}\,,\qquad
\sum_I x^I_S= 2 \kappa   +\frac{(-1)^{t_S}}{kn_S}\,.
\end{align}
From the expression for the fluxes in 
\eqref{eq:pi_xit}, using \eqref{eq:m_pole_relation} and \eqref{eq:scI_sum} we now have
two constraints on the flux:
\begin{align}\label{constfluxes}
 \frac{p_R}{n_N n_S} &= \frac{(-1)^{\tN+1}}{n_N} + \frac{(-1)^{\tS+1}}{n_S}\,,\nn\\
 \frac{p_B}{n_N n_S} &= \frac{(-1)^{\tN}\hm_N}{n_N} + \frac{(-1)^{\tS}\hm_S}{n_S}\,.
\end{align}
The flavour flux $p_F\equiv p^1-p^2$, is given by
\begin{align}\label{eq:pi_xitf}
 \frac{p_F}{n_N n_S} = k \left( x_{F,N} -  x_{F,S} \right)\,,
\end{align}
where $x_{F}\equiv x^1- x^2$.
When the hyperscalar is non-vanishing we have two conserved charges as in \eqref{twoconschges}:
\begin{align}\label{twoconschgesag}
\frac{1}{2\kappa}\cE_R &= \mathcal{F}(x^I)\left[ \left(\frac{1}{x^1}+\frac{1}{x^2}+\frac{2}{x^3} \right)-4\kappa \right]_{N,S}\,,\\ \nn
\frac{1}{2\kappa}\cE_F &= \mathcal{F}(x^I) \left(\frac{1}{x^1}-\frac{1}{x^2} \right)|_{N,S}\,.
\end{align}

So, for given spindle data $n_{N,S}$, $(-1)^{t_{N,S}}$, $\hm_{N,S}$ and freely specified flux $p_F$,
we have seven algebraic equations: two from \eqref{eq:m_pole_relation}, two from \eqref{eq:scI_sum}, one from
\eqref{eq:pi_xitf} and two from \eqref{twoconschgesag}. These can be used to solve for $ x^I_{N,S}$ and $k$
and hence obtain the central charge, without solving the BPS equations, just assuming they exist.
Closed form expressions analogous to \eqref{stufullsoln} are given in section \ref{eqlocalsec}
(see \eqref{localcetc}). We further comment on non-coprime spindles below.

We again have the following necessary conditions for the existence of solutions
\begin{align}
\label{eq:stu_positivity2}
(-1)^{t}x^{1,2,3}|_{N,S} > 0 , \qquad c > 0\,.
\end{align}
In general, using  
\eqref{eq:m_pole_relation} and \eqref{eq:scI_sum} we deduce that at either pole
\begin{align}\label{chrisrel}
\kappa(-1)^t=\frac{1}{2\hm}[ (1+\hm)(-1)^tx^1+(\hm+1)(-1)^tx^2+(\hm-1)(-1)^tx^3 ]\,.
\end{align}
 Since $(-1)^tx^1>0$, focussing on $\kappa=+1$, we see that for each pole $\hm\ge 1$ implies that $t=0$.
 Thus, anti twist solutions, for which $t_N\ne t_S$, necessarily require either $\hm_N=0$ or
 $\hm_S=0$. 

To solve the BPS equations, we can specify $n_{N,S}\in\mathbb{N}$, signs $(-1)^{t_{N,S}}$, $\hm_{N,S}\in\mathbb{Z}_{\ge 0}$, 
$p^I\in \mathbb{Z}^3$ (constrained by \eqref{constfluxes}, so only $p_F$ is undetermined)
and this is then sufficient to determine the boundary conditions at both poles of all of the fields, except the hyperscalar. This allows us to solve the ODEs by performing a search over the values of the leading coefficient of the hyperscalar at one of the poles. 
In the non-coprime case, we should also 
specify the values of $m_N^I$, coprime to $n_N$, and $m_S^I$, coprime to $n_S$, which can then be freely added
after solving the BPS equations; while 
the central charge will not depend on $m_N^I$, $m_S^I$, the solutions do depend
on them and they uplift
to different type IIB solutions. The $m_{N,S}^I$ must
also satisfy \eqref{rsymms}, to ensure the spinors are well defined, and
\eqref{eq:zeta_scI_m}, \eqref{deethetassecsix} to ensure the hyperscalar is a section of a line bundle.

\bigskip
{\bf Summary:}
We now summarise the constraints imposed on the regular hyperscalar solutions that we have just derived.
The dressed scalars satisfy the constraints:
\begin{align}\label{eq:m_pole_relation2}
 \sum_I x^I_N&= 2 \kappa -\frac{(-1)^{t_N} }{kn_N}\,,\qquad
\sum_I x^I_S= 2 \kappa   +\frac{(-1)^{t_S}}{kn_S}\,,\nn\\
\zeta_I x^I_N&=\frac{(-1)^{t_N}\hm_N}{kn_N}\,,\qquad 
\zeta_I x^I_S=\frac{(-1)^{t_S+1}\hm_S}{kn_S}\,.
\end{align}
The fluxes are given by
\begin{align}\label{condfluxsmooth2sum}
p^I=n_Nm^I_S -n_Sm^I_N  +\gamma^I n_N n_S \in\mathbb{Z}\,,
\end{align}
and smoothness of the orbibundle requires 
\begin{align}\label{coprimecondsmssum}
\mathrm{hcf}(m^I_N,n_N)=1 \qquad \text{and}\qquad \mathrm{hcf}(m^I_S,n_S)=1\,,
\end{align}
which implies $\mathrm{hcf}(p^I,n_N)=\mathrm{hcf}(p^I,n_S)=\mathrm{hcf}(n_N,n_S)$.
For coprime spindles with $\text{hcf}(n_N,n_S)=1$, the $m^I_{N,S}$
are uniquely specified given the fluxes $p^I$, but this is not the case for non-coprime spindles.
Supersymmetry implies the fluxes are constrained via
\begin{align}\label{constfluxes2}
 \frac{p_R}{n_N n_S} &= \frac{(-1)^{\tN+1}}{n_N} + \frac{(-1)^{\tS+1}}{n_S}\,,\nn\\
 \frac{p_B}{n_N n_S} &= \frac{(-1)^{\tN}\hm_N}{n_N} + \frac{(-1)^{\tS}\hm_S}{n_S}\,.
\end{align}
For the hyperscalar to be a smooth section of a line bundle with fall-off at the poles as in \eqref{falloffhs}, we require
\begin{align}\label{eq:zeta_scI_m2}
\frac{\hm_N}{n_N} (-1)^{t_N}=\bar\theta_N-\frac{\zeta_I m^I_N}{n_N}\,,\qquad
\frac{\hm_S}{n_S} (-1)^{t_S+1}=\bar\theta_S-\frac{\zeta_I m^I_S}{n_S}\,,
\end{align}
with $\bar\theta_{N,S}\in\mathbb{Z}$ and so $ {\hm_{N,S}}\in\mathbb{Z}_{\ge 0}$;
for coprime spindles these are automatically satisfied given $p_B$ is as in \eqref{constfluxes2}.
For the spinors to be well defined we require
\begin{align}\label{rsymms2}
\frac{m^R_N}{n_N}-\bar s_N=\frac{(-1)^\tN}{n_N}\,,
\qquad
\frac{m^R_S}{n_S}-\bar s_S=-\frac{(-1)^\tS}{n_S}\,,
\end{align}
where $m^R\equiv \sum_Im^I$ and $\bar s_{N,S}\in\mathbb{Z}$;
for coprime spindles these are automatically satisfied given $p_R$ is as in \eqref{constfluxes2}.
This concludes the summary.
\bigskip

 Using the boundary conditions consistent with these constraints, we have explicitly solved the BPS equations numerically, with non-vanishing hyperscalars, for various 
 spindles.   For the twist class with $t_N=t_S=0$ or $t_N=t_S=1$ we have found no explicit solutions 
for any values of $r_{N,S}\ge 0$.
For the case $r_N=r_S=0$, this is in agreement with \cite{Arav:2022lzo}, where it was shown
that the algebraic conditions \eqref{eq:stu_positivity2} eliminate the possibility of twist solutions. 
For other values of $r_{N,S}$ the algebraic conditions \eqref{eq:stu_positivity2} do not eliminate
the possibility of such solutions, but we have not found any
and we strongly suspect that they don't exist.
This would be in alignment with the fact that there are no relevant modes
in the STU twist solutions, as we saw at the end of the previous section.

However, for the anti-twist class we have found a rich landscape of numerical solutions, which we summarise in
section \ref{exasmplessec} (see also appendix \ref{plotssols}).
Without loss of generality, for definiteness we will set 
\begin{align}
\kappa=+1\,,
\end{align} 
and focus on anti-twist solutions with
\begin{align}\label{assumptionshscsol}
t_N=1\,, \qquad t_S=0\,,\qquad \Rightarrow \qquad \hm_N=0\,,
\end{align}
from \eqref{chrisrel}.
The solutions of most interest satisfy the conditions discussed in section \ref{smoothupliftssec} and above,
which ensure that after uplifting the solution on $S^5$
we obtain a smooth\footnote{Relaxing some of these conditions leads to solutions with orbifold singularities on $Y_7$, which may be of interest too.} $AdS_3\times Y_7$ solution of type IIB supergravity, with well-defined spinors, however
we briefly comment below on examples where some of the regularity conditions are not satisfied.
We note here, though, that if $n_N$ is even, then there are no smooth hyperscalar solutions. To see this,
observe that the smoothness condition
\eqref{coprimecondsmssum} implies that $m^I_N$ are all odd and hence $\zeta_I m^I_N$ is odd. But then we see
\eqref{eq:zeta_scI_m2} cannot be satisfied with $r_N=0$ and $\bar\theta_N\in\mathbb{Z}$. We thus conclude that smooth hyperscalar solutions (with \eqref{assumptionshscsol}) require
\begin{align}\label{ennoddconc}
\text{$n_N$ odd}\,.
\end{align}

A further necessary condition for the existence of solutions is given by the algebraic conditions \eqref{eq:stu_positivity2}.
We find that these are not sufficient. However, we find that if we supplement these conditions with the extra requirement that
the scaling dimension $\Delta=2-\delta$ of the linearised perturbation about the STU solution with same magnetic fluxes, with
$\delta$ given in \eqref{eq:delta_final_general}, satisfies
\begin{align}\label{deltacond}
0<\Delta<2\,,
\end{align}
and hence is associated with a relevant operator in the SCFT dual to the STU solution,
then an extensive numerical search indicates this is then sufficient. Moreover, we also find that in all
cases where a hyperscalar solution exists then the central charge of the $AdS_3\times \Sigma$ solution with non-vanishing 
hyperscalar and magnetic fluxes $p^I$ (satisfying \eqref{constfluxes})
is smaller than that of the central charge the $AdS_3\times \Sigma$ solutions of the STU model with
the same magnetic fluxes $p^I$. This strongly suggests that there is always
an RG flow between the two fixed points, starting in the
UV from the STU solution and ending with a hyperscalar solution in the IR, with the same values of
$(n_N,n_S,p^I)$ and $m_{N,S}^I$ (in the non-coprime case). We shall refer to this as the ``RG scenario" below.

Note that the anti-twist hyperscalar solutions can have enhanced flavour symmetries. The solutions with 
$p_1=p_2$ (i.e. $p_F=0$) will have $SU(2)_F$ symmetry after uplifting on $S^5$, provided that
$m^1_{N,S}=m^2_{N,S}$ so that the solutions have $A^1=A^2$. The condition $m^1_{N,S}=m^2_{N,S}$
is, potentially, an extra condition that one needs to impose in the non-coprime case. Alternatively, there
are non-coprime solutions with $p_1=p_2$ and $m^1_{N,S}\ne m^2_{N,S}$, which then break $SU(2)_F$ in a subtle way (analogous to our discussion of the STU solutions).
We also note that the anti-twist hyperscalar solutions which have $SU(2)_F$ symmetry and in addition $p_B=0$, have $r_N=r_S=0$ and the solutions have the value of the scalars fixed to their LS values \eqref{LSvacvals} (as noticed in
\cite{Arav:2022lzo}); 
when in addition $m^1_{N,S}+m^2_{N,S}-m^3_{N,S}=0$ the solutions are in fact solutions of minimal gauged supergravity obtained
using the truncation associated with the LS vacuum (with $A_B=A_F=0$).

\section{Examples}\label{exasmplessec}
In this section we summarise some results for various examples of $AdS_3\times \Sigma$ solutions.
For the anti-twist case, we discuss examples of such solutions with non-vanishing hyperscalar, which we have constructed numerically, for both coprime and non-coprime spindles, and give the central charge. In appendix \ref{plotssols} we present some plots of the metric, scalar and gauge-field functions; we only present plots for the coprime case since the non-coprime solutions
can easily be obtained by scaling the metric functions and appropriately inserting discrete fluxes. 
In this section we also present some results regarding the central charge and
spectrum of BPS hyperscalar fluctuations around the corresponding STU solutions which, in particular, are consistent
with the RG scenario mentioned above.
In the tables \ref{table1}-\ref{table6} given below, we emphasise that they are all associated with 
smooth, supersymmetric $AdS_3\times Y_7$ solutions of type IIB supergravity, arising from an anti-twist STU solution or
a hyperscalar solution (unless otherwise noted).

We have not found any hyperscalar $AdS_3\times \Sigma$ solutions in the anti-twist class 
when $p^1=p^2=p^3$ and we conjecture that they don't exist. 
In section \ref{atmgs} we summarise some aspects of STU solutions in this class, associated with
smooth, supersymmetric $AdS_3\times Y_7$ solutions of type IIB supergravity,
that either preserve or break the $SU(3)$ flavour symmetry. We also present
the spectrum of BPS hyperscalar fluctuations.

We have also not found any hyperscalar $AdS_3\times \Sigma$ solutions in the twist class and we conjecture that they don't exist.
In section \ref{texampleshsfs} we also summarise the spectrum of 
hyperscalar fluctuations for some examples of STU solutions in the twist class, both spindles
and $h$-fold flux quotients of $S^2$,
which again are
associated with smooth, supersymmetric $AdS_3\times Y_7$ solutions of type IIB supergravity. 

For all solutions we have chosen $\kappa=+1$.

\subsection{Hyperscalar non-vanishing at both poles, $p_B=0$}
With the hyperscalar non-vanishing at both poles, the hyperscalar $AdS_3$ solutions necessarily have $r_N=r_S=0$ and $p_B=0$.
This class of anti-twist solutions was discussed in \cite{Arav:2022lzo}. Here, we further clarify the regularity
of the solutions in
the coprime case and also discuss new solutions in the non-coprime case. We also present the values of
$\Delta$ for the linearised perturbation about the STU solution providing evidence for the RG scenario, discussed above.
Note that we give the values of $(r_N,r_S)$ for the linearised perturbation, as in \eqref{rmrels2},
and we find in practise that the linearised mode associated with the relevant deformation with $(r_N,r_S)=(0,0)$
has either $n_{KK}=0$ or $n_{KK}=-1$.

\subsubsection{Coprime spindles, $\mathrm{hcf}(n_N,n_S)=1$}
Some examples of such solutions are summarised in table \ref{table1}. 
We present the central charge of the STU solution, $c_{STU}$, as well
as the central charge of the hyperscalar solution $c_H$. We also include the scaling dimension of
the relevant mode about the STU solution, which can generate an RG flow from the STU solution to the 
corresponding hyperscalar solution.
For coprime spindles, the solutions are uniquely fixed by $(n_N,n_S, p^I)$, with $m_{N,S}^I \in\mathbb{Z}^3_{n_{N,S}}$ uniquely determined.
The smooth solutions require, for \eqref{coprimelamndannns}, each individual flux $p^I$ to be coprime to both $n_N$ and $n_S$.  Notice that the values of the central charge and $\Delta$ are consistent
with the RG scenario.
In fact, we recall (from below \eqref{rnrszerocondsstu}) that for STU anti-twist solutions there
are always relevant modes with $r_N=r_S=0$.
 \begin{table}[h!]
 \begin{center}
\begin{tabular}{ |c|c|c|c|c|c|c| } 
 \hline
 $(n_N, n_S)$& $(p^1,p^2,p^3)$ &  $\hm_N$ &$\hm_S$ &$\Delta$&$\frac{1}{N^2}c_{STU}$& $\frac{1}{N^2}c_H$\\ 
 \hline
$(1, 5)$ & $(1,1,2)$& $n_{KK}$ &$5n_{KK}$& $\frac{9}{5}+ 3n_{KK}$&$\frac{3}{25}$& $\frac{18}{155}$\\ 
$(1, 7)$ &  $(2,1,3)$&$n_{KK}$ &$7n_{KK}$& $\frac{16}{9}+\frac{28  }{9}n_{KK}$&$\frac{1}{7}$& $\frac{9}{65}$\\ 
$(1,9)$ &  $(2,2,4)$&$n_{KK}$ & $9n_{KK}$&$\frac{50}{29}+ \frac{90 }{29} n_{KK}$&$\frac{16}{87}$& $\frac{16}{91}$\\ 
$(3, 7)$ &  $(1,1,2)$&$3n_{KK}'$ & $7n_{KK}'$&$\frac{25}{13} + \frac{105}{13}n_{KK}'$&$\frac{1}{91}$& $\frac{6}{553}$\\ 
$(3, 11)$ &  $(2,2,4)$&$3n_{KK}'$ & $11n_{KK}'$&$\frac{98}{53} + \frac{462}{53}n_{KK}'$&$\frac{16}{583}$& $\frac{48}{1793}$\\ 
 \hline
\end{tabular}
\caption{Examples of anti-twist coprime hyperscalar spindle solutions with $r_N=r_S=0$ and hence $p_B=0$.
The KK spectrum of the BPS hyperscalar perturbation of the STU solution is given, with either $n_{KK}\ge 0$ (first 3) 
or $n_{KK}\ge -1$ (last 2), and the minimum value giving the relevant deformation of the STU solution to
the hyperscalar solution; 
for compactness, $n_{KK}'\equiv n_{KK}+1$.
Also given are the central charges of the STU solution and the hyperscalar solution.
\label{table1}
}
\end{center}
 \end{table}
In \cite{Arav:2022lzo} an example with $(n_N,n_S)=(1, 9)$ and $p^I=(3,1,4)$ was presented; since
$p^1=3$ is not coprime to $n_S=9$ this is, in fact, not a smooth solution, in contrast to what was implicitly assumed in \cite{Arav:2022lzo}. 
Numerical plots for the solution with $(n_N, n_S)=(1, 7)$ in table 
\ref{table1} were given in \cite{Arav:2022lzo}.

\subsubsection{Non-coprime spindles}
For non-coprime spindles, in addition to specifying $(n_N,n_S, p^I)$ we also need to specify 
$m_{N,S}^I\in\mathbb{Z}^3_{n_{N,S}}$.
The smooth solutions require each individual flux $p^I$ 
satisfy 
$\mathrm{hcf}(p^I,n_N)=\mathrm{hcf}(p^I,n_S)=\mathrm{hcf}(n_N,n_S)$ as in
\eqref{coprimelamndannns}. But
we also require 
$\mathrm{hcf}(m_N^I,n_N)=1$ and $\mathrm{hcf}(m_S^I,n_S)=1$. Furthermore, we
also must satisfy the conditions
\eqref{eq:zeta_scI_m2}, \eqref{rsymms2}.
 Some solutions are summarised
in tables \ref{table2p1}-\ref{tablehfoldsu2}.

There are certainly examples of $h$-fold flux quotients of coprime spindles with non-trivial hyperscalars, 
all of which are smooth. Recalling the examples of anti-twist STU solutions discussed at the end of section 
\ref{sec:stusols}, we note that there is a hyperscalar solution $(n_N,n_S)=3(1, 5)$ with $(p^1,p^2,p^3)=
3(1,1,2)$, summarised in table \ref{table2p1}, which is the 3-fold flux quotient of the coprime solution with $(n_N,n_S)=(1, 5)$ with $(p^1,p^2,p^3)=
(1,1,2)$ in table \ref{table1}. 
In the table we give the values of $m_{N,S}^I$, $r_{N,S}$ and the tower
of KK modes about the STU solution; the mode with $n_{KK}=0$, or $n_{KK}=-1$ if the solution is labelled by $n_{KK}'$, with $n_{KK}'\equiv n_{KK}+1$, is the relevant mode
which can generate an RG flow from the STU solution to the hyperscalar solution and notice that the central charges
are consistent with the RG scenario.

   \begin{table}[h!]
 \begin{center}
 $(n_N,n_S)=3(1,5),~p^I = 3(1,1,2)$ \\ \smallskip
\begin{tabular}{ |c|c|c|c|c|c|c| } 
 \hline
 $m_N^I$ &  $m_S^I$ & $r_N$&$r_S$&$\Delta$&$\frac{1}{N^2}c_{STU}$& $\frac{1}{N^2}c_H$\\ 
 \hline 
$(2,2,1)$ & $(11, 11,7)$ & $3 n_{KK}'$&$15 n_{KK}'$& $\frac{9}{5}+9n_{KK}'$&$\frac{1}{25}$& $\frac{6}{155}$\\ 
 \hline
\end{tabular}
\\ \bigskip
$(n_N,n_S)=5(1,5),~p^I=5(1,1,2)$ \\ \smallskip
\begin{tabular}{ |c|c|c|c|c|c|c| } 
 \hline
 $m_N^I$ &  $m_S^I$ & $r_N$&$r_S$& $\Delta$& $\frac{1}{N^2}c_{STU}$&$\frac{1}{N^2}c_H$\\ 
 \hline 
$(1,1,2)$ & $(6, 6,12)$ & $5 n_{KK}$&$25 n_{KK}$& $\frac{9}{5}+15n_{KK}$&$\frac{3}{125}$& $\frac{18}{775}$\\ 
$(3,3,3)$ & $(16,16,17)$ & $3+5 n_{KK}$&$15+25 n_{KK}$&  $\frac{54}{5}+15n_{KK}$&$\frac{3}{125}$& $-$\\ 
$(4, 4,1)$ & $(21,21,7)$ & $2+5n_{KK}'$&$10+25n_{KK}'$& $\frac{39}{5}+15n_{KK}'$&$\frac{3}{125}$& $-$\\ 
\hline
\hline
$(4,3,2)$ & $(21,16,12)$ & $5n_{KK}'$&$25n_{KK}'$&$\frac{9}{5}+15n_{KK}'$&  $\frac{3}{125}$& $\frac{18}{775}$\\ 
$(3,2,4)$ & $(16,11,22)$ & $1+5n_{KK}$&$5+25n_{KK}$& $\frac{24}{5}+15n_{KK}$& $\frac{3}{125}$& $-$\\ 
$(4,1,4)$ & $(21,6,22)$ & $1+5n_{KK}$&$5+25n_{KK}$& $\frac{24}{5}+15n_{KK}$& $\frac{3}{125}$&$-$\\ 
$(2,1,1)$ & $(11,6,7)$ & $2+5n_{KK}$&$10+25n_{KK}$& $\frac{39}{5}+15n_{KK}$& $\frac{3}{125}$& $-$\\ 
$(4,2,3)$ & $(22,11,17)$ & $3+5n_{KK}$&$15+25n_{KK}$& $\frac{54}{5}+15n_{KK}$& $\frac{3}{125}$&$-$\\ 
$m_1 \leftrightarrow m_2 $ & $\dots$ & $\dots$&$\dots$&$\dots$& $\dots$& $\dots$\\ 
 \hline
\end{tabular}
\caption{
Examples of non-coprime hyperscalar solutions with $p_B=0$.
The KK spectrum of the BPS hyperscalar perturbation of the STU solution is given, as well
as the central charges of the STU solution and the hyperscalar solution.
$n_{KK}'\equiv n_{KK}+1$: the spectra have $n_{KK}\ge 0$ (those labelled by $n_{KK}$) or  $n_{KK}\ge -1$ (those labelled by $n_{KK}'$).
Top table: $(n_N, n_S)=3(1,5)$ and $p^I=3(1,1,2)$.
Bottom table: 
$(n_N, n_S)=5(1,5)$ and $p^I=5(1,1,2)$, so $p_B=0$. 
The top 3 STU solutions preserve $SU(2)$ flavour symmetry, while the bottom 10 solutions  break the flavour symmetry to $U(1)$ (only 5 solutions given, the other 5 are obtained by exchanging $m_1 \leftrightarrow m_2$).
\label{table2p1}}
\end{center}
 \end{table}

We next highlight that if we start with the smooth coprime hyperscalar solution $(n_N,n_S)=(1, 5)$ with $(p^1,p^2,p^3)=(1,1,2)$
 in table \ref{table1}, we might wonder about the existence of a 2-fold flux quotient, non-coprime solution with
$(n_N,n_S)=2(1, 5)$ with $(p^1,p^2,p^3)=2(1,1,2)$. Like for the STU solutions, this case is obstructed
since $n_N$ and $n_S$ are both odd and some of the fluxes are odd in the coprime solution, so there is no
regular 2-fold flux quotient solution. In fact there appears to be no hyperscalar solution with
$(n_N,n_S)=h(1, 5)$ with $(p^1,p^2,p^3)=h(1,1,2)$, for any even $h$.

Another interesting illustrative example is the case 
 $(n_N, n_S)=5(1,5)$ with fluxes given by $p^I=5(1,1,2)$, which is a 5-fold flux quotient
 of the coprime hyperscalar spindle and summarised in table \ref{table2p1}. It has $p_B=0$ and also $p_F=0$. Since $p_F=0$ the corresponding
 STU solution
 preserves an $SU(2)$ flavour symmetry (after uplift on $S^5$).
 We find that there are 64 regular uplifts, satisfying the coprime conditions \eqref{coprimecondsmssum};
 however only 13 are supersymmetric, satisfying \eqref{rsymms2}.
 Of these 13 regular supersymmetric solutions, 3 also preserve the $SU(2)$ flavour symmetry, since $m^1_N=m^2_N$ and $m^1_S=m^2_S$, but
 the other 10 do not. Notice that of all these 10 STU solutions, only some of the choices of 
 $m_{N,S}^I$ are compatible with having a hyperscalar solution. Furthermore, for these cases 
 the STU solutions have a relevant hyperscalar deformation and the central charges of the STU solutions
 and the hyperscalar solutions have central charges 
  in exact alignment with the RG scenario.

Some additional non-coprime solutions with $p_B=0$ are summarised in table \ref{tablehfoldsu2}, 
which are all $h$-fold flux quotients of teardrop spindles. Observe
the absence of hyperscalar solutions with $n_N$ even, as deduced in \eqref{ennoddconc}.
 \begin{table}[h!]
 \begin{center}
\begin{tabular}{ |c|c|c||c|c|} 
 \hline
 $(n_N, n_S)$ &STU solutions& $SU(2)$ invt.& hyper solutions & $SU(2)$ invt.\\ 
 \hline 
$\text{even}(1, 5)$ & 0   &-& 0&- \\ 
$3(1, 5)$ & 1 & $1$&1 &1\\ 
$5(1, 5)$ & 13 & $3$& 3&1\\ 
$7(1, 5)$ & 19& $3$& 3&1\\ 
$9(1, 5)$ & 9 & $3$&3 &1\\ 
\hline
\hline
$2(1, 9)$ & 1 &1& 0&-\\ 
$3(1, 9)$ & 3 &1& 1&1\\ 
$4(1, 9)$ & 4 &2& 0&-\\
$5(1, 9)$ & 5 &3& 3&1\\
$6(1, 9)$ & 3 &1& 0&-\\
 \hline
\end{tabular}
\caption{Examples of non-coprime hyperscalar spindle solution with $p_B=0$ associated with
the $(n_N,n_S)=(1,5)$, $p^I=(1,1,2)$ and $(n_N,n_S)=(1,9)$, $p^I=(2,2,4)$ solutions of table \ref{table1}.
We have enumerated the number of STU and hyperscalar solutions that exist, and whether or not they
preserve the $SU(2)_F$ symmetry.\label{tablehfoldsu2}}
\end{center}
 \end{table}

\subsection{Hyperscalar vanishing at one of the poles, $p_B\ne 0$ }\label{onepole}
We now consider hyperscalar solutions that are non-vanishing at one of the poles and hence
have $p_B\ne0$.
Recalling that we are considering \eqref{assumptionshscsol} we have $r_N=0$ and
so we are necessarily considering solutions with $r_S\ge 1$.

\subsubsection{Coprime spindles}
We first consider the case of hyperscalar solutions for coprime spindles.
Some solutions, including associated STU solutions, are summarised in\footnote{Note, in particular, that we have found no smooth, supersymmetric teardrop spindle hyperscalar solutions with $(n_N,n_S)=(1,n_S)$ with $n_S<16$.} table \ref{table3}. 
For the two cases of $(n_N, n_S)=(1,16)$ and $(n_N, n_S)=(1,17)$ numerical plots of
the solutions are given in 
figures \ref{solnexample1}, \ref{solnexample2} in appendix \ref{plotssols}.
Notice the absence of hyperscalar solutions with $n_N$ even in table \ref{table3}, as deduced in \eqref{ennoddconc}.
All solutions in the table are consistent with 
the RG scenario. 
\begin{table}[h!]
 \begin{center}
\begin{tabular}{ |c|c|c|c|c|c|c| } 
 \hline
 $(n_N, n_S)$ & $(p^1,p^2,p^3)$&  $\hm_N$&$\hm_S$&$\Delta$& $\frac{1}{N^2}c_{STU}$ &$\frac{1}{N^2}c_H$\\ 
 \hline
$(1, 16)$& $(5,3,7)$ & $n_{KK}$ &$1+16n_{KK}$&$\frac{160}{87}+\frac{272}{87} n_{KK}$& $\frac{105}{464}$& $\frac{12 \, 915}{57\,586}$\\ 
$(1, 17)$ & $(5,4,7)$& $n_{KK}$ &$2+17n_{KK}$&$\frac{97}{50}+\frac{153}{50} n_{KK}$& $\frac{21}{85}$& $\frac{18\,060}{73 \,253}$\\ 
$(1, 17)$& $(6,3,7)$ & $n_{KK}$ &$2+17n_{KK}$&$\frac{97}{49}+\frac{153}{49}n_{KK}$& $\frac{27}{119}$& $\frac{17\,010}{74\, 987}$\\ 
$(3, 43)$ & $(13,10,17)$& $3n_{KK}'$ &$2+43n_{KK}'$&$\frac{649}{325}+\frac{2967 }{325}n_{KK}'$& $\frac{17}{215}$& $\frac{576\,810}{7\,294\,993}$\\ 
$(3, 47)$& $(14,11,19)$ & $3n_{KK}'$ & $2+47n_{KK}'$&$\frac{757}{385}+\frac{3525}{385} n_{KK}'$&$\frac{19}{235}$&$\frac{939\,246}{11\,624\,933}$\\ 
$(3, 49)$& $(13,13, 20)$ & $3n_{KK}'$ &$2+49n_{KK}'$&$\frac{37}{19}+ \frac{1911}{209} n_{KK}'$&$\frac{845}{10\, 241}$& $\frac{59\,319}{720\,104}$ \\ 
 \hline
\end{tabular}
\caption{Examples of anti-twist coprime hyperscalar solutions with $r_N=0$, $r_S\ge 1$ and $p_B\ne 0$.
The KK spectrum of the BPS hyperscalar perturbation of the STU solution is given as well
as the central charge of the STU solution and the hyperscalar solution.
The first three examples have modes starting at $n_{KK}=0$ and the latter three at $n_{KK}=-1$,
 and $n_{KK}'\equiv n_{KK}+1$.\label{table3}}
\end{center}
 \end{table}

\subsubsection{Non-coprime spindles}
We now turn to the non-coprime case and $p_B\ne 0$. We can illustrate with various $h$-fold
flux quotients of the $(n_N,n_S)=(1,16)$ case (first line in table \ref{table3}). For hyperscalar solutions
we necessarily have $h$ odd consistent with \eqref{ennoddconc}.
The number of smooth STU solutions and associated hyperscalar solutions, when they exist, are summarised in table \ref{table6}
and all are consistent with 
the RG scenario.
 \begin{table}[h!]
 \begin{center}
\begin{tabular}{ |c|c|c|c|c| } 
 \hline
 $(n_N, n_S)$ &  STU solutions & hyperscalar solutions\\ 
 \hline 
$2(1, 16)$ & 1  & $0$ \\ 
$3(1, 16)$ & 1  & $1$ \\ 
$4(1, 16)$ & 4  & $0$ \\ 
$5(1, 16)$ & 8  & $3$ \\ 
$6(1, 16)$ & 1  & $0$ \\ 
$7(1, 16)$ &22  & $4$ \\ 
 \hline
\end{tabular}
\caption{Examples of anti-twist non-coprime solutions with $p_B\ne 0$.
With $(n_N, n_S)= h(1, 16)$, we have
$c_{STU}=\frac{1}{h}c_{STU}(1,16)$ and when a hyperscalar solution exists
$c_H=\frac{1}{h}c_H(1,16)$ and $\Delta=\Delta(1,16)=160/87$.
\label{table6}}
\end{center}
 \end{table}

\subsection{Hyperscalar modes for STU solutions in the anti-twist class with equal $p^I$}\label{atmgs}

We now discuss some examples of anti-twist STU solutions with $p^1=p^2=p^3\equiv p$
that were discussed at the end of section \ref{sec:stusols} and also in appendix \ref{appCmingaugedsugra}. These solutions have $p=(n_S-n_N)/3$. 
If $n_N$ and $n_S$ are both not divisible by 3, after uplift on $S^5$, there are unique smooth, supersymmetric solutions that preserve $SU(3)$ flavour symmetry with 
$m^1_N=m^2_N=m^3_N\equiv m_N$
and $m^1_S=m^2_S=m^3_S\equiv m_S$ (which are then solutions of minimal gauged supergravity with $A^1=A^2=A^3$). 
However, for certain such $n_N,n_S$, in the non-coprime case, there can
also be smooth, supersymmetric solutions that break the 
$SU(3)$ flavour symmetry. In addition, if $n_S$ and $n_N$ are
divisible by 3, so there are no $SU(3)$ invariant solutions, for
certain $n_N,n_S$ there can be $SU(3)$ non-invariant solutions. 
For example, for $(n_N,n_S)=3(1,4)$ there are no smooth supersymmetric solutions
but there are $SU(3)$ breaking solutions for $(n_N,n_S)=3(1,10)$.
Some examples are presented in tables \ref{tableat1}, \ref{tableat2}.  We also
give the spectrum of hyperscalar modes and we recall from the end of section \ref{seclinfluct}
that there are never
any relevant hyperscalar modes for this class.
\begin{table}[h!]
 \begin{center}
\begin{tabular}{ |c|c|c|c|c|c|c| } 
 \hline
 $(n_{N},n_S)$ &  $p^I$ & $r_N$&$r_S$&$\Delta$ &$\frac{1}{N^2}c_{STU}$\\ 
 \hline 
$(1,4)$ & $(1,1,1)$ & $n_{KK}$&$1+4n_{KK}$& $\frac{16}{7} + \frac{20}{7} n_{KK}$ &$\frac{3}{28}$\\ 
$(2,5)$ & $(1,1,1)$ & $1+2n_{KK}$&$3+3n_{KK}$& $\frac{64}{13} + \frac{70}{13} n_{KK}$ &$\frac{3}{130}$\\ 
$(1,7)$ & $(2,2,2)$ & $n_{KK}$&$2+7n_{KK}$& $\frac{44}{19} + \frac{56}{19} n_{KK}$ &$\frac{24}{133}$\\ 
$(5,11)$ & $(2,2,2)$ & $3+5n_{KK}$&$7+11n_{KK}$& $\frac{676}{67} + \frac{880}{67} n_{KK}$ &$\frac{24}{3685}$\\ 
$(1,10)$ & $(3,3,3)$ & $n_{KK}$&$3+10n_{KK}$& $\frac{86}{37}+\frac{110}{37} n_{KK}$&$\frac{81}{370}$\\ 
$(2,11)$ & $(3,3,3)$ & $1+2n_{KK}$&$7+11 n_{KK}$& $\frac{256}{49}+\frac{286}{49} n_{KK}$&$\frac{81}{1078}$\\ 
 \hline
\end{tabular}
\caption{Examples of smooth STU solutions in the anti-twist class with
$p^1=p^2=p^3\equiv p$ and coprime $n_N$, $n_S$. They have a unique smooth
uplift on $S^5$ which is $SU(3)$ invariant.
We also give the spectrum of hyperscalar modes. 
\label{tableat1}}
\end{center}
 \end{table}

   \begin{table}[h!]
 \begin{center}
 $(n_N,n_S)=2(1,4),~p^I=2(1,1,1)$\\ \smallskip
\begin{tabular}{ |c|c|c|c|c|c|c| } 
 \hline
 $m_N^I$ &  $m_S^I$ & $r_N$&$r_S$& $\Delta$&$\frac{1}{N^2}c_{STU}$\\ 
 \hline 
$(1,1,1)$ & $(5,5,5)$ & $1+2n_{KK}$&$5+8n_{KK}$& $\frac{36}{7}+\frac{40}{7} n_{KK}$&$\frac{3}{56}$\\ 
 \hline
\end{tabular}
\\ \bigskip
 $(n_N,n_S)=5(1,4),~p^I=5(1,1,1)$\\ \smallskip
\begin{tabular}{ |c|c|c|c|c|c|c| } 
 \hline
 $m_N^I$ &  $m_S^I$ & $r_N$&$r_S$& $\Delta$&$\frac{1}{N^2}c_{STU}$\\ 
 \hline 
$(3,3,3)$ & $(13,13,13)$ & $3+5n_{KK}$&$13+20n_{KK}$& $\frac{76}{7}+\frac{100}{7}n_{KK}$&$\frac{3}{140}$\\ 
 \hline
 \hline
$(4,3,2)$ & $(17,13,9)$ & $5n_{KK}'$&$1+20 n_{KK}'$& $\frac{16}{7}+\frac{100}{7} n_{KK}'$&$\frac{3}{140}$ \\
$(4,2,3)$ & $(17,9,13)$ & $3+5n_{KK}$&$13+20 n_{KK}$& $\frac{76}{7}+\frac{100}{7} n_{KK}$&$\frac{3}{140}$ \\
$(3,2,4)$ & $(13,9,17)$ & $1+5n_{KK}$&$5+20 n_{KK}$& $\frac{36}{7}+\frac{100}{7} n_{KK}$&$\frac{3}{140}$ \\
$m_1 \leftrightarrow m_2$ & $\ldots$ & $\ldots$&$\ldots$& $\ldots$&$\ldots$ \\
\hline
\end{tabular}
\\ \bigskip
 $(n_N,n_S)=3(1,10),~p^I=3(3,3,3)$\\ \smallskip
\begin{tabular}{ |c|c|c|c|c|c|c| } 
 \hline
 $m_N^I$ &  $m_S^I$ & $r_N$&$r_S$& $\Delta$&$\frac{1}{N^2}c_{STU}$\\ 
 \hline 
$(2,2,1)$ & $(23,23,13)$ & $3n_{KK}'$&$3+30 n_{KK}'$& $\frac{86}{37}+\frac{330}{37} n_{KK}'$& $\frac{27}{370}$ \\
$(2,1,2)$ & $(23,13,23)$ & $1+3n_{KK}$&$13+30n_{KK}$& $\frac{196}{37}+\frac{330}{37} n_{KK}$&$\frac{27}{370}$ \\
$(1,2,2)$ & $(13,23,23)$ & $1+3n_{KK}$&$13+30n_{KK}$& $\frac{196}{37}+\frac{330}{37} n_{KK}$&$\frac{27}{370}$ \\
\hline
\end{tabular}
\caption{Examples of non-coprime anti-twist STU solutions 
with $p^1=p^2=p^3\equiv p$.
Top table: there is a unique $SU(3)$ invariant solution. Middle table: there is both
an $SU(3)$ invariant solution and $SU(3)$ breaking solutions. 
Bottom table: there are no $SU(3)$ invariant solutions, but there are $SU(3)$ breaking
solutions (that preserve an $SU(2)$ flavour symmetry).
\label{tableat2}}
\end{center}
 \end{table}

\subsection{Hyperscalar modes for STU solutions in the twist class}\label{texampleshsfs}
We conclude this section by considering the spectrum of hyperscalar modes
about some STU solutions in the twist class. In all cases
the modes are dual to irrelevant operators and, consistent with the RG scenario,
we find no hyperscalar $AdS_3\times\Sigma$ solutions in the twist class.

In table \ref{table7} we present some examples of smooth coprime STU twist solutions and the hyperscalar modes. 
We require that
the hyperscalar mode has $r_N\ge 0$ and $r_S\ge 0$ and this allows us to count the
finite number of hyperscalar modes, which we denote by $\#(\Delta)$ in the table.
The last entry in table \ref{table7} is interesting is that we find modes with
$r_N=7n_{KK}-9$ and 
$r_S=13-8n_{KK}$, and so there is no integer $n_{KK}$ which gives rise
to $r_N\ge 0$ and $r_S\ge 0$ and hence there are no BPS hyperscalar modes for this STU solution.
The first entry in table \ref{table7} is an $AdS_3\times S^2$ (topological twist) solution. 
The values of $\Delta$ for the 7 hyperscalar modes are all given by $\Delta=3$ and this degeneracy
arises from the $SO(3)$ isometry of the round $S^2$ (recall the discussion at the end
of section \ref{seclinfluct} and \eqref{deltas2case}).

In table \ref{3folds2} we summarise some $h$-fold quotients of $AdS_3\times S^2$ solutions, with
$(n_N,n_S)=h(1,1)$,
that were discussed at the end of section \ref{sec:stusols}. Here we also include
the spectrum of hyperscalar modes. Again we observe the degeneracy of modes with the same value of $\Delta$ arising from \eqref{deltas2case}, with a possible reduction of the degeneracy compared to the case of $h=1$.
\begin{table}[h!]
 \begin{center}
\begin{tabular}{ |c|c|c|c|c|c|c| } 
 \hline
 $(n_{N},n_S)$ &  $p^I$ & $r_N$&$r_S$&$\Delta$ &$\#(\Delta)$&$\frac{1}{N^2}c_{STU}$\\ 
 \hline 
$(1,1)$ & $(1,1,-4)$ & $n_{KK}$&$6-n_{KK}$& $3$ &$7$&$\frac{3}{2}$\\ 
$(1,2)$ & $(1,1,-5)$ & $n_{KK}$&$7-2n_{KK}$& $\frac{2}{11}(19-n_{KK})$&$4$&$\frac{15}{22}$\\ 
$(2,3)$ & $(1,1,-7)$ & $2n_{KK}-1$&$6-3n_{KK}$& $\frac{2}{19}(35-3n_{KK})$&$2$&$\frac{7}{38}$\\ 
$(4,5)$ & $(1,1,-11)$ & $4n_{KK}-3$&$7-5n_{KK}$& $4-\frac{20}{41}n_{KK}$&$1$&$\frac{33}{820}$\\ 
$(5,6)$ & $(1,1,-13)$ & $5n_{KK}-5$&$9-65n_{KK}$& $\frac{232}{55}-\frac{30}{55}n_{KK}$&$1$&$\frac{13}{550}$\\ 
$(7,8)$ & $(1,1,-17)$ & $-$&$-$& $-$&$0$&$\frac{51}{4984}$\\ 
 \hline
\end{tabular}
\caption{Examples of smooth STU solutions in the twist class and their spectrum of BPS hyperscalar modes. In the last example, there are no supersymmetric hyperscalar modes.
\label{table7}}
\end{center}
 \end{table}

   \begin{table}[h!]
 \begin{center}
 $(n_N,n_S)=3(1,1),~p^I=3(1,1,-4)$\\ \smallskip
\begin{tabular}{ |c|c|c|c|c|c|c| } 
 \hline
 $m_N^I$ &  $m_S^I$ & $r_N$&$r_S$& $\#(\Delta)$&$\frac{1}{N^2}c_{STU}$\\ 
 \hline 
$(1,1,2)$ & $(2,2,1)$ & $3n_{KK}$&$6-3n_{KK}$& $3$&$\frac{1}{2}$\\ 
 \hline
\end{tabular}\\ \bigskip
 $(n_N,n_S)=5(1,1),~p^I=5(1,1,-4)$\\ \smallskip
\begin{tabular}{ |c|c|c|c|c|c|c| } 
 \hline
 $m_N^I$ &  $m_S^I$ & $r_N$&$r_S$& $\#(\Delta)$&$\frac{1}{N^2}c_{STU}$\\ 
 \hline 
$(2,2,2)$ & $(3,3,3)$ & $5n_{KK}-2$&$8-5n_{KK}$& $1$&$\frac{3}{10}$\\ 
 \hline
 \hline
$(3,2,1)$ & $(4,3,2)$ & $5n_{KK}-4$&$10-5n_{KK}$& $2$&$\frac{3}{10}$ \\
$(3,1,2)$ & $(4,2,3)$ & $5n_{KK}-2$&$8-5n_{KK}$& $1$&$\frac{3}{10}$ \\
$(2,1,3)$ & $(3,2,4)$ & $5n_{KK}$&$6-5n_{KK}$& $2$&$\frac{3}{10}$ \\
$m_1 \leftrightarrow m_2$ & \ldots &  \ldots & \ldots & \ldots &\ldots \\
\hline
\end{tabular}
\caption{Examples of non-coprime twist solutions associated with $h$-fold
flux quotients of the round $S^2$
solution with $p^I=(1,1,-4)$. The top table summarises the single smooth supersymmetric solution
for $(n_N,n_S)=3(1,1)$ and $p^I=3(1,1,-4)$ while the bottom table 
summarises the 7 smooth supersymmetric solutions
for $(n_N,n_S)=5(1,1)$ and $p^I=5(1,1,-4)$.
In all cases the (degenerate) operators all have dimension $\Delta=3$ as in the round $S^2$ case.
\label{3folds2}}
\end{center}
 \end{table}

\section{Equivariant Localization}\label{eqlocalsec}
In this section we change tack and show how the central charge for the $AdS_3\times \Sigma$ solutions
with non-vanishing hyperscalars can be obtained using localization without
solving the BPS equations, just assuming they exist. This is a direct generalisation of the results of 
\cite{BenettiGenolini:2024kyy}. At the end of the section
we also make a connection with field theory via
an analysis of anomaly polynomials, generalising the results of \cite{Arav:2022lzo}.

We begin by recalling some key points of the analysis of \cite{BenettiGenolini:2024kyy}.
We work in conformal gauge with metric as in \eqref{confgaugemetric}:
\begin{align}\label{confgaugemetric2}
 \dd s_5^2 &= \ex^{2V}\left(\dd s^2(AdS_3)+ds^2(\Sigma)\right),\qquad ds^2(\Sigma)=dy^2+ g^2 dz^2\,,
\end{align}
where $g=g(y)$. We assume the Killing spinor $\epsilon$ has form $\epsilon = \psi\otimes e^{V/2}\zeta$,
where $\psi$ is a Killing spinor on $AdS_3$ and
$\zeta$ is a spinor on $\Sigma$ (as in \eqref{epsilon3plus2}, \eqref{spinorsbps}).
The $D=3$ spinor satisfies $D_i \psi = \frac{\kappa}{2} \beta_i\psi$, as in \eqref{eq:3d_cks_eqt},
and now we restrict in this section to
\begin{align}\label{kappachoice}
\kappa=+1\,.
\end{align}
A $D=2$ action on $\Sigma$ can be obtained by substituting the ansatz \eqref{confgaugemetric2} into the $D=5$ action \eqref{d3lagoverall}, and this then gives rise to the correct $D=2$ equations of motion. After using the trace of the $D=5$ Einstein equations, this leads to a ``partially off-shell action'' given by 
\begin{equation}\label{actionLS}
    S_2 |_{\text{POS}}=\frac{2}{3}\int_{\Sigma}\Big[\ex^{5V}\mathcal{V} \, \vol-\frac{1}{2}\sum_{I=1}^3\ex^V(X^I)^{-2}F^I_{12} F^I\Big]\,,
\end{equation}
where $F^I\equiv F^I_{12}\vol$ and $\vol$ is the volume form on $\Sigma$. This can be used to define 
a trial central charge function
\begin{align}\label{cLScaseitofa}
c=-\frac{3}{\pi}a_{\mathcal{N}=4}\, S_2 |_{\text{POS}}\,,
\end{align}
which, on-shell, is the central charge of the $AdS_3$ solution.

We can define\footnote{Writing the Killing spinors as in \eqref{spinorsbps}, with a fixed normalisation,
these spinor bilinears are given by \eqref{5dbilinears}. To compare with previous sections,
with this normalisation, we have $S=1$, $P=-\cos\xi$ and $P_{N,S}=(-1)^{t_{N,S}+1}$. We will also find it convenient in this section to write $k=1/b_0$, since $b_0$ was the notation used in \cite{BenettiGenolini:2024kyy}.
We also highlight that the important conditions
\eqref{eq:gauge_bpst},\eqref{eq:gauge_bpst2} are equivalent to the existence of the equivariantly closed form \eqref{gfieldecf}, discussed below. }
the following real bilinears in $\zeta$:
\begin{equation}\label{5dbilinearssec6}
    S = \zeta^\dagger\zeta\,, \qquad P = \zeta^\dagger\gamma_3\zeta\,, 
    \qquad \xi^\mu = -\ii\zeta^\dagger\gamma^\mu\gamma_3\zeta\,.
\end{equation}
They satisfy various algebraic and differential conditions and, in particular, $\xi^\mu$ 
is a Killing vector on $\Sigma$.  The one-form $\xi^\flat$
dual to the Killing vector satisfies 
\begin{align}\label{dxiflattext5d}
\dd \xi^\flat & = -2\Big(2+PS^{-1}\ex^{V}W\Big)P\, \vol\,.
\end{align}

There are several canonically defined equivariantly closed forms which can be constructed from
the bilinears and the supergravity fields. Associated with the action \eqref{actionLS}, the polyform 
\begin{align}\label{LSACTFORM}
    \Phi^S
        =& \frac{2}{3}\Big[\ex^{5V}\mathcal{V}\, \vol-\frac{1}{2}  \sum_{I=1}^3\ex^V(X^{I})^{-2}F^I_{12}
    {F^{I}}\Big]
    +\Big[2e^{3V} P+\frac{2}{3}e^{4V}W S^{-1}(P^2-S^2)\Big]\,,
\end{align}
is equivariantly closed, $d_\xi \Phi^{{S}}=0$, where $d_\xi\equiv (d-\xi \hook)$.
Associated with the gauge fields, we find
\begin{align}\label{gfieldecf}
    \Phi^{{I}}
    &\equiv F^I+x^I\,,
\end{align}
is also equivariantly closed, $d_\xi\Phi^{{I}}=0$, where the dressed scalar fields are
\begin{align}
x^I\equiv -X^{I} \ex^V  P\,.
\end{align}
In addition, the presence of the hyperscalar implies that
the linear combination of field strengths associated with $U(1)_B$ satisfies
 \begin{align}\label{LSdthetacon}
 \zeta_I\Phi^{I}=  - d_\xi D\theta\,.
\end{align}
If $D\theta$ is a globally defined one-form then $ \zeta_I\Phi^{I}$ is clearly equivariantly exact. This
is the case if the hyperscalar is no-where vanishing. However, we will be interested
in cases when the hyperscalar vanishes at one or both of the poles of the spindle and in this case
$D\theta$ is not globally defined.

\subsection{Localization}
The globally defined Killing vector is taken to be $\xi=b_0\partial_z$, with $\Delta z=2\pi$ and $b_0$ a constant.
We take $y\in [y_N,y_S]$ with the poles of the spindle, located at $y=y_{N,S}$, being fixed points of the action of $\xi$.
We assume they have $\mathbb{Z}_{n_{N,S} }$ orbifold singularities with $n_{N,S} >0$. This can be achieved by assuming
$g\to \frac{1}{n_N}(y-y_N)$ and $g\to \frac{1}{n_S}(y_S-y)$ as one approaches the poles. Taking $\vol= g dy\wedge dz$ 
the weights of the action of the Killing vector at the poles of the spindle, defined by
$d\xi^\flat |_{N,S}=2\epsilon_{N,S}\vol$,
are given by $\epsilon_N=b_0/n_N$ and
$\epsilon_S=-b_0/n_S$.

The BVAB theorem allows us to evaluate the integral of the two-form component
of an equivariantly closed form in terms of the zero form component evaluated on the two fixed points of $\xi$.
The bilinear $S$ is a positive constant and we now
normalise $S=1$ (as in previous sections). 
At a fixed point it can be shown that $P^2=S^2=1$, so $P=\pm1$ at a fixed point and, from \eqref{5dbilinearssec6},
the sign is associated with the chirality of the spinor at the fixed point. To use the BVAB formula we need the weights of the action of the Killing vector at the poles of the spindles as given above.
We leave the signs of the spinor chirality at the poles, $P_{N,S}$, arbitrary. 

Applying the BVAB formula to \eqref{LSACTFORM} and using the definition
of the dressed scalars $x^I$ in \eqref{gfieldecf}, we find
the action can be written in the form
\begin{equation}\label{LSSPSt}
\begin{aligned}
    S_2|_{\text{POS}} &=\frac{4\pi}{b_0} \left[\mathcal{F}(\newc^I_S)-\mathcal{F}(\newc^I_N)\right]\,,
\end{aligned}
\end{equation}
where $\mathcal{F}$ is the prepotential in \eqref{prepotdef} and $x^I_{N,S}$ are the values of the
dressed scalars at the poles. Thus, the partially off-shell action and hence the central charge can
be computed just by knowing $b_0$ and the values of  $x^I_{N,S}$. 

Applying the BVAB formula to \eqref{gfieldecf} we also obtain the following expression for the three magnetic fluxes
through the spindle:
\begin{align}\label{LSBVFLUX}
\frac{p^I}{n_Nn_S}&\equiv \int_{\Sigma} \frac{F^{I}}{2\pi} =-\frac{1}{b_0}(\newc^I_S-\newc^I_N)\,.
     \end{align}
As discussed in section \ref{smoothupliftssec},
regularity of the uplifted solution requires that $p^I$ are all integers, and
in addition that $\mathrm{hcf}(p^I,n_N)=\mathrm{hcf}(p^I,n_S)=\mathrm{hcf}(n_N,n_S)$.
When we have a non-coprime spindle with $\mathrm{hcf}(n_N,n_S)\ne 1$ we also need to specify
the integers $m_{N,S}^I\in\mathbb{Z}^3_{n_{N,S}}$ satisfying \eqref{condfluxsmooth2}, \eqref{coprimecondsms}, as well as
additional conditions, mentioned below.

The value of the superpotential $W$ \eqref{superpottextLS} at the poles only depends
on the scalars in the vector multiplets:
\begin{align}\label{Wexppoles}
W|_{N,S} =\sum_{I=1}^3 X^I |_{N,S}=-\sum_{I=1}^3 e^{-V} P x^I |_{N,S}\,.
\end{align}
This is obvious when $\rho=0$ at the poles, but is also true when $\rho$ is non-zero at the poles, because in that case one can show
$\zeta_IX^I|_{N,S}=0$. We next evaluate \eqref{dxiflattext5d} (with $S=1$) at the poles.
Recalling that $\dd \xi^\flat |_{N,S}=2\epsilon_{N,S}\vol$, with the weights given by
 $\epsilon_N=b_0/n_N$ and
$\epsilon_S=-b_0/n_S$, we obtain
\begin{align}\label{RsymcsLSt}
\sum_{I=1}^3\newc^I_N=2+\frac{b_0}{n_N}P_N\,,\qquad
\sum_{I=1}^3\newc^I_S=2-\frac{b_0}{n_S}P_S\,.
\end{align}
From \eqref{LSBVFLUX}, this implies the R-symmetry flux is given by
\begin{align}\label{RsymconstLSt}
p_R={n_NP_S+ n_S P_N}\,.
\end{align}
In the non-coprime case we must also demand that  $m_{N,S}^I$ satisfy
\eqref{rsymms} for well defined spinors.

In previous work \cite{BenettiGenolini:2024kyy}, the hyperscalar was assumed to be non-vanishing at both poles. In this case 
$D\theta$ is globally defined and then \eqref{LSdthetacon} implies the broken flux is necessarily zero, $p_B=0$.
Here we are interested in cases where, instead, the hyperscalars smoothly go to zero,
in the orbifold sense, at one or both of the poles. 
Recalling the discussion below \eqref{hsregconds}
we must have
\begin{align}\label{kpmcond1t}
D\theta|_N=-P_N\frac{\hm_N}{n_N}dz\,,\qquad
D\theta|_S=P_S\frac{\hm_S}{n_S}dz\,,
\end{align}
 with $r_{N,S}\in \mathbb{Z}_{\ge 0}$.  The zeroth component of \eqref{LSdthetacon} then gives the constraints
\begin{align}\label{brokencsLSt}
\zeta_I \newc^I_N=-b_0 P_N\frac{\hm_N}{n_N}\,,\qquad
\zeta_I \newc^I_S=b_0 P_S\frac{\hm_S}{n_S}\,.
\end{align}
From \eqref{LSBVFLUX}, this then implies that the broken flux is constrained to be
\begin{align}\label{brokefluxcondt}
p_B=-\big(P_S\hm_S n_N+P_N \hm _N n_S\big)\,.
\end{align}
We also observe that if we set $\hm_{N,S}=0$ then 
\eqref{kpmcond1t}-\eqref{brokefluxcondt} are exactly the conditions used in \cite{Arav:2022lzo,BenettiGenolini:2024kyy} when the hyperscalar is non-vanishing at both poles of
the spindle. 
In the non-coprime case we must also demand that  $m_{N,S}^I$ satisfy \eqref{eq:zeta_scI_m2}.
We also highlight that the conditions are invariant under the following symmetry
\begin{align}\label{symmextconds}
x^I_N\leftrightarrow x^I_S\,,\quad
P_N\leftrightarrow P_S\,,\quad
n_N\leftrightarrow n_S\,,\quad
\hm_N\leftrightarrow \hm_S\,,\quad
b_0\to-b_0\,,
 \end{align}
 with the fluxes left unchanged.
 
We now take stock. We consider spindle data given by $n_{N,S}$, $P_{N,S}$, and $\hm_{N,S}$; in the non-coprime case we can specify the $m^I_{N,S}$ after the following.
This data determines the R-symmetry flux $p_R$ and the broken flux $p_B$ via \eqref{RsymconstLSt}
and \eqref{brokefluxcondt}, so there is just one independent flux $p_F$. We can then use
\eqref{LSBVFLUX} to determine $x^I_N$, while
the constraints \eqref{RsymcsLSt} and \eqref{brokencsLSt} 
determine two of the $x^I_S$.  Thus, for given spindle data $n_{N,S}$, choice of spinor chiralities 
at the poles $P_{N,S}$, independent flux parameter $p_F$ and behaviour of the hyperscalar as we approach the poles, 
$\hm_{N,S}$, the action \eqref{LSSPSt} depends on 
two variables, which we can take to be, for example, $\newc_S^1$  and $b_0$.

Recalling that the action is only partially on-shell, we should extremize the action over these
two variables to get the final on-shell action and hence the central charge.
Extremizing also gives $b_0$, and hence the weights of the R-symmetry Killing vector, as well as the
values at the poles of the warp factor and vector multiplet scalars, using
$e^{3V}=-P x^1 x^2 x^3$ and $X^I=-Pe^{-V} x^I$.
Notice, however, that this extremization procedure does not fix the value of the hyperscalar $\vpvar$ at the poles of the spindle:
if the hyperscalar vanishes then we need to input how it vanishes via the data $\hm_{N,S}$. 
On the other hand if the complex scalar is non-vanishing at both poles, its value does not enter into the evaluation of the on-shell action at all \cite{Arav:2022lzo,BenettiGenolini:2024kyy}. In particular, this extremization gives
the following result for the central charge and $b_0$:
\begin{align}\label{localcetc}
c&=\frac{3\gamma}{32\alpha}N^2    \,,\nn\\
b_0&=\frac{\beta}{\alpha}\,,
\end{align}
where 
\begin{equation}
\resizebox{0.93\textwidth}{!}{$
\begin{aligned}
\gamma=&
\left[ (P_Sn_S(r_N-1)+P_Nn_N(r_S-1) )^2-4p_F^2\right]
\left[P_Sn_S(r_N+1)+P_N n_N(r_S+1)                    \right]\\
&
\times \left[  -n_S^2(r_N-1)(5r_N+3)-n_N^2(r_S-1)(5r_S+3)+
2P_SP_N n_S n_N(3+r_S+r_N+3r_Sr_N)  +4p_F^2 \right]\,,\\
\alpha=&P_SP_Nn_Sn_N
\Big[n_S^4(r_N^2-1)^2
+n_N^4(r_S^2-1)^2
\\
&\qquad\qquad\qquad
+P_SP_Nn_Sn_N(r_S+1)(r_N+1)
\left(
n_S^2(r_N-1)^2+
n_N^2(r_S-1)^2
-4 p_F^2\right) \Big]\,,\\
\beta=&-n_S^2n_N^2\Big[
(P_Sn_S(r_N+1)+P_Nn_N(r_S+1))
(-n_S^2(r_N-1)(r_N+3)+n_N^2(r_S-1)(r_S+3))
\\
&\qquad\qquad+4p_F^2\left(P_S n_S(1+r_N)-P_N n_N (1+r_S)\right)
\Big]\,.
\end{aligned}
$}
\end{equation}

These are the correct expressions, provided that the solution actually exists.
Clearly, there are some immediate necessary conditions for a solution to exist, namely
\begin{align}\label{cxpmconsts}
c>0\,,\qquad -P_{N,S} x^I_{N,S}>0\,.
\end{align}
These are rather strong constraints but, by explicitly solving the BPS equations numerically,  as we discussed
in the previous section, we find they are not sufficient.
There we also presented strong numerical evidence for a conjecture for sufficient conditions
based on the possibility of having RG flows from $AdS_3\times \Sigma$ solutions of the STU model
(i.e. with vanishing hyperscalar).

We also highlight that the equivariant localisation procedure did not use the
conserved quantities \eqref{threeconschges}, for STU solutions, or \eqref{threeconschges}, for
hyperscalar solutions. This is the reason why an off-shell expression was obtained. Alternatively, if
we do use this additional information, then one can obtain an explicit on-shell
expression for the central charge without extremization, as we saw in the analysis of the BPS equations in previous sections. 
Finally, we reiterate that the expression for the central charge is valid for both
coprime and non-coprime spindle solutions and in the non-coprime one also
must specify $m^I_{N,S}$ consistent with the constraints discussed in previous sections.

\subsection{Connection with field theory}

We expect\footnote{For the anti-twist class some subtleties are discussed in \cite{Ferrero:2020twa} and in section 6 of \cite{Ferrero:2021etw}.}
the $AdS_3\times \Sigma$ solutions we are considering (when they exist) are dual to $d=2$ SCFTs that
can be obtained by compactifying $\cN=4$ $d=4$ SYM theory on a spindle. We also need
to switch on appropriate magnetic fluxes $p^I$ through the spindle with $p_R$ constrained by \eqref{RsymconstLSt}
and, for solutions with non-vanishing hyperscalar, $p_B$ constrained by \eqref{brokefluxcondt}.
For non-coprime spindles, as we have discussed, in order to specify the orbibundle
we also need to give the integers $m_{N,S}^I$.
 In the UV
the source for the operator dual to the hyperscalar itself must satisfy the appropriate smoothness conditions at the poles, as discussed in section \ref{seclinfluct},
but is otherwise arbitrary. It may be possible that the $d=2$ SCFTs can also arise from the $\cN=1$ Leigh-Strassler SCFT in $d=4$ after reducing on a spindle for solutions with
$p_B=0$; when $p_B\ne 0$ this is less clear as the LS SCFT does not have a $U(1)_B$ symmetry.

For the case of STU solutions (with vanishing hyperscalar) it was shown how 
a consideration of the anomaly polynomial of $\cN=4$ $d=4$ SYM theory, suitably reduced on the spindle, would give
rise to an anomaly polynomial in $d=2$ \cite{Ferrero:2021etw}. Furthermore, $c$-extremization \cite{Benini:2013cda} 
can then be used to determine
the correct $d=2$ R-symmetry, as a linear combination of the $U(1)^3$ symmetry of $\cN=4$ $d=4$ SYM theory
and the geometric symmetry rotating the spindle, and hence an expression for the central charge.  
The result was found to be in exact agreement with the twist and anti-twist solutions of the STU model. 
A similar computation was performed in \cite{Arav:2022lzo}, when $p_B=0$, starting with the 
 anomaly polynomial of the $\cN=1$ Leigh-Strassler SCFT in $d=4$.

Here we can show that the equivariant localization result in the previous subsection
can be recast in a form where one can make a direct
comparison with the field theory approach.
Suppose we write 
\begin{align}
x^I_N=b_0f^I_N+\Delta^I\,,
\qquad
x^I_S=b_0f^I_S+\Delta^I\,,
\end{align}
where $f^I_{N,S}$ are arbitrary constants, subject to the constraints
\begin{align}\label{bcsrhoI}
\sum_If^I_N&=\frac{P_N}{n_N}\,,
\qquad
\sum_If^I_S=-\frac{P_S}{n_S}\,,
\nn\\
\sum_I\zeta_If^I_N&=-\frac{P_N\hm_N}{n_N}\,,
\qquad
\sum_I\zeta_If^I_S=\frac{P_S\hm_S}{n_S}\,,
\end{align}
and $\Delta^I$ are constrained via
\begin{align}\label{twodeltas}
\sum_I\Delta^I=2\,,\qquad \sum_I\zeta_I \Delta^I=0\,.
\end{align}
The fluxes can then be written in the form
\begin{align}
p^I=n_N n_S(f^I_N -f^I_S )\,,
\end{align}
with $p_R$ and $p_B$ constrained as in \eqref{RsymconstLSt}, \eqref{brokefluxcondt}.
The trial central charge in gravity, i.e. \eqref{cLScaseitofa} and \eqref{LSSPSt}
can then be written in the form
\begin{align}\label{entchgerewrite}
c=3N^2\Big[   (\Delta_1\Delta_2I_3+\Delta_2\Delta_3 I_1+\Delta_3 \Delta_1 I_2)
+(\Delta_1I_4+\Delta_2I_5+\Delta_3  I_6)b_0
+I_7b_0^2
\Big]\,,
\end{align}
where
\begin{align}
I_I=f^I|^N_S\,,\quad
I_4= (f^2f^3)|^N_S\,,\quad
I_5= (f^3f^1)|^N_S\,,\quad
I_6= (f^1f^2)|^N_S\,,\quad
I_7= (f^1f^2f^3)|^N_S\,.
\end{align}

For the STU case, with vanishing hyperscalar, one should set $\zeta_I=0$ in the above expressions.
In this case there is just a single constraint on the $\Delta^I$ and so one should vary $c$ over
$b_0$ and the two independent $\Delta^I$. This is in precise alignment\footnote{The functions denoted
by $\rho_I(y)$ in
\cite{Ferrero:2021etw} can be taken to be arbitrary smooth functions that satisfy the boundary conditions in 
\eqref{bcsrhoI} with $\rho_I(y_{N,S})=f^I_{N,S}$.} with the off-shell field theory expression obtained in section 3.4 of \cite{Ferrero:2021etw}.
For non-vanishing hyperscalar, there are two constraints on the $\Delta^I$ and so one should vary $c$ over
$b_0$ and the single independent $\Delta^I$. The second constraint on the $\Delta^I$ in \eqref{twodeltas}
is associated with the fact that the $U(1)_B$ is a broken and hence, from the field theory point of view, does
not enter $c$-extremization. Notice that for non-coprime spindles this computation is independent
of the values for $m_{N,S}^I$.

 \section{Final comments}\label{fincomms}
 We have studied supersymmetric $AdS_3\times Y_7$
 solutions of type IIB supergravity, where $Y_7$ is a smooth $S^5$ bundle over a
 spindle $\Sigma(n_N,n_S)$. 
We have used a $D=5$ gauged supergravity theory consisting of the $U(1)^3$ 
STU model coupled to a complex scalar field which comprises  half of a hypermultiplet and is charged under $U(1)_B$.
 
 We focussed on two new features. First, we allowed for $(n_N,n_S)$ to be non-coprime integers, including when $n_N=n_S$ which are
 orbifolds of the round $S^2$,
 and second, for solutions with non-trivial hyperscalar, we allowed for boundary conditions
 where the hyperscalar vanishes at the poles and hence allowing for non-vanishing flux $p_B\ne0$.  For both of these,  
 a careful analysis was required to determine the conditions needed to have smooth uplifted solutions, with well-defined spinors.
In addition, for the STU solutions, we also determined the
spectrum of BPS modes associated with the hyperscalar.
 
\begin{figure}[h!]
	\centering
	\includegraphics[scale=0.6]{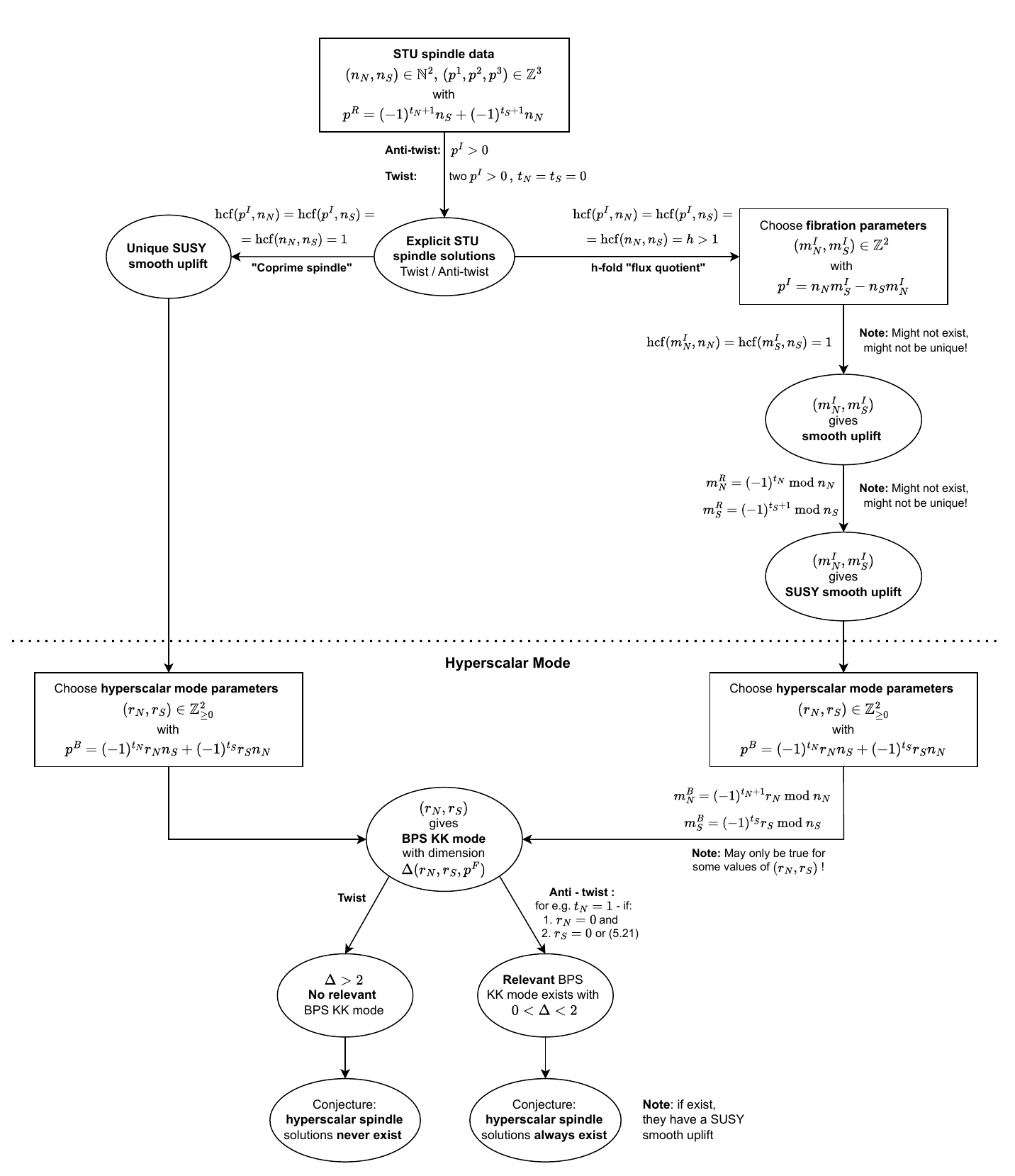}
	\caption{Summary of results. We have set $\kappa=+1$.}
	\label{bigfigure}
\end{figure}

 We have summarised the overall story in figure \ref{bigfigure}.
  To simplify the discussion here, without loss of generality, we set the parameter $\kappa=+1$.
 The $D=5$ STU solutions are analytically known \cite{Ferrero:2021etw}. Initially, we can consider the data to be 
 $(n_N, n_S)$, the fluxes $p^I$ and the signs $t_N,t_S$, which determine whether we have a twist or an anti-twist solution.
 The $D=5$ BPS solutions necessarily have a constraint on the R-symmetry flux: $p^R=p^1+p^2+p^3$.
 The $D=5$ STU solutions exist in the twist class, $t_N=t_S$, provided that $t_N=t_S=0$ and two $p^I>0$.
 They also exist in the anti-twist class, $t_N\ne t_S$, provided that $p^I>0$. 
 In the coprime case, this spindle data completely determines the uplifted solution, which moreover has a smooth
 supersymmetric uplift. In the non-coprime case, however, additional discrete flux data at the poles,
 given by integers $m_{N,S}^I$, 
is needed to determine the uplifted solution; they can be trivially added to the known analytic solution.
 The  $m_{N,S}^I$ are constrained in order to have a smooth uplift, and there are additional constraints 
 to have well defined
 spinors and hence giving rise to a smooth and supersymmetric uplifted solutions.

For the STU solutions we analysed the BPS spectrum associated with linearised 
fluctuations of the hyperscalar. We allowed the hyperscalar to be a non-vanishing constant at the poles,
as in \cite{Arav:2022lzo}, as well as vanishing like $\sim(y-y_{N,S})^{r_{N,S}}$, for integers $r_{N,S}\ge 0$. 
In order that the hyperscalar is a section of a smooth bundle there are additional constraints on
the broken flux, $p_B=p^1+p^2-p^3$, and, in the non-coprime case, also on the $m^I_{N,S}$. We determined
the  conformal dimension, $\Delta$, of the operator in the SCFT dual to the STU solution. In the twist class we find that there are no relevant
hyperscalar modes, while in the anti-twist class there can be. We also found that the existence of these relevant modes
appears to be precisely correlated with the existence of fully back-reacted hyperscalar spindle solutions, which we constructed numerically. Furthermore, the
central charge of the hyperscalar solutions is always less than that of the STU solutions with the same 
$n_N,n_S$, $p^I$, $m^I_{N,S}$. These results strongly suggest that there is a BPS RG flow 
that starts in the UV at the STU solution and ends up at the hyperscalar solution in the IR. It would
be interesting to verify this, however constructing such RG flows will necessarily involve solving PDEs.

We have examined the BPS spectrum of hyperscalar fluctuations for the STU solutions within the truncation of maximal gauged supergravity that we have utilised
in this paper. It would be interesting to extend  this to more general truncations such as, for example, including additional hypermultiplets as in \cite{Bobev:2010de}. It would also be of interest to examine the BPS spectrum for the new $AdS_3$ solutions
with non-vanishing hyperscalars that we have constructed numerically.

Associated with the non-coprime STU solutions with the same fluxes but different values of the discrete fluxes $m^I_{N,S}$, we have $d=2$ SCFTs with the same central charges (in the large $N$ limit). While the BPS spectra of the hyperscalar are different, it is related in a specific way as illustrated in table \ref{table2p1}, for example. It would be interesting to have a better understanding of this and also to know whether or not similar phenomena have been observed for other classes of $d=2$ CFTs.
The exact equality of the central charge has been derived here in the context of two-derivative gravity.
Based on the discussion in \cite{Kraus:2006wn} we anticipate that the equality is also valid if one were to include
higher-derivative corrections. However, it would be surprising if the equality were to be valid at sub-leading order in the
large $N$ expansion, since one-loop contributions can lift the degeneracy. Indeed this might be expected since 
the uplifted $AdS_3$ solutions have a different topology and a different spectrum of fluctuations.
It would be interesting to study this further and also to obtain some further insight directly from the field theory point of view.

The STU solutions uplifted on $S^5$ give rise to $AdS_3\times Y_7$ solutions with $Y_7$ a GK geometry \cite{Kim:2005ez,Gauntlett:2007ts}. For solutions
 in the coprime class the GK geometry was analysed in detail in \cite{Boido:2022mbe} and
 it would be interesting to generalise this to the non-coprime class.
 
It is possible to generalise the results of this paper to construct analogous $AdS_2\times Y_9$ solutions of M-theory.
One can use a $D=4$ gauged supergravity theory consisting of the $U(1)^4$ 
STU model coupled to a complex scalar field which comprises half of a hypermultiplet.  $AdS_2\times \Sigma$
solutions of this theory can then be uplifted on $S^7$ to find $AdS_2\times Y_9$ solutions of  $D=11$ supergravity.
The general STU spindle solutions have been constructed in \cite{Ferrero:2021etw}. In addition, some hyperscalar solutions
with constant hyperscalars at the poles
were constructed in \cite{Suh:2022pkg} and further clarified in \cite{BenettiGenolini:2024kyy}. 
As in this paper, we find that there are rich classes of new solutions when we allow for non-coprime spindles and also when we
allow for the hyperscalar to vanish at the poles. We will report on this soon \cite{Arav:2026unc}.

\section*{Acknowledgements}

\noindent 
We thank P. Ferrero and Jaeha Park for discussions.
This work was supported in part by STFC grant ST/X000575/1;
in the framework of H.F.R.I. call “Basic research Financing
(Horizontal support of all Sciences)” under the National Recovery and
Resilience Plan “Greece 2.0” funded by the European Union –
NextGenerationEU (H.F.R.I. Project Number: 15384);
by FWO projects G094523N and G003523N, as well as by KU Leuven grant C16/25/010.
This work was performed in part at Aspen Center for Physics, 
which is supported by National Science Foundation grant PHY-2210452.
JPG is supported as a Visiting Fellow at the Perimeter Institute.

\appendix

\section{BPS equations, hyperscalar modes and $AdS_3\times S^2$ solutions}\label{appbpseqns}
\subsection{The BPS equations}
To derive the BPS equations for the $AdS_3\times \Sigma$ ansatz \eqref{ansatz5d2}
we use Poincar\'e coordinates on $AdS_3$ and the $D=5$ orthonormal frame given by
\begin{align}
\label{eq:5d_frame_ansatz2}
e^0=e^V \frac{dt}{u},\quad
e^1=e^V \frac{d\phi}{u},\quad
e^2=e^V \frac{du}{u},\quad
e^3=f dy\,,\quad
e^4=h dz\,,
\end{align}
with $f,h>0$. 
We also use the following explicit basis for the $D=5$ Clifford algebra
\begin{align}
\Gamma^i &= \beta^i\otimes \s^3, \qquad  
\Gamma^3 = 1 \otimes \s^1 , \qquad \Gamma^4 =1 \otimes \s^2,
\end{align}
where 
$\beta^i = (-i\sigma^2, \sigma^3, \sigma^1)$ is a basis for the $D=3$ Clifford algebra,
with $\g^4 = -i\g^0 \g^1 \g^2 \g^3$ and so with regard to footnote \ref{footconvs} we 
have taken $c_3=+1$.

We write the $D=5$ Dirac spinors 
\begin{align}
\epsilon = \psi \otimes \chi\,,
\end{align}
where $\psi$ is a complex spinor on $AdS_3$ satisfying
\begin{align}\label{eq:3d_cks_eq}
D_i \psi = \frac{\kappa}{2} \beta_i \psi\,,
\end{align}
with $\kappa=\pm 1$.
Concretely, 
we have two Poincar\'e 
Killing spinors given by
\begin{align}\label{eq:ads3_q_proj}
\psi_Q =u^{-1/2}  \psi_0, \qquad \beta_2 \psi_0 = - \kappa \psi_0\,,
\end{align}
associated with preserving $\mathcal{N}=(0,2)$ (or $\mathcal{N}=(2,0)$) supersymmetry.
For the  Poincar\'e Killing spinors we can write the projection condition on the $D=5$ spinors as
\begin{align}\label{eq:Q_spinor_proj_5d}
\Gamma^2 \epsilon_Q = i \kappa \Gamma^{34} \epsilon_Q\,.
\end{align}
There are also two superconformal 
Killing spinors given by
\begin{align}
\psi_S =u^{-1/2} (\beta_0 t + \beta_1 \phi + \beta_2 u) \psi_0\,, \qquad \beta_2 \psi_0 = +\kappa \psi_0\,.
\end{align}

Similar\footnote{ 
Compared with \cite{Arav:2022lzo,Arav:2024exg} we have redefined the phase proportional to $z$ by a factor of two and
changed the signature.} to \cite{Arav:2022lzo,Arav:2024exg}, for the complex spinor $\chi$ on the spindle
we take 
\begin{align}
\chi = e^{\frac{i\hs z}{2}}e^\frac{V}{2} \begin{pmatrix}\sin\frac\xi 2 \\ \cos \frac \xi 2 \end{pmatrix}\,,
\end{align}
with
\begin{align}\label{eq:genl_spinor_proj}
\sin\xi \Gamma^3\epsilon +i \cos\xi \Gamma^{34} \epsilon = \epsilon\,, \qquad
\sin\xi \Gamma^4 \epsilon + i \cos\xi \epsilon = -\Gamma^{34} \epsilon\,.
\end{align}
Analysing in the same way as in \cite{Arav:2022lzo,Arav:2024exg}, we find the following set of BPS equations 
\begin{align}
\label{eq:bps_onederive_genl_h_xi}
f^{-1}{\xi'} = & -W \cos\xi + 2 \kappa e^{-V}\,,\nn \\
f^{-1}{V'} = & -\frac{W}{3}\sin\xi\,, \nn \\
(fh)^{-1}{h'}\sin\xi =&  2 \kappa e^{-V} \cos\xi - \frac{W}{3} (1+2 \cos^2\xi)\,,\nn \\
f^{-1}{\varphi_i'} =& 2 \partial_{\varphi_i} W \sin\xi\,,\nn   \\
f^{-1}{\rho'}= & \frac{2 \partial_\rho W}{\sin\xi}\,, 
\end{align}
along with the constraint equations 
\begin{align}\label{eq:bps_Q_constraint_1}
(\hs-Q_z)\sin\xi =& W h \cos\xi - 2\kappa h e^{-V}\,,\nn  \\
\partial_\rho W \cos\xi +&  \partial_\rho Q_z \frac{\sin\xi}{h} = 0\,.
\end{align}
Furthermore, in the $D=5$ orthonormal frame, the field strengths for the gauge fields take the form (no sum on $I$)
\begin{align}\label{eq:flux_bps}
F^I_{34} = X^I \left( 2 \kappa e^{-V}-\frac{2}{3} \cos\xi \left(W + 3 \vec{a}^I \cdot \partial_{\vec\varphi} W \right) \right) \,,
\end{align}
where $X^I = \exp[\vec{a}^I \cdot \vec{\varphi}]$ and
\begin{align}
\vec{a}^1 = (-\frac{1}{\sqrt 6},-\frac{1}{\sqrt 2})\,,\quad
\vec{a}^2 = (-\frac{1}{\sqrt 6},\frac{1}{\sqrt 2})\,,\quad
\vec{a}^3 = (2\times6^{-1/2},0)\,.
\end{align}

Crucially, we can integrate to find the following expression for $h$:
\begin{align}\label{eq:h_int_soln}
h=ke^V \sin\xi\,,
\end{align}
for some constant $k$. This simplifies  \eqref{eq:bps_Q_constraint_1} to
\begin{align}\label{eq:sQz_no_h}
\hs-Q_z = k \left(We^V \cos\xi -2 \kappa \right)\,.
\end{align}
Substituting \eqref{eq:h_int_soln} into \eqref{eq:flux_bps} also gives
\begin{align}\label{eq:gauge_bps}
(a^I)' =(\cI^I)'\,,
\end{align}
where 
\begin{align}\label{eq:gauge_bpsexp}
 \cI^I\equiv -k e^V \cos\xi  X^I \,.
\end{align}

Finally, the gauge equations of motion yield
\begin{align}
\left( \frac{e^{3V}}{fh}\left( \frac{a_1'}{(X^1)^2}- \frac{a_2'}{(X^2)^2} \right) \right) ' = 0\,,  \nonumber \\
\left( \frac{e^{3V}}{fh}\left( \frac{a_1'}{(X^1)^2}+ \frac{a_2'}{(X^2)^2}+2\frac{a_3'}{(X^3)^2} \right) \right) ' = 0\,,  \nonumber \\
\left( \frac{e^{3V}}{fh}\left( \frac{a_1'}{(X^1)^2}+ \frac{a_2'}{(X^2)^2}-2\frac{a_3'}{(X^3)^2} \right) \right) ' =-4 e^{3V} \frac{f}{h} \sinh^2 \rho (D\theta)_z\,,
\end{align}
where $D\theta_z = \bar\theta - \zeta_I a^I$. Using the BPS equations we then obtain constants of motion
as given in \eqref{threeconschges} and \eqref{twoconschges}.

\subsection{Linearised perturbation of the hyperscalar about an STU solution}\label{linpertapp}
We analyse linearized perturbations of the hyperscalar $\rho e^{i\theta}$ about an STU background
of the form
\begin{align}\label{linpert}
\rho=w(y)u^\delta\,,\qquad
\theta = \bar\theta z\,,
\end{align}
Here the $u$ is the Poincar\'e AdS radial coordinate, as in \eqref{eq:5d_frame_ansatz}.
The gauge invariant one-form $D\theta$ is given by $D\theta=(\bar\theta- \zeta_I a^I) dz$.

From \eqref{superpottextLS}, the superpotential is $W=W_{STU}+\cO(\rho^2)$
and hence the $\cO(\rho)$  parts of the gravitino and gaugino variations in \eqref{5d_susy1} vanish.
Next, noting that
$\partial_\rho W = \frac{\zeta_I X^I }{2}\rho+\cO(\rho^2)$, $\partial_\rho Q_\mu = - \frac{D_\mu \theta}{2}\rho+\cO(\rho^2)$, 
the linearised hyperino variation reduces to
\begin{align}
\left[ 
\delta e^{-V} w \Gamma^2 + \frac{w'}{f}\Gamma^3 - \zeta_I X^I w - i \frac{D\theta_z}{h} w \Gamma^4
\right]  \epsilon =0 \,.
\end{align}
We assume that we preserve the Poincar\'e supersymmetry which implies that $\epsilon$
satisfies the projection condition \eqref{eq:Q_spinor_proj_5d} as well 
as \eqref{eq:genl_spinor_proj}. Using this we deduce 
\begin{align}
\left[
\left( 
\frac{w'}{f\sin\xi} -\frac{D\theta_z}{h\tan\xi}w - \zeta_I X^I w
\right)
+
i\left( 
\kappa e^{-V} \delta w +\frac{D\theta_z}{h\sin\xi}w - \frac{w'}{f \tan\xi}
\right) \Gamma^{34}
\right] \epsilon = 0\,.
\end{align}
With no further breaking of supersymmetry, $\epsilon$ and $\Gamma^{34} \epsilon$ are independent, and so we conclude
\begin{align}\label{eq:delta_equation_h}
\delta &= \kappa e^V \left( \zeta_I X^I \cos\xi - \frac{D\theta_z}{h}\sin\xi \right)\,,\nn\\
\frac{w'}{w} &= f \zeta_IX^I \sin\xi + D\theta_z \frac{f}{h} \cos\xi\,.
\end{align}
We can simplify these by replacing $h$ via \eqref{eq:h_int_soln}, and the definition
of $\cI$ \eqref{eq:gauge_bps} to get
\begin{align}\label{eq:delta_from_Dtheta_zetaI}
\delta &=-\frac{\kappa}{k} ({D\theta_z+\zeta_I \cI^I })\,,\nn\\
\frac{w'}{w} &= \frac{f e^{-V}}{k} \left( \frac{D\theta_z}{\tan\xi} - \zeta_I \cI^I \tan\xi \right).
\end{align}

Note that it will be useful to analyze this at the poles of the spindle, using
the results of section \ref{somebcs}, 
to determine the behaviour of the spinor in those regions. Near the poles in conformal gauge $f=e^V$,
we find
\begin{align}
\frac{1}{k} \left( \frac{D\theta_z}{\tan\xi} - \zeta_I \cI^I \tan\xi \right) &= \frac{(-1)^{t_N} n_N D\theta_z(y_N)}{y-y_N} + \cO(1)\,,\nn\\
\frac{1}{k} \left( \frac{D\theta_z}{\tan\xi} - \zeta_I \cI^I \tan\xi \right) &= \frac{(-1)^{t_S} n_S D\theta_z(y_S)}{y_S-y} + \cO(1)\,,\
\end{align}
at the two poles.
From \eqref{eq:delta_from_Dtheta_zetaI} we can immediately determine that as we approach a pole we have
\begin{align}\label{eq:Dtheta_polesapp}
w \sim (y-y_{N,S})^{\hm_{N,S}},
\qquad
D\theta_z|_N=(-1)^\tN \frac{\hm_N}{n_N}\,,
\qquad
D\theta_z|_S=-(-1)^\tS \frac{\hm_S}{n_S}\,,
\end{align}
with $r_{N,S}\ge 0$.
From the results for STU solutions \eqref{stufullsoln} we deduce that at the poles, 
\begin{align}\label{eq:zetaI_poles}
\zeta_I \cI^I_N &=\frac{(-1)^\tN 6 \nS p_B - (-1)^\tS 2 \nN p_B + (-1)^{\tN+\tS}2 \nN \nS + \nN^2+\nS^2 +p_B^2-4p_F^2}{8 \nN \nS  \left((-1)^\tS \nN - (-1)^\tN \nS \right)} , \nonumber \\
\zeta_I \cI^I_S &=\frac{(-1)^\tS 6 \nN p_B - (-1)^\tN 2 \nS p_B + (-1)^{\tN+\tS}2 \nN \nS + \nN^2+\nS^2 +p_B^2-4p_F^2}{8 \nN \nS  \left((-1)^\tS \nN - (-1)^\tN \nS \right)} \,.
\end{align}
By differentiating \eqref{eq:Dtheta_polesapp} and then integrating, we deduce that for given $p_B$ flux in the STU solution, $\hm_N$ and $\hm_S$ must be constrained via
\begin{align}\label{eq:pb_constraintapp}
p_B = (-1)^\tN \hm_N \nS + (-1)^\tS \hm_S \nN\,.
\end{align}
Using this, we obtain 
\begin{align}\label{eq:delta_final_generala}
\delta = \frac{1}{4s}\left[p_R^2+p_B^2-4p_F^2-2(\hm_S n_N^2+3(-1)^{t_N+t_S+1}n_N n_S(\hm_N+   \hm_S)+\hm_N n_S^2)
\right]\,,
\end{align}
with $s$ defined in \eqref{defs} and $p_R$ as in \eqref{conststuflux}.

\subsection{$AdS_3\times S^2$ solutions}\label{spherecasebps}
For the STU model, $AdS_3\times S^2$ solutions, with a round metric on $S^2$ and supersymmetry preserved with a topological twist, were constructed in \cite{Benini:2013cda}.
These solutions lie within the ansatz \eqref{ansatz5d2} with $h=f\sin y$, $a^I=-\frac{p^I}{2}\cos y$, 
where $y\in [0,\pi]$, and $V,f$ constants. They preserve supersymmetry
with Killing spinors as in \eqref{spinorsbps} with $\cos\xi=(-1)^t$ and so $\sin\xi=0$, that are either chiral or antichiral on
the $S^2$: $i\Gamma^{34}\epsilon=(-1)^t \epsilon$. The analysis of the BPS equations is similar but simpler than in section \ref{secbpseqs}. Using the variables
$x^I=(-1)^t e^V X^I$, we find the BPS equations imply
\begin{align}
Q_z-\bar s&=(-1)^t\cos y \quad\Rightarrow\quad \sum_Ip^I=2(-1)^{t+1}\,,\nn\\
\sum_Ix^I&=2\kappa\,,\nn\\
p^I&=4(-1)^tf^2e^{-2V}(\kappa x^I-(x^I)^2)\,.
\end{align}
These can be solved in terms of the $p^I$, satisfying $\sum_Ip^I=2(-1)^{t+1}$, 
to give
\begin{align}
x^I&=-\frac{2\kappa}{s}((-1)^t+p^I)p^I\,,\nn\\
e^V f^2&=\frac{\kappa p^1 p^2 p^3}{2s}=\frac{\kappa}{4}(p^1x^2 x^3 +p^2 x^1x^3 +p^3 x^1 x^2 )\,,
\end{align}
where here $s=2-\sum_I(p^I)^2$ and $e^{3V}=(-1)^tx^1 x^2 x^3$. The central charge is given by
\begin{align}
c=12N^2e^Vf^2\,.
\end{align}
If $\kappa=+1$, for example, we have solutions with $(-1)^t x^I>0, c>0$, provided that either $p^1,p^2>0$ or
$p^1,p^3>0$ or $p^2,p^3>0$.

We can also analyse linear perturbations of the hyperscalar about these STU solutions as in section \ref{linpertapp}.
We consider perturbations as in \eqref{linpert}. 
Preserving Poincar\'e supersymmetries on the $AdS_3$ we find 
\begin{align}
\delta=\kappa(x^1+x^2-x^3)\,,
\end{align}
and also
\begin{align}
\frac{w'}{w} &= (-1)^t\frac{D\theta_z}{\sin y}\,.
\end{align}
Thus, near the poles we have 
\begin{align}\label{eq:Dtheta_polesapp2}
w \sim (y-y_{N,S})^{\hm_{N,S}},
\qquad
D\theta_z|_N=(-1)^t{\hm_N}\,,
\qquad
D\theta_z|_S=-(-1)^t {\hm_S}\,.
\end{align}
From this we deduce
\begin{align}
p_B=(-1)^t(r_N+r_S)\,.
\end{align}
One can show that on the space of solutions we necessarily have $\delta<0$ and hence
$\Delta=2-\delta >2$. Thus, for the $AdS_3\times S^2$ solutions
there is no relevant mode for the hyperscalar for any $r_N$, $r_S$ which could trigger an RG flow. In particular, there is no homogeneous relevant deformation with $r_N=r_S=0$ and
this is consistent with the fact that there is no homogeneous $AdS_3\times S^2$ solution with non-vanishing hyperscalar \cite{Bobev:2014jva}.

\section{Circle fibrations over spindles}\label{app:WCP}
We are interested in uplifting solutions on $S^5$ to obtain solutions of type IIB supergravity. The
regularity of these uplifted solutions can be deduced by first considering the regularity of circle
orbibundles over spindles. To do so, we revisit the discussion of section 2.2 of \cite{Ferrero:2021etw}.
A new feature we incorporate here is the analysis of cases when $n_{N,S}$ are not coprime.

Before we begin we briefly recall some aspects of Lens spaces, which arise as the total spaces of the circle orbibundles.
Let ($z_1,z_2)$ be complex coordinates on $\mathbb{C}^2$ and consider $S^3$ embedded via $|z_1|^2+|z_2|^2=1$.
For any pair of coprime integers $(p,q)$ with $p>0$ the Lens space $L(p,q)$ is the quotient of
the three sphere $S^3\subset \mathbb{C}^2$ under the free $\mathbb{Z}_p$ action that is generated
by $(z_1,z_2)\to (e^{2\pi i/p}z_1,e^{2\pi iq/p}z_2)$. Notice that $L(1,0)=S^3$.
Also, notice that if $q'$ satisfies $q' q=1$ mod $p$, then this action
can be written $(z_1,z_2)\to (e^{2\pi i q'/p}z_1,e^{2\pi i/p}z_2)$. By interchanging $z_1$ and $z_2$ this displays the fact that 
$L(p,q)\simeq L(p,q')$. In fact it 
is known that we have the orientation preserving diffeomorphism $L(p,q)\simeq L(p,q')$ if and only if $q=q'$ mod $p$ or
$qq'=1$ mod $p$:
\begin{align}
L(p,q)\simeq L(p,q'),\qquad \Leftrightarrow\qquad \text{$q=q'$ mod $p$}\qquad \text{or}\qquad  \text{$qq'=1$ mod $p$}\,.
\end{align}
We will also find it useful to utilise the description of Lens spaces in terms of Seifert fibrations which include, as a special
subclass, circle orbibundles over spindles e.g. \cite{orlik,JankinsNeumann,scott,geiges2017seifertfibrationslensspaces}.

We now follow the analysis of circle orbibundles over a spindle as presented in \cite{Ferrero:2021etw}.
We let $\Sigma$ be a spindle with an azimuthal symmetry\footnote{The coordinate $z$ should be identified with
$\varphi$ in \cite{Ferrero:2021etw}.} $\partial_z$ with $\Delta z=2\pi$. At the poles the spindle has orbifold singularities of
the form $\mathbb{R}^2/\mathbb{Z}_{n_N}$ and $\mathbb{R}^2/\mathbb{Z}_{n_S}$, with $n_N,n_S$ positive integers,
$n_N, n_S>0$, which are \emph{not necessarily coprime}. To describe the $U(1)$ orbibundle we use $N$ and $S$ patches on the spindle (topologically discs) 
and supplement them with coordinates $\psi_{N,S}$ to parametrise the circle fibre with
$\Delta\psi_{N,S}=2\pi$. In these coordinates, a connection one-form, $A$, for the $U(1)$ orbibundle will not be regular at the poles
of each patch, but instead will have some flat connection pieces that capture the
orbifold data. In particular, evaluating the one-form at the poles in the two patches we have
\begin{align}\label{appendixamnrels}
A|_N\to \frac{m_N}{n_N}dz\,,\qquad
A|_S\to \frac{m_S}{n_S}dz\,,
\end{align}
with $m_N\in\mathbb{Z}_{n_N}$, $m_S\in\mathbb{Z}_{n_S}$. 
Furthermore, the gauge fields in the two patches are related by a 
$U(1)$ gauge transformation on the overlap of the patches via
\begin{align}\label{gtgfsapb}
A|_{\text{$N$ patch}}=A|_{\text{$S$ patch}}+\gamma dz\,,
\end{align}
with $\gamma\in \mathbb{Z}$. 
On the total space of the orbibundle $(d\psi+A)$ is a smooth global one-form; the gauge transformation
\eqref{gtgfsapb} is implemented by identifying the angular coordinates $(\psi_N,z)$ with $(\psi_S-\gamma z,z)$ on the overlap (and reversing the orientation).
Using Stokes' theorem, the flux of this gauge field through the spindle is
\begin{align}\label{infbrflconapb}
\frac{1}{2\pi}\int _\Sigma F\equiv \frac{p}{n_Nn_S}\,,
\end{align}
with $p\in \mathbb{Z}$ given by\footnote{Note that $p^{here}=\lambda^{there}$ and $\gamma^{here}=p^{there}$ in
\cite{Ferrero:2021etw}.}
\begin{align}\label{condfluxsmooth2apb}
p=n_Nm_S -n_Sm_N  +\gamma n_N n_S \in\mathbb{Z}\,.
\end{align}
Importantly, for the total space of the circle orbibundle to be smooth it is necessary and sufficient 
that the coprime conditions 
\begin{align}\label{coprimemncond}
\mathrm{hcf}(m_N,n_N)=1 \qquad \text{and}\qquad \mathrm{hcf}(m_S,n_S)=1\,,
\end{align}
are satisfied \cite{Ferrero:2021etw}.
From \eqref{condfluxsmooth2apb}, \eqref{coprimemncond} we have
$\mathrm{hcf}(p,n_S)=\mathrm{hcf}((\gamma n_N-m_N )n_S+m_S n_N,n_S)=
\mathrm{hcf}(m_S n_N,n_S)=\mathrm{hcf}(n_N,n_S)$.
We can similarly argue for $\mathrm{hcf}(p,n_N)$
and so we have
\begin{align}\label{coprimelamndannnsapb}
\mathrm{hcf}(p,n_N)=\mathrm{hcf}(p,n_S)=\mathrm{hcf}(n_N,n_S)\,.
\end{align}

It will be convenient in the following to now consider an equivalent description of this data by taking 
$m_{S},m_{N}\in\mathbb{Z}$, satisfying
$\mathrm{hcf}(m_N,n_N)=\mathrm{hcf}(m_S,n_S)=1$, which allows one to absorb
$\gamma$ into either $m_N$ or $m_S$ and we write 
\begin{align}\label{condfluxsmooth3}
p=\det\begin{pmatrix}n_N&n_S \\ m_N&m_S \end{pmatrix}
=m_Sn_N - m_Nn_S\,.
\end{align}
Clearly $m_N, m_S$ are not unique since if $m_N, m_S$
satisfies \eqref{condfluxsmooth3}, then so does $(m_N+l n_N,m_S+ln_S)$ for integer $l$.

We pause to state the following\\
{\bf Lemma:}  Let $a,b$ be two coprime integers and $c$ an arbitrary integer. Then there exist
integers $x,y$ such that
\begin{itemize}
\item $c=xa+yb$\,
\item
$\mathrm{hcf}(x,b)=\mathrm{hcf}(c,b)$ and $\mathrm{hcf}(y,a)=\mathrm{hcf}(c,a)$
\item $x$ is unique mod $b$ and 
$y$ is unique mod $a$
\end{itemize}
The proof essentially follows from B\'ezout's identity.

We  first consider ``coprime spindles" with $\mathrm{hcf}(n_N,n_S)=1$. In this case smoothness 
of the orbibundle is equivalent
to the condition that $\mathrm{hcf}(p,n_S)=\mathrm{hcf}(p,n_N)=1$. 
To see this, first note that clearly given $\mathrm{hcf}(n_N,n_S)=1$, then $\mathrm{hcf}(p,n_S)=\mathrm{hcf}(p,n_N)=1$ follows from \eqref{coprimelamndannnsapb}.
Conversely, from the lemma, given $\mathrm{hcf}(n_N,n_S)=1$ all values of $p$, with
$\mathrm{hcf}(p,n_S)=\mathrm{hcf}(p,n_N)=1$ have solutions $m_N,m_S$ to
\eqref{condfluxsmooth3} and moreover, 
$\mathrm{hcf}(m_N,n_N)=1$, with
$m_N$ unique mod $n_N$, and
$\mathrm{hcf}(m_S,n_S)=1$, with $m_S$ unique mod $n_S$.
This completes the argument.
The total spaces of the smooth circle orbibundles over coprime spindles
are Lens spaces of a type discussed in appendix A 
of \cite{Ferrero:2020twa}, which we recall below. In particular, for coprime $n_N,n_S$, 
the Lens space is uniquely specified by 
$p$, which is coprime to both $n_N$ and $n_S$.

We next consider circle bundles over ``non-coprime spindles" with 
$h=\mathrm{hcf}(n_N,n_S)$ and $h\ne 1$. From \eqref{coprimelamndannnsapb} we conclude
that the smooth orbibundles
are necessarily $h$-fold ``flux quotients" of coprime spindles in the following sense: we must have 
$(n_N,n_S,p)=h(\hat n_N, \hat n_S,\hat p)$, with 
$(\hat n_N, \hat n_S,\hat p)$ specifying a smooth orbibundle for a coprime spindle
with $\mathrm{hcf}(\hat n_N,\hat n_S)=\mathrm{hcf}(\hat p, \hat n_S)=\mathrm{hcf}(\hat p, \hat n_N)=1$. Notice that the flux of this non-coprime spindle can be expressed as
\begin{align}
\frac{1}{2\pi}\int _\Sigma F=\frac{p}{n_Nn_S}=\frac{\hat p}{h \hat n_N \hat n_S}\,,
\end{align}
which is the origin of the term ``flux quotient".

However, not every $h$-fold flux quotient of a smooth orbibundle for a coprime spindle will be smooth.
Indeed, the following class is always obstructed:
\begin{align}\label{obstructedclass}
\text{Obstructed:}\qquad \text{$\hat n_N$, $\hat n_S$ and $\hat p$ are all odd and $h$ is even}\,.
\end{align}
To see this we argue as follows.
$(n_N, n_S,p)=h (\hat n_N, \hat n_S,\hat p)$ with $h$ even, so $n_N, n_S$ are both even and hence
$m_N, m_S$ are both odd. The condition we need to solve is 
$p=n_N m_S-m_N n_S$ which is equivalent to $\hat p=\hat n_N m_S-m_N \hat n_S$. But since
$\hat n_N$ and $m_S$ are both odd, then so is $\hat n_N m_S$. Similarly $\hat n_S m_N$ is odd
and thus $\hat n_N m_S-m_N \hat n_S$ is even, and so we cannot have 
$\hat p=\hat n_N m_S-m_N \hat n_S$ with $\hat p$ odd.
In fact, we believe that $h$-fold flux quotients of smooth orbibundles over coprime spindles 
$(\hat n_N, \hat n_S,\hat p)$ will always have at least one choice of $m_{N,S}$ that
will give a smooth uplift except in the special case of the obstructed class
\eqref{obstructedclass}.
It would be interesting to prove this.

Continuing with the non-coprime case,  
for smoothness of the orbibundle we want to solve
$p=n_Nm_S-n_Sm_N$ with $\mathrm{hcf}(m_N,n_N)=1$ and $\mathrm{hcf}(m_S,n_S)=1$. This is equivalent to
solving $\hat p=\hat n_Nm_S-\hat n_Sm_N$ with $\mathrm{hcf}(m_N,h \hat n_N)=1$ and $\mathrm{hcf}(m_S,h \hat n_S)=1$. Now for example,
$\mathrm{hcf}(m_N,h \hat n_N)=1$ is equivalent to $\mathrm{hcf}(m_N,\hat n_N)=1$ and $\mathrm{hcf}(m_N,h)=1$. So we want to solve
$\hat p=\hat n_Nm_S-\hat n_Sm_N$ with $\mathrm{hcf}(m_N,\hat n_N)=1$,  $\mathrm{hcf}(m_S,\hat n_S)=1$ as well as 
$\mathrm{hcf}(m_N,h)=1$, $\mathrm{hcf}(m_S,h)=1$. 
From the coprime analysis above, for any $\hat p$ there
is a solution to $\hat p=\hat n_Nm_S-\hat n_Sm_N$ with 
$\mathrm{hcf}(m_N,\hat n_N)=1$, $\mathrm{hcf}(m_S,\hat n_S)=1$ and
moreover, $m_N$ is unique mod $\hat n_N$ and 
$m_S$ is unique mod $\hat n_S$. What is left to prove is that we can choose a specific representative solution for $m_N, m_S$ that also satisfies $\mathrm{hcf}(m_N,h)=1$, $\mathrm{hcf}(m_S,h)=1$.
We believe that this is always possible 
provided that we are not in the obstructed class \eqref{obstructedclass}. 
We also emphasise that such $m_N$ and $m_S$ are not guaranteed to be unique
mod $n_N$ and $n_S$, respectively, and this is why for the non-coprime case, in contrast to the coprime
case, one can obtain several inequivalent uplifts.

Having discussed regularity from the point of view of the analysis in \cite{Ferrero:2021etw}, 
we now want to determine the Lens space which is associated with the total
space of the circle orbibundle. We obtain the result for both coprime and non-coprime spindles
using results on Seifert fibrations;  in the coprime case we recover the same result 
as appendix A of \cite{Ferrero:2020twa} which was obtained slightly differently.
The key point is that the data for the $U(1)$ orbibundle $(n_N,m_N)$, $(n_S,m_S)$ with
$\mathrm{hcf}(m_N,n_N)=\mathrm{hcf}(m_S,n_S)=1$
(with $p$ given by \eqref{condfluxsmooth3}), precisely\footnote{\label{ftnoteappseif}Consider the discussion of \cite{Ferrero:2021etw}.
In the $N$ patch for example, we first write $ \phi=\frac{1}{n_N} z$ and move from the coordinates
$(\psi_N,\phi)$ to coordinates $(\chi_N,\hat \phi)$ via
$\chi_N=\psi_N+{m_N}\phi$, 
$\hat \phi=\phi$. The global identifications are then characterised by starting with the solid torus 
$D\times S^1$, with $D$ a disc with angular coordinate $\hat\phi$ and $S^1$ a circle parametrised 
by $\chi_N$ with $\Delta\chi_N=\Delta \hat\phi=2\pi$, which parametrises the covering space for the $S^1$ orbibundle in this patch,
and then making the $\mathbb{Z}_{n_N}$ identification
$(\chi_N, \hat\phi)\sim (\chi_N+\frac{2\pi m_N}{n_N}, \hat\phi+  \frac{2\pi}{n_N})$. 
Equivalently, since $m_N$, $n_N$ are coprime, 
we can define $b_N$ so that $b_N m_N=1$ mod $n_N$. The $\mathbb{Z}_{n_N}$ identification is then
equivalent to $(\chi_N, \hat\phi)\sim (\chi_N+\frac{2\pi}{n_N}, \hat\phi+  \frac{2\pi b_N}{n_N})$ on the covering space. That is,
the global identification corresponds to
identifying the bottom disc of the solid torus (at $\chi_N=0$) with the top disc 
(at $\chi_N=2\pi/n_N$) with a $\frac{2\pi b_N}{n_N}$ twist. Thus $(n_N, b_N)$ are the ``orbit invariants" and
$(n_N, -m_N)$ are the Seifert invariants in the conventions of \cite{JankinsNeumann,geiges2017seifertfibrationslensspaces}.
We can then glue the $S$ patch to the $N$ patch being mindful of an orientation change.} 
give the ``Seifert invariants" 
$(n_N,-m_N)$, $(n_S,m_S)$ which specify a smooth Seifert manifold. Here we are using the notation of
\cite{JankinsNeumann,geiges2017seifertfibrationslensspaces} and the difference in sign between 
$-m_N$ and $m_S$ can be traced back to the fact that we
are using a single coordinate $z$ in the $N$ and $S$ patches described above.

According to theorem 4.4. of \cite{JankinsNeumann,geiges2017seifertfibrationslensspaces}, then, the Lens space
that is associated with the $U(1)$ orbibundle $(n_N,m_N)$, $(n_S,m_S)$ 
is given by $L(p,q)$ with 
$p$ as in \eqref{condfluxsmooth3},
\begin{align}\label{condfluxsmooth4}
p=\det\begin{pmatrix}n_N&n_S \\ m_N&m_S \end{pmatrix}
=m_Sn_N - m_Nn_S\,,
\end{align}
and 
$q$ identified using the following prescription.
Since $\mathrm{hcf}(m_N,n_N)=1$, $\mathrm{hcf}(m_S,n_S)=1$, the B\'ezout lemma implies we can find integers 
$\alpha$, $\beta$ satisfying
\begin{align}\label{sumseifu1}
1=\det\begin{pmatrix}n_S&\alpha  \\ m_S&\beta \end{pmatrix}\,.
\end{align}
The total space is then the Lens space $L(p, q)$ with $q$ given by
\begin{align}\label{sumseifu2}
q=\det\begin{pmatrix}n_N&\alpha \\ m_N&\beta \end{pmatrix}\,.
\end{align}
We now make some additional observations regarding the Lens spaces for the total spaces of the orbibundles for coprime and non-coprime spindles.

Consider first, a coprime spindle specified by $(n_N,n_S, p)$ with
$\mathrm{hcf}(n_N,n_S)=\mathrm{hcf}(n_N,p)=\mathrm{hcf}(n_S,p)=1$.
There exist
integers $m_N, m_S$ satisfying
\begin{align}\label{lambdadef}
p=\det\begin{pmatrix}n_N&n_S \\ m_N&m_S \end{pmatrix}\,,
\end{align}
with $\mathrm{hcf}(m_N,n_N)=\mathrm{hcf}(m_S,n_S)=1$. Since we have a coprime spindle we also have 
$\mathrm{hcf}(p, n_Nn_S)=1$ and hence, by B\'ezout's lemma, there exists integers $\tilde a, b$ such that
\begin{align}\label{atildeb}
\tilde a n_N n_S+bp=1\,.
\end{align}
At this point we pause to highlight that
if we define $a=\tilde a n_S$ then $a,b$ are integers that are used in the construction of the Lens space in appendix A of \cite{Ferrero:2020twa}. Specifically the Lens space in \cite{Ferrero:2020twa} is given by $L(p, q)$ where
$q=an_S$ mod $p$. We now continue and show that we get the same Lens space (up to diffeomorphism)
using the prescription summarised above. 
We have $\tilde a n_N n_S+bp=1$ and with $p$ given in \eqref{lambdadef}
we can write this in the form 
\begin{align}
1=\det\begin{pmatrix}n_S& -bn_N \\ m_S&\tilde a n_N-b m_N \end{pmatrix}\,.
\end{align}
From \eqref{sumseifu1}, \eqref{sumseifu2}
we see that we have a Lens space $L(p,q)$ with $q$ given by
\begin{align}\label{lambdadef2}
q=\det\begin{pmatrix}n_N& -bn_N \\ m_N&\tilde a n_N-b m_N \end{pmatrix}=\tilde an_N^2\,.
\end{align}
Taking this $q$ we notice that $q (an_S)=(\tilde a n_N n_S)^2$. From \eqref{atildeb} we have
$\tilde a n_N n_S=1$ mod $p$ and hence $(\tilde a n_N n_S)^2=1$ mod $p$. 
Thus, $q (an_S)=1$ mod $p$ and so the Lens space $L(p,q=\tilde an_N^2)$
is diffeomorphic to $L(p,q=an_S)$ in agreement with the result of
\cite{Ferrero:2020twa} highlighted above. We emphasise, in particular, that
for coprime spindles the Lens space is uniquely specified by the data $(n_N,n_S, p)$ and does not
depend on the specific choice of non-unique integers\footnote{For the coprime case $m_N$ and $m_S$ are unique modulo $n_N$ and $n_S$.} $(m_N,m_S)$ in \eqref{lambdadef}.

We now consider circle orbibundles for non-coprime spindles with smooth total spaces. In this case, as explained above, we necessarily have 
$(n_N,n_S, p)=h(\hat n_N, \hat n_S, \hat p)$ where $(\hat n_N, \hat n_S, \hat p)$ is associated with 
a smooth orbibundle for a coprime spindle. In this case, the construction of appendix A of \cite{Ferrero:2020twa} does not apply and
instead we determine the topology of the total space of the smooth circle orbibundle again using the
results of Seifert fibrations
\cite{JankinsNeumann,geiges2017seifertfibrationslensspaces}. As above, this requires first finding
integers $m_N, m_S$ satisfying
\eqref{lambdadef}
with $\mathrm{hcf}(m_N,n_N)=1$, $\mathrm{hcf}(m_S,n_S)=1$. Using B\'ezout, we then find integers 
$\alpha$, $\beta$ satisfying \eqref{sumseifu1} and then we obtain
a Lens space $L(p, q)$ with $q$ given by \eqref{sumseifu1}.
Importantly, and in contrast to the coprime case, the data
$(n_N,n_S, p)$ is not sufficient to uniquely fix the value of $q$ in the Lens space and the specific choice
of $(m_N, m_S)$ is required in general. For example consider $(n_N,n_S, p)=(35,14, 7)$. If we take 
$(m_N, m_S)=(12,5)$ we get $L(7,1)$,
if $(m_N, m_S)=(2,1)$ we get $L(7,2)\simeq L(7,4)$ and
if $(m_N, m_S)=(32,13)$ we get $L(7,3)\simeq L(7,5)$.

We also highlight another feature in the non-coprime case. Suppose we have chosen a specific $(m_N,m_S)$ leading to a Lens space $L(p,q)$.
Then we know there exists $(\alpha,\beta)$ with
\begin{align}
1=\det\begin{pmatrix}n_S&\alpha  \\ m_S&\beta \end{pmatrix}
\qquad\Rightarrow\qquad
1=\det\begin{pmatrix}h\hat n_S&\alpha  \\ m_S& \beta \end{pmatrix}
=\det\begin{pmatrix}\hat n_S&\alpha  \\ m_S&h \beta \end{pmatrix}\,.
\end{align}
Furthermore, we have
\begin{align}
q=\det\begin{pmatrix}n_N&\alpha \\ m_N&\beta \end{pmatrix}=
\det\begin{pmatrix}h \hat n_N&\alpha \\ m_N&\beta \end{pmatrix}=
\det\begin{pmatrix}\hat n_N&\alpha \\ m_N&h \beta \end{pmatrix}\,.
\end{align}
We thus see that there will be an associated coprime spindle
$(\hat n_N, \hat n_S,\hat p)$, leading to a Lens space with the same value of $q$: $L(\hat p=p/h,q)$. Of course for the 
non-coprime spindle $q$ is defined modulo $p$, whereas in the associated 
coprime case it is defined modulo $\hat p$. In the above examples with 
$(n_N,n_S, p)=(35,14, 7)$, associated with the coprime spindle 
$(\hat n_N, \hat n_S,\hat p)=(5,2,1)$ is the three-sphere $L(1,0)$.

The non-coprime analysis also extends to orbifolds of the sphere when
we take $n_N=n_S$. Then, \eqref{condfluxsmooth4} implies that $n_N$ should divide $p$.
For example, if we consider the particular case that the flux is
$p=n_N=n_S$, then since
$m_{N,S}$ are coprime to $n_N=n_S$, we deduce that $n_N$ would necessarily have to be odd
(if it was even $m_{N,S}$ would be odd and then we couldn't satisfy \eqref{condfluxsmooth4}).
Alternatively, we can deduce that $n_N$ is odd from \eqref{obstructedclass}.
With e.g. $n_N=n_S=p=5$, we can take: $m_N=4$, $m_S=3$, giving the Lens space $L(5,2)$;
$m_N=3$, $m_S=2$, giving the Lens space $L(5,4)$ or
$m_N=2$, $m_S=1$, giving the Lens space $L(5,3)\simeq L(5,2)$.
We can take $n_N=n_S$ to be even, provided that $p= n_N \hat p$ with $\hat p$
a non-zero even integer (from \eqref{obstructedclass}).

We now consider $S^5$ bundles over a spindle that arise by uplifting the solutions of $D=5$ gauged supergravity to type IIB.
By construction these have an isometric $U(1)^3$ action. In the case of coprime spindles, the analysis above
shows that the $U(1)$ orbibundles
are classified by the Chern number $p$ and the associated complex line bundle is
denoted $O({p})$. Thus we can pick three Chern numbers $(p_1,p_2,p_3)$ to get
the direct sum of line bundles $O(p_1,p_2,p_3)=O(p_1)\oplus O(p_2)\oplus O(p_3)$.
We may then form the associated $S^5$ bundle over $\mathbb{WCP}^1_{[n_N, n_S]}$ by using the same $U(1)^3$ 
transition functions of $O(p_1,p_2,p_3)$.
From the above analysis, we conclude that the seven dimensional total space of this $S^5$ bundle will be a smooth manifold if
both $n_N$ and $n_S$ are coprime to each of $p_1,p_2$ and $p_3$. 

In the non-coprime case, the main novelty is that, in general, 
the $U(1)$ orbibundles is specified not just by the Chern number $p$, but also by the data $m_N,m_S$ with
$\text{hcf}(n_N,m_N)=1$ and $\text{hcf}(n_S,m_S)=1$ and $p$ as in \eqref{condfluxsmooth3}; the associated
complex line bundle might be denoted $O( (n_N, m_N); (n_S,m_S))$.
For given $(p_1,p_2,p_3)$ and suitable $m_N^I$, $m_S^I$
we can then form an associated $S^5$ bundle, as above, by using the transition functions for the direct sum of the
three line bundles associated with the three $U(1)$'s.
For non-coprime spindles, in order for the smooth uplift on $S^5$ to preserve supersymmetry, we must further
demand that the $m_N^I$, $m_S^I$ satisfy \eqref{rsymms} (for the coprime spindles the smoothness
conditions automatically imply \eqref{rsymms}).

\section{Minimal gauged supergravity}\label{appCmingaugedsugra}
 Here we expand upon some comments made in the text regarding STU solutions
 with $p^1=p^2=p^3\equiv p$, which only exist in the anti-twist class. Solutions that also have
 $m_{N,S}^1=m_{N,S}^2=m_{N,S}^3\equiv m_{N,S}$ are then solutions of minimal $D=5$ gauged supergravity. Such solutions were first discussed in 
 \cite{Ferrero:2020laf} in the case of coprime spindles, where they were related to the type IIB solutions of
 \cite{Gauntlett:2006af}  (see also \cite{Ferrero:2020twa}).
 Here we can further comment on the case of non-coprime spindles and also show that these
 solutions have no relevant hyperscalar deformation modes. 
For simplicity, we focus on uplifts of the solutions on $S^5$, as we have been discussing throughout this paper, but uplifting solutions of minimal gauged supergravity on other regular Sasaki-Einstein can also be achieved using \cite{Ferrero:2020laf}
and suitably modifying the analysis here.

We again assume
\begin{align}\label{atassumpappc}
\kappa=1\,,\qquad t_N=1\,, \qquad t_S=0\,,
\end{align}
for definiteness and no essential loss of generality. The R-symmetry flux condition 
\eqref{conststuflux} reads 
\begin{align}\label{poneappc}
p=\frac{1}{\beta}(n_S-n_N)\,,
\end{align} 
with $\beta=3$, which places a
constraint on $n_{N,S}$ for smooth uplifts on $S^5$ as highlighted\footnote{Note the typo below eq. (19) of  \cite{Ferrero:2020laf}, where instead of $n_-=5,9,\dots$, it should
say $n_- =5,11,\dots$ and there $n_\pm$ were assumed coprime.
For uplifting on other regular Sasaki-Einstein spaces, one should consider other
values of $\beta$.} in \cite{Ferrero:2020laf}. 
Uplifted solutions which also have $m_{N,S}^1=m_{N,S}^2=m_{N,S}^3\equiv m_{N,S}$ will preserve $SU(3)$ flavour symmetry (i.e.  $SU(3)\times U(1)$ symmetry); we will focus on this class, which corresponds to the solutions considered in \cite{Gauntlett:2006af,Ferrero:2020laf}, but here we allow non-coprime spindles too (there are certainly STU solutions 
with $p=\frac{1}{\beta}(n_S-n_N)$ that break the $SU(3)$ flavour symmetry as illustrated in the text). Smoothness of the solutions
requires
\begin{align}\label{ptwoappc}
p=n_Nm_S -n_S m_N\,,
\end{align}
with $\text{hcf}(m_N,n_N)=1$, $\text{hcf}(m_S,n_S)=1$ (as in \eqref{coprimemncond}),
and, recalling \eqref{rsymms}, for well defined spinors we need
\begin{align}\label{modmnappc}
\beta m_N=-1\quad \text{mod $n_N$}\,,\qquad \beta m_S=-1\quad \text{mod $n_S$}\,.
\end{align} 
When $n_N$, $n_S$ are coprime, 
\eqref{modmnappc} is automatically satisfied; when
$n_N$, $n_S$ are non-coprime, then \eqref{modmnappc} is an additional extra condition
to be satisfied given $p$ satisfying \eqref{poneappc}, \eqref{ptwoappc}.

Notice that if we write $\beta m_S=-1+l n_S$ for some integer $l$, then 
$\beta m_N=-1+ln_N$ (with the same $l$). To see this, note that
 $\beta m_N=\beta(n_N m_S-p)/n_S=(n_N (-1+l n_S)-(n_S-n_N))/n_S$ and hence $\beta m_N=-1+ln_N$. This result (or directly from \eqref{modmnappc}) implies that
 smooth BPS solutions require 
 \begin{align}\label{condcfive}
 \text{hcf}(n_N,\beta)=1\,,\qquad \text{hcf}(n_S,\beta)=1\,.
\end{align}

Conversely, if $\text{hcf}(n_S,\beta)=1$ then from the Lemma in appendix \ref{app:WCP}, we can choose integers $l, m_S$ with $\beta m_S=-1+l n_S$ and 
$\text{hcf}(m_S,n_S)=1$. Then, with $\beta p=n_S-n_N$, we have
$\beta(p-m_S n_N)=n_S(1-n_N l)$. But since $\text{hcf}(n_S,\beta)=1$ we must have $p-m_S n_N$ is
divisible by $n_S$ and so we can define $m_N\equiv (m_S n_N -p)/n_S$, so that \eqref{ptwoappc} is satisfied, and it also follows that 
$\beta m_N=-1+l n_N$, so \eqref{modmnappc} is satisfied, and it then also follows that
$\text{hcf}(m_N,n_N)=1$ so \eqref{coprimemncond} is satisfied, as also required (for a smooth uplift).

We thus conclude that for $n_N, n_S$, not necessarily coprime, there is
a unique\footnote{In the non-coprime case, using \eqref{condcfive} the Lemma in appendix \ref{app:WCP} implies that $m_N$ is unique mod $n_N$ and $m_S$ is unique mod $n_S$.} $SU(3)$ invariant and supersymmetric uplift, on $S^5$ with $3 p=n_S-n_N$ if and only if $n_N$ and $n_S$ are
not divisible by $3$ and $n_S-n_N$ is divisible by 3 . From appendix \ref{app:WCP}
we can also determine the Lens space associated with the circle bundle over the spindle. From $p=n_Nm_S -n_S m_N$
and \eqref{condfluxsmooth4} we have a Lens space $L(p,q)$, for some $q$. We can show that $q=1$: indeed we have $1=l n_S -\beta m_S$
and hence from \eqref{sumseifu1} and \eqref{sumseifu2} we have $q=n_N l-m_N\beta=1$.
For the coprime case this conclusion aligns with final conclusion of
appendix A of \cite{Ferrero:2020twa}.

Thus, after uplifting on $S^5$ we obtain $AdS_3\times Y_7$ solutions of type IIB
with $Y_7$ an $L(p,1)$ bundle over $CP^2$, both in the coprime case, as in
\cite{Ferrero:2020laf}, and also in the non-coprime case. We can make a 
further connection with \cite{Ferrero:2020laf,Gauntlett:2006af}: we should identify
$n_S=n_-$, $n_N=n_+$, $I^{there}=\beta=3$, $k^{there} =1$ (to uplift on $S^5$),
$p^{there}=n_N$, $q^{there}=p=\frac{1}{3}(n_S-n_N)$. 
We have argued that $n_N$, $n_S$
are not divisible by $3$, which means that when $n_N, n_S$ are coprime
$p^{there}$ and $q^{there}$ are also coprime, so that as in \cite{Gauntlett:2006af},
the total space $Y_7$ is simply connected.
On the other hand when  $n_N, n_S$ are non-coprime then $p^{there}$ and $q^{there}$
are not coprime and $Y_7$ is not simply connected.

Finally, we note that for the STU solutions with $p^1=p^2=p^3$ (and no restrictions on $m^I_{N,S}$), there
are no relevant hyperscalar modes. 
Indeed, from \eqref{cons2hsmodes}, we should take $v=0$ and $u=1+3r_S$,
but the latter is not consistent with the first condition in
\eqref{cons1hsmodes}.

\section{Non-trivial hyperscalars}\label{appa}

We are interested in obtaining the constraints that are needed to ensure that the complex hyperscalar is smooth, in the orbifold sense. 
The case when it is non-vanishing at both poles was discussed 
in \cite{Arav:2022lzo} (for the coprime case). Here we allow for the hyperscalar to vanish at
one or both of the poles.
We are especially interested in solutions that solve the BPS equations and this imposes
additional constraints over and above regularity.

The complex hyperscalar is charged with respect to the $U(1)_B$ gauge field $A_B\equiv \zeta_I A^I$. 
We follow the discussion and notation of \cite{Ferrero:2021etw}, where more details can be found. 
We consider two patches on the spindle, covering the $N$ and $S$ poles as in the previous appendix.
We want the
complex scalar field, which we write as $\rho e^{i\theta}$,
 to be a smooth section of a line bundle with $A_B$ a connection one-form on this bundle.
We assume that $\theta =\bar \theta z$, for constant $\bar \theta$. In the two patches we take (with $\rho>0$)
\begin{align}\label{degens}
\rho e^{i\theta}|_{\text{$N$ patch}}&=\rho(y)e^{iQ_Nz},\qquad \rho\sim (y-y_N)^{\alpha_N}\,,\quad y\to y_N\,,\nn\\
\rho e^{i\theta}|_{\text{$S$ patch}}&=\rho (y)e^{iQ_Sz},\qquad \rho\sim (y_S-y)^{\alpha_S}\,,\quad y\to y_S\,,
\end{align}
with $\alpha_{N,S}{\ge 0}$ and 
recalling that the spindle is behaving at the poles like $dy^2+\frac{1}{n_{N,S}^2}(y-y_{N,S})^2 dz^2$ with $\Delta z=2\pi$, 
smoothness
implies that we should take $\alpha_{N,S}\in\mathbb{Z}_{\ge 0}$.
Notice that $D\theta=d\theta-A_B$ behaves at the poles as follows:
\begin{align}
D\theta|_N=(Q_N-\frac{m_N}{n_N})dz\,,\qquad
D\theta|_S=(Q_S-\frac{m_S}{n_S})dz\,,
\end{align}
in the gauge for $A_B$ used in \eqref{appendixamnrels}.
The two patches are patched together at the equator of the spindle using the $U(1)_B$ gauge transformation $e^{i \gamma z}$ leading to 
\begin{align}\label{QNQSgam}
Q_S=Q_N-\gamma\,,
\end{align}
(associated with \eqref{gtgfsapb} so that $D\theta$ is gauge invariant). Using \eqref{infbrflconapb} this is equivalent
to
\begin{align}\label{qpmexp1}
(Q_S-\frac{m_S}{n_S})=(Q_N-\frac{m_N}{n_N})-\frac{p_B}{n_Nn_S}\,.
\end{align}

We now need to determine the condition for regularity of the complex scalar at the poles. 
To do so we should use the coordinates for the $U(1)$ bundle as given in
footnote \ref{ftnoteappseif}. 
As discussed in
\cite{Ferrero:2021etw,Arav:2022lzo} we consider smooth functions $F$ on the total space of the 
bundle with definite unit charge ${\mathtt{r}=1}$ under $\partial_\psi$ and  then
in its dependence on $\theta,z$ it gives a section of $L^{\mathtt{r}=1}$. In the north pole patch
we find
\begin{align}\label{appexpcapf1}
F\to e^{i\mathtt{r}\chi_N}(y-y_N)^{\alpha_N}e^{i(Q_N-\frac{\mathtt{r}m_N}{n_N})n_N\hat \phi}\,,
\end{align}
with $\mathtt{r}=1$.
On the covering space of the orbibundle we have
$\Delta\chi_N=\Delta \hat\phi=2\pi$, and then the orbibundle is obtained by
making the $\mathbb{Z}_{n_N}$ identification
$(\chi_N, \hat\phi)\sim (\chi_N+\frac{2\pi m_N}{n_N}, \hat\phi+  \frac{2\pi}{n_N})$ on the covering space. We can proceed similarly in the south pole patch.
This analysis shows that we must have $Q_{N,S}\in \mathbb{Z}$ to be consistent with the orbifold identifications (a point not emphasised in \cite{Arav:2022lzo})
and we note that this is consistent with \eqref{QNQSgam}. We can thus write 
$n_NQ_N-m_N\equiv \sigma_N{\hm_N}$ and
$n_SQ_S-m_S\equiv \sigma_S{\hm_S}$ 
where $r_{N,S}\in\mathbb{Z}_{\ge 0}$ and $\sigma_{N,S}$ are signs, with
\begin{align}\label{kpmcond1}
D\theta|_N=\sigma_N\frac{\hm_N}{n_N}dz\,,\qquad
D\theta|_S=\sigma_S\frac{\hm_S}{n_S}dz\,.
\end{align}
Furthermore, with the degeneration
of $\rho$ given in \eqref{degens}, we also conclude from \eqref{appexpcapf1} that regularity implies 
 \begin{align}\label{regbd}
r_N\le\alpha_N\,,\qquad 
r_S\le\alpha_S\,.
\end{align}
Notice also from \eqref{qpmexp1} that we deduce the flux $p_B$ can be expressed as
\begin{align}\label{fluxkpmcon}
\frac{p_B}{n_Sn_N}=\sigma_N\frac{\hm_N}{n_N}-\sigma_S\frac{\hm_S}{n_S}\,.
\end{align}
Observe that if we set $\hm_{N,S}=0$ then 
\eqref{kpmcond1}-\eqref{fluxkpmcon} are exactly the conditions obtained in \cite{Arav:2022lzo} when the complex scalar is non-vanishing at both poles of
the spindle.

In this paper we are interested in solutions that solve the BPS equations, which we have not yet used in this section.
For example, from (A.13) and (A.15) in\cite{BenettiGenolini:2024kyy}
 (correcting a sign in the second term) we have
\begin{align}\label{spmPpmeq}
S\dd\rho=\left(-e^{V}\zeta_I X^I (\xi\hook\vol)-P *D\theta\right)\sinh\rho\,.
\end{align}
We now evaluate this at the two poles. In the $S$ patch we have $\vol=\frac{1}{n_S}(y_S-y)dy \wedge dz$ and in the $N$ patch we have
$\vol=\frac{1}{n_N}(y-y_N) dy \wedge dz$.
If we now evaluate \eqref{spmPpmeq} at the poles with $S=1$, we find that only the second term on the right hand side is relevant and
we conclude that the bound in \eqref{regbd} is saturated
 \begin{align}\label{regbdsat}
r_N=\alpha_N\,,\qquad 
r_S=\alpha_S\,,
\end{align}
and also that there is a correlation of the signs $\sigma_{N,S}$ with the chirality of the spinor
at the poles:
\begin{align}
\sigma_N= -P_N\,,\qquad 
\sigma_S=P_S\,.
\end{align}
Notice then that \eqref{fluxkpmcon} reads
\begin{align}
p_B=-P_N({\hm_N}{n_S}\pm{\hm_S}{n_N})\,,
\end{align}
where the upper sign is for the twist and the lower sign is for the anti-twist.

\section{The case of $S^3$}\label{s3app}
To illustrate some aspects of the regularity of complex scalars that we discussed in the previous appendix,
we consider the special case of $U(1)$ orbibundles over a \emph{coprime} spindle $\Sigma(n_N, n_S)$
with flux equal to $\frac{1}{n_N n_S}$. The
total space is then $S^3$ and we can be very explicit about sections of line orbibundles over the spindle, which can be described by functions on $S^3$. 

We begin by improving some of the comments made in footnote 18 of \cite{Ferrero:2020twa}. Consider the metric
on a round $S^3$ given by
\begin{align}
ds^2= d\theta^2+\cos^2\theta d\phi_1^2+\sin^2\theta d\phi_2^2\,,
\end{align}
with $\theta\in [0,\pi/2]$ and $\Delta\phi_i=2\pi$. To see that this is a $U(1)$ orbibundle over $\Sigma(n_N,n_S)$ we consider
the weighted $U(1)$ action $V=n_N\partial_{\phi_1}+n_S\partial_{\phi_2}$. To do this we consider new coordinates
\begin{align}
\begin{pmatrix}
\phi_1\\\phi_2
\end{pmatrix}
=\begin{pmatrix}
n_N&m_N\\
n_S&m_S\\
\end{pmatrix}
\begin{pmatrix}
\psi\\ z
\end{pmatrix}\,,
\end{align}
with $m_N$, $m_S$ integers that are coprime to $n_N$, $n_S$, respectively.
We also demand that the coordinates are related by an $SL(2,\mathbb{Z})$ transformation with
\begin{align}\label{detcondnapp}
n_N m_S-n_Sm_N=1\,,
\end{align}
which exist from B\'ezout's lemma,
and so $\Delta\psi=\Delta{z}=2\pi$. In the new coordinates the metric can be written
\begin{align}
ds^2=\Lambda(d\psi+\mathcal{A} d{z})^2+d\theta^2+\frac{\cos^2\theta \sin^2\theta}{\Lambda}d{z}^2\,,
\end{align}
with $\Lambda=n_N^2\cos^2\theta+n_S^2\sin^2\theta$ and
 $\mathcal{A}=\frac{n_S}{n_N}\frac{\sin^2\theta}{\Lambda}
+\frac{m_N}{n_N}$. As $\theta\to 0$ we have
\begin{align}
ds^2\to n_N^2(d\psi+\frac{m_N}{n_N} d{z})^2+d\theta^2+\frac{1}{n_N^2}\theta^2d{z}^2\,,
\end{align}
while as  $\theta\to \pi/2$:
\begin{align}
ds^2\to n_S^2(d\psi+\frac{m_S}{n_S} d{z})^2+d\theta ^2+\frac{1}{n_S^2}(\theta-\pi/2)^2d{z}^2\,.
\end{align}

Clearly we have precisely the behaviour discussed in \cite{Ferrero:2021etw} and in appendix \ref{app:WCP}, 
but notice that the coordinate $\psi$ is valid in
the $N$ and the $S$ patch (that is we do not need to do a $U(1)$ gauge transformation to go from
the $N$ to the $S$ patch, so $\gamma=0$ in \eqref{condfluxsmooth2apb}). For example, at the north pole
we can introduce the coordinate $\phi={z}/n_N$. The covering space of this local patch would take coordinates $(\psi, \phi)$ with $\Delta\phi=2\pi$ and then the above local patch of the orbifold is obtained by
taking a $\mathbb{Z}_{n_N}$
quotient just acting on $\phi$ i.e. $\phi\to \phi+2\pi/n_N$.
In these coordinates though, the gauge field is singular. This is remedied by introducing new coordinates in this patch
obtained by the  $SL(2,\mathbb{Z})$ transformation $\chi=\psi+\frac{m_N}{n_N} {z}$, $\hat\phi=\phi$.
On the covering space of the patch we have
$\Delta\chi=\Delta\hat\phi=2\pi$ and the orbifold identification now acts on both coordinates via
\begin{align}\label{orbiid}
(\chi,\hat\phi)\sim (\chi+2\pi\frac{m_N}{n_N},\hat\phi+\frac{2\pi}{n_N})\,.
\end{align}
In these coordinates the metric then behaves
as $\theta\to 0$ like
\begin{align}
ds^2\to n_N^2(d\chi)^2+d\theta^2+\theta^2d\hat\phi^2\,.
\end{align}
This reveals that on the covering space of the $N$ patch, with $\Delta\chi=\Delta\hat\phi=2\pi$ ,
both the gauge field and the metric are regular.

Now lets consider a section $\zeta$ of a line bundle $L^\mathtt{r}$, with $A$ a connection one-form on $L$ and $\mathtt{r}$ is the charge of the scalar. As discussed in
\cite{Ferrero:2021etw,Arav:2022lzo} this is equivalent
to considering a complex function on $S^3$ with definite charge $\mathtt{r}$ under $\partial_\psi$ and then
in its dependence on $\theta,z$ it gives a section of $L^\mathtt{r}$. We are also interested in sections with
definite charge under $\partial_{z}$.
Let's explore this in more detail. 
Consider in the original coordinates functions on $S^3$ of the form
\begin{align}\label{s3perspective}
F=f(\theta)e^{ik_1\phi_1+ik_2\phi_2}\,,
\end{align}
with $k_i\in\mathbb{Z}$; regularity at the poles is discussed below. 
This can also be written as
\begin{align}
F=e^{i\mathtt{r}\psi}\zeta\,,\qquad \zeta=f(\theta)e^{iQ_N{z}}\,,
\end{align}
with $\zeta$ a section of $L^\mathtt{r}$, with
\begin{align}\label{rQvals}
\mathtt{r}=k_1n_N+k_2 n_S\,,\qquad  Q_N=k_1m_N+k_2 m_S\,,
\end{align}
and implicitly in this setup we have $Q_S=Q_N$ as the coordinates
$\psi,{z}$ are used in both the $N$ and the $S$ patch, as noted above. We can also write this in
the $(\chi,\hat \phi)$ coordinates as
\begin{align}\label{appexpcapf}
F=e^{i\mathtt{r}\chi}f(\theta)e^{i(Q_N-\frac{\mathtt{r}m_N}{n_N})n_N\hat \phi}\,.
\end{align}
For this to be consistent with the orbifold identification \eqref{orbiid} we need $\mathtt{r}\in \mathbb{Z}$ {\it as well as} $Q_N\in \mathbb{Z}$, consistent with \eqref{rQvals}.

Let's consider what happens if we demand that at the $N$ pole $f\ne 0$. In the $\chi,\hat\phi$ coordinates on the covering space of this patch,
the gauge field is regular at the $N$ pole and so we should demand, as in \cite{Arav:2022lzo}, that 
\begin{align}\label{appeqncond}
Q_N-\frac{\mathtt{r}m_N}{n_N}=0\,,
\end{align}
since $\hat\phi$ is not well defined at $\theta=0$. Let us see how this compares with the more
straightforward global analysis associated with functions on $S^3$ as in 
\eqref{s3perspective}. Notice that using \eqref{rQvals} and \eqref{detcondnapp} we have
$Q_N-\frac{\mathtt{r}m_N}{n_N}=k_2/n_N$ and so \eqref{appeqncond} is equivalent to $k_2=0$.
On the other hand, at $\theta=0$ the coordinate $\phi_2$ is not well defined and so from \eqref{s3perspective}
we should indeed demand that $k_2=0$.  One can highlight that since $m_N$ is coprime to $n_N$ (in order for the $U(1)$ orbibundle over the coprime spindle
to be smooth) then the condition that $Q_N\in\mathbb{Z}$ can only be achieved if $\mathtt{r}$ is an integer multiple of $n_N$. That is, only certain values of the charge $\mathtt{r}$ of the scalar field are compatible with having $f\ne0$ at the $N$ pole: for example
if $n_N\ne 1$ then it is not possible to have $\mathtt{r}=1$, which is also very clear from \eqref{rQvals}.
 
 A similar story unfolds if we demand that at the $S$ pole $f\ne 0$. Then with $Q_N=Q_S$
 we should have
 \begin{align}\label{appeqncond2}
Q_N-\frac{\mathtt{r}m_S}{n_S}=0\,.
\end{align}
 As above, this is again consistent with considering regularity of functions as in
 \eqref{s3perspective}. Notice, then, that it is not possible to have $f\ne 0$ at both poles unless
 $\mathtt{r}=0=Q_N=Q_S=0$. To
 see this, with $Q_N=Q_S$  
 we would need both \eqref{appeqncond} and \eqref{appeqncond2}, so we would require
 $\frac{\mathtt{r}m_S}{n_S}=\frac{\mathtt{r}m_N}{n_N}$ but this contradicts \eqref{detcondnapp} unless $\mathtt{r}=0$.
 Equivalently, we must have $k_1=k_2=0$ in \eqref{s3perspective}.

We next consider what happens if we demand that at the $N$ pole $f\to \theta^\alpha$, $\alpha>0$. In the  $\phi_1,\phi_2$ coordinates in \eqref{s3perspective} we can write 
\begin{align}\label{s3perspective2}
F\to \theta^{\alpha -|k_2|}[\theta e^{ i\text{sign}(k_2)\phi_2}]^{|k_2|}e^{ik_1\phi_1}\,,
\end{align}
and so regularity requires that $|k_2|\le \alpha$ or equivalently
 \begin{align}
|Q_N-\frac{\mathtt{r}m_N}{n_N}|\le \frac{\alpha}{n_N}\,,
\end{align}
so that as $\theta\to 0$
\begin{align}\label{appexpcapf2}
F\to e^{i\mathtt{r}\chi}\theta^\alpha e^{i(Q_N-\frac{\mathtt{r}m_N}{n_N})n_N\hat \phi}\,,
\end{align}
is regular. In the special case that this regularity bound is saturated, we have
 \begin{align}
F\to e^{i\mathtt{r}\chi}(\theta e^{\pm i\hat \phi})^\alpha\,.
\end{align}
In this case we see that the section $\zeta$ depends
holomorphically (or anti-holomorphically) on a natural complex coordinate on the spindle
at the $N$ pole. In the solutions of interest in this paper, the BPS equations impose
this extra condition with, moreover, the holomorphicity/anti-holomorphicity correlated
with the chirality of the spinor.

\section{Some plots for solutions}\label{plotssols}

We have numerically constructed various examples of $AdS_3\times \Sigma$ hyperscalar solutions
in the anti-twist class (we have found none in the twist class).
The system of BPS equations that we need to solve are given in \eqref{bulkbps}, \eqref{bpsconsts} with the metric function
$h$
given by \eqref{eq:h_int_solnt} and the gauge fields determined from \eqref{expsforFs}. We work in conformal gauge \eqref{confgauge}, $f=e^V$.
Recall, from below \eqref{twoconschgesag}, for given spindle data $n_{N,S}$, $(-1)^{t_{N,S}}$, $\hm_{N,S}$ and freely specified flux $p_F$, we have seven algebraic constraints which can be used to determine $x^I$ and $k$ at the poles 
and hence, in particular, the value of the scalars $\varphi_1,\varphi_2$ as well $e^V$ at the poles. 
Specifically,  we have two BPS constraints at each pole, \eqref{eq:m_pole_relation}, \eqref{eq:scI_sum},
two from conserved quantities, \eqref{twoconschgesag} and one from the expression for $p_F$ in
\eqref{eq:pi_xitf}.

We consider solutions which have the hyperscalar non-vanishing at one of the poles which,
without loss of generality, we take to be the $N$ pole (i.e. $r_N=0$). For the anti-twist class this covers all
cases in which the algebraic constraints have a solution (recall the comment below \eqref{chrisrel}; we work with
$\kappa=+1$, $t_N=1$, $t_S=0$). 
The only boundary condition that is left unspecified is for $\rho$. We then specify a trial value of the scalar $\rho$ at the $N$ pole, $\rho_N$, and then integrate the BPS equations. 
We find that this can lead to solutions that become singular
and also to solutions which approach a $S$ pole (i.e. $h$ vanishes for a finite value
of the $y$ coordinate).  To ensure that we obtain a \emph{bona fide} spindle solution, we need to demand that such a solution satisfies the algebraic values of $\varphi_1,\varphi_2, e^V$ at the $S$ pole, for
quantised values of $n_S$, $r_S$, $p_F$. In practise, we do this by varying $\rho_N$ and searching for solutions with matched values of $\varphi_1$.

\begin{figure}[h!]
	\centering
	\includegraphics[scale=0.45]{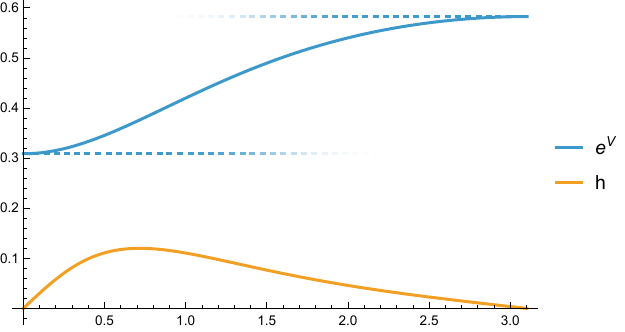}~
	\includegraphics[scale=0.45]{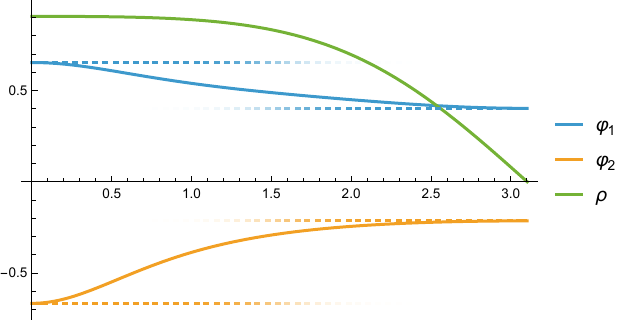}~
	\includegraphics[scale=0.45]{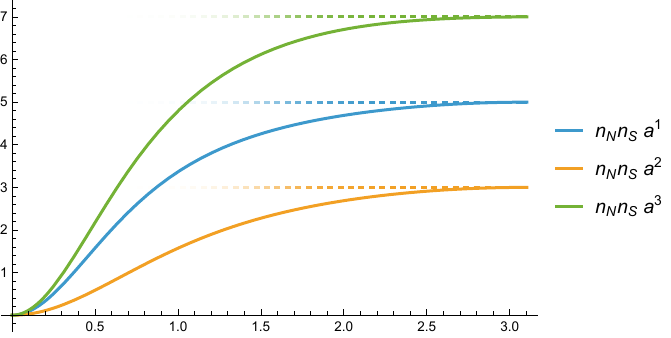}
	\caption{Metric, scalar functions and gauge fields for the solution with $(n_N, n_S)=(1,16)$ in table \ref{table3}.}
	\label{solnexample1}
	\end{figure}

\begin{figure}[h!]	
 \centering
		\includegraphics[scale=0.45]{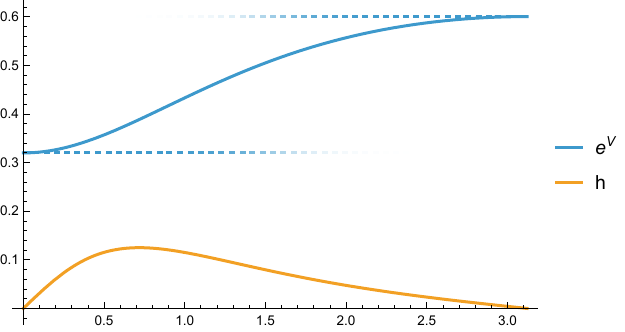}~
	\includegraphics[scale=0.45]{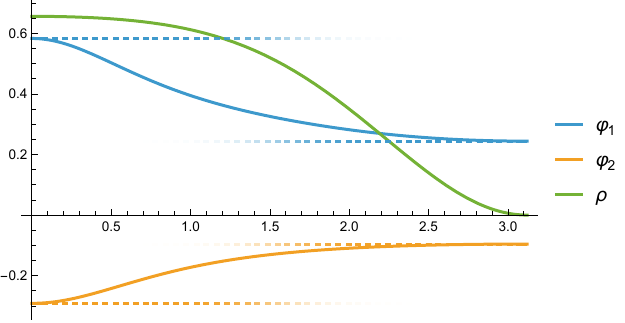}~
	\includegraphics[scale=0.45]{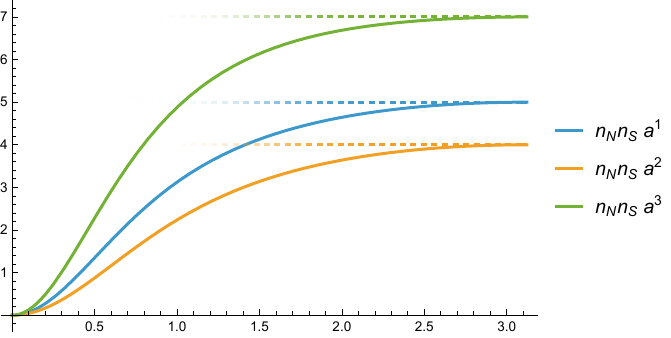}
	\caption{Metric, scalar functions and gauge fields for the solution with $(n_N, n_S)=(1,17)$ in table \ref{table3}.}
	\label{solnexample2}
\end{figure}

For all of the examples summarised in table \ref{table1}-\ref{table6} (and more), we have solved
the BPS equations numerically in the above fashion. Here we illustrate by plotting the metric functions
$e^V,h$, the scalars $\varphi_1,\varphi_2$ and $\rho$, as well as the three gauge fields $a^I$, 
for two representative examples in figures \ref{solnexample1} and \ref{solnexample2}. The $N$ pole is taken to be located at $y=0$ and the location of the
$S$ pole is determined numerically. Recalling \eqref{eq:h_int_solnt}, \eqref{confgaugemetric}, \eqref{firstbcsec}
the behaviour of the function $h$ is consistent with the values of $(n_N,n_S)$. 
The dashed lines in the plots associated 
with the metric function $e^V$ as well as the scalars $\varphi_1,\varphi_2$ are those
determined algebraically in studying the BPS equations. The gauge fields are plotted in a gauge
where $a^I=0$ at the $N$ pole; this then allows us to easily compare the behaviour
at the $S$ pole with the fluxes $p^I$, as indicated by dashed lines in the plots.
One can easily move to the gauge used in the text and appendices when discussing smooth
$S^1$ orbibundles, by suitably adding discrete fluxes at the poles and then gluing
the gauge fields at the equator of the spindle with a $U(1)$ gauge transformation
(recall \eqref{appendixamnrels}, \eqref{gtgfsapb}).

The case $(n_N,n_S)=(1,16)$ with $p^I=(5,3,7)$ of table \ref{table3}
is presented in figure \ref{solnexample1}. 
This hyperscalar $AdS_3$ solution has $r_N=0$ and $r_S=1$; we see in the figure
that $\rho$ vanishes at the $S$ pole linearly in $y$. We find that the non-zero value
for $\rho$ at the $N$ pole is given by $\rho_N\sim 0.90662 $.

The case $(n_N,n_S)=(1,17)$ with $p^I=(5,4,7)$ of table \ref{table3}
is presented in figure \ref{solnexample2}.
This hyperscalar $AdS_3$ solution has $r_N=0$ and $r_S=2$; we see in the figure
that $\rho$ vanishes at the $S$ pole quadratically in $y$. We find that the non-zero value
for $\rho$ at the $N$ pole is given by $\rho_N\sim 0.65631$.


\providecommand{\href}[2]{#2}\begingroup\raggedright\endgroup

\end{document}